\documentclass[format=sigconf]{acmart}

\usepackage{subfigure}
\usepackage{graphicx}

\usepackage{url}

\usepackage{verbatim}
\usepackage{color}
\usepackage{multirow}
\usepackage{rotating} 
\usepackage{amsmath}
\usepackage{amssymb}
\usepackage{gensymb}

\usepackage[ruled,noend,linesnumbered]{algorithm2e}

\newcommand{\commentout}[1]{}

\usepackage[normalem]{ulem}
\newcommand{\revrev}[2]{#2}

\newcommand{\revtwo}[2]{#2}

\newcommand{\revthree}[2]{#2}

\newcommand{\revtomm}[2]{#2}

\renewcommand\footnotetextcopyrightpermission[1]{}
\begin{document}

\title{The Prefetch Aggressiveness Tradeoff in 360$\degree$ Video Streaming}\titlenote{This paper is an extended version of
  our original {\em ACM MMSys} 2018 paper~\cite{AAK+18}.  It is posted here by permission of ACM for your personal use, not for redistribution.
  Please cite our original paper (with the same title) published in {\em ACM Multimedia Systems (MMSys)} '18,
  Amsterdam, Netherlands, June 2018. \url{https://dx.doi.org/10.1145/3204949.3204970}}

\author[Almquist]{Mathias Almquist}\affiliation{ \institution{Link\"oping University, Sweden}}
\author[Almquist]{Viktor Almquist}\affiliation{ \institution{Link\"oping University, Sweden}}
\author[Krishnamoorthi]{Vengatanathan Krishnamoorthi}\affiliation{ \institution{Link\"oping University, Sweden}}
\author[Carlsson]{Niklas Carlsson}\affiliation{ \institution{Link\"oping University, Sweden}}
\author[Eager]{Derek Eager}\affiliation{ \institution{University of Saskatchewan, Canada}}

\begin{abstract}

With 360$\degree$ video, only a limited fraction of the full view is displayed at each point in time. This has prompted
the design of streaming delivery techniques that allow alternative playback qualities to be delivered for each candidate viewing direction.
However, while prefetching based on the user's expected viewing direction is best done close to playback deadlines, large buffers
are needed to protect against shortfalls in future available bandwidth.  This results in conflicting goals and
an important
prefetch aggressiveness tradeoff problem regarding how far ahead in time from the current playpoint prefetching should be done.
This paper presents the first characterization of this tradeoff.
The main contributions include
an empirical characterization of head movement behavior based on
data from viewing sessions of four different categories
of 360$\degree$ video, an optimization-based comparison of the prefetch aggressiveness tradeoffs seen for these video
categories, and a data-driven discussion of further optimizations,
which include a novel system design that allows both
tradeoff objectives
to be targeted simultaneously.
By qualitatively and quantitatively analyzing the above tradeoffs,
we provide insights into how to best design tomorrow's
delivery systems for 360$\degree$ videos,
allowing content providers to reduce bandwidth costs and improve users' playback experiences.
  
\end{abstract}



\maketitle

\section{Introduction}

Interactive video streaming is becoming increasingly popular, with 360$\degree$ video leading the way.
For example, since 2015 both YouTube
\revthree{(Mar. 2015\footnote{https://youtube-creators.googleblog.com/2015/03/a-new-way-to-see-and-share-your-world.html})}{(Mar. '15)}
and Facebook
\revthree{(Sep. 2015\footnote{\url{https://newsroom.fb.com/news/2015/09/introducing-360-video-on-facebook/}})}{(Sep. '15)}
offer a rapidly growing selection of 360$\degree$
videos that can be viewed in the browser on PCs, on smartphones, on tablets,
or with head mounted displays (HMDs).
360$\degree$ videos are typically recorded using an
omnidirectional camera that captures every direction or by a collection of cameras whose video
streams are stitched together into a single video~\cite{Naya97}.  When viewing these videos, users can freely
chose to look in any viewing direction (e.g., by moving their head while wearing an HMD).
The flexibility in the users' choice of view (or region of interest)
provides users an enriched
viewing experience as they can explore a scene
similar to as if they were at the location of the filming.  However, this flexibility comes at the cost of significant bandwidth
consumption
\revthree{for content providers}{when}
delivering these services over the internet.

360$\degree$ videos are typically significantly larger than regular videos and can therefore consume
a lot of bandwidth.
However, similar to in everyday life, users have
a limited field of view, resulting in only a small fraction (e.g., 20-30\%) of the video data being needed
for the actual viewing.
Recently, this observation has prompted research into delivery techniques that allow
alternative playback qualities to be delivered for each candidate viewing direction~\cite{CSDC17,vengat179,LFL+17}.

Clearly, delivery solutions that ignore the users' current field of view are likely to waste a lot of bandwidth
delivering data that corresponds to scene data outside of this field of view or delivering higher
than necessary data quality for scenes at the periphery of the field of view.
In the ideal case,
a content provider
would be able to
perfectly predict both the
\revtwo{head-movements}{head movements}
and future bandwidths so
that it could deliver only the data that the user views, at the highest possible quality allowed by the
time-varying available bandwidth.
Unfortunately, neither head-movement prediction nor bandwidth prediction is perfect.  Content providers
wanting to deliver these services effectively over the internet therefore face the following two challenges.

First, content providers must take into account the uncertainty in the
user viewing directions
and the impact that
changes in viewing direction
may have on the perceived playback quality.
Second, as with regular streaming, video data must be buffered at the clients so to protect
against playback stalls caused by (future) bandwidth variations.
Maintaining a reasonably large playback buffer is particularly
important when using HTTP-based Adaptive Streaming (HAS) solutions such as
those used by
the most popular international streaming services (e.g., YouTube, Facebook, Apple, Netflix)
and most regional streaming providers, and that are currently being standardized though MPEG-DASH.
Such services typically download 2-5 second chunks using HTTP(S) and try to maintain
significant buffers (e.g., 10-120 seconds, depending on service and type of network device)
to account for
today's networks being best-effort with
significant bandwidth variations being common.

While much work has been done to study quality adaptation algorithms for regular non-360$\degree$ videos~\cite{SES+15},
for which the user viewing direction is fixed,
to the best of our knowledge,
no prior work has considered the
problem of
prefetching aggressiveness
(i.e., how far ahead in time from the current play point should prefetching be done),
for 360$\degree$ videos where the user viewing direction may vary. 
In this work we present the first such study.
This is an important problem for wide-area delivery techniques,
since here the need to protect against shortfalls in future available bandwidth and
the need to quickly respond to changes in user viewing direction result in conflicting goals.
To see this, note that
the need to protect against future bandwidth drops motivates building up a large buffer
(i.e., aggressively prefetching well-ahead of the current play point).  On the other hand,
prefetched data can  be worthwhile only if the data will be within the user's field of view.
Use of a small prefetch buffer is motivated by the observation that prediction of the user
viewing direction is most accurate over short time scales.

\begin{figure*}[t]
  \centering
  \subfigure[$T$=2s]{
    \includegraphics[trim = 0mm 2mm 0mm 2mm, clip, width=0.48\textwidth]{{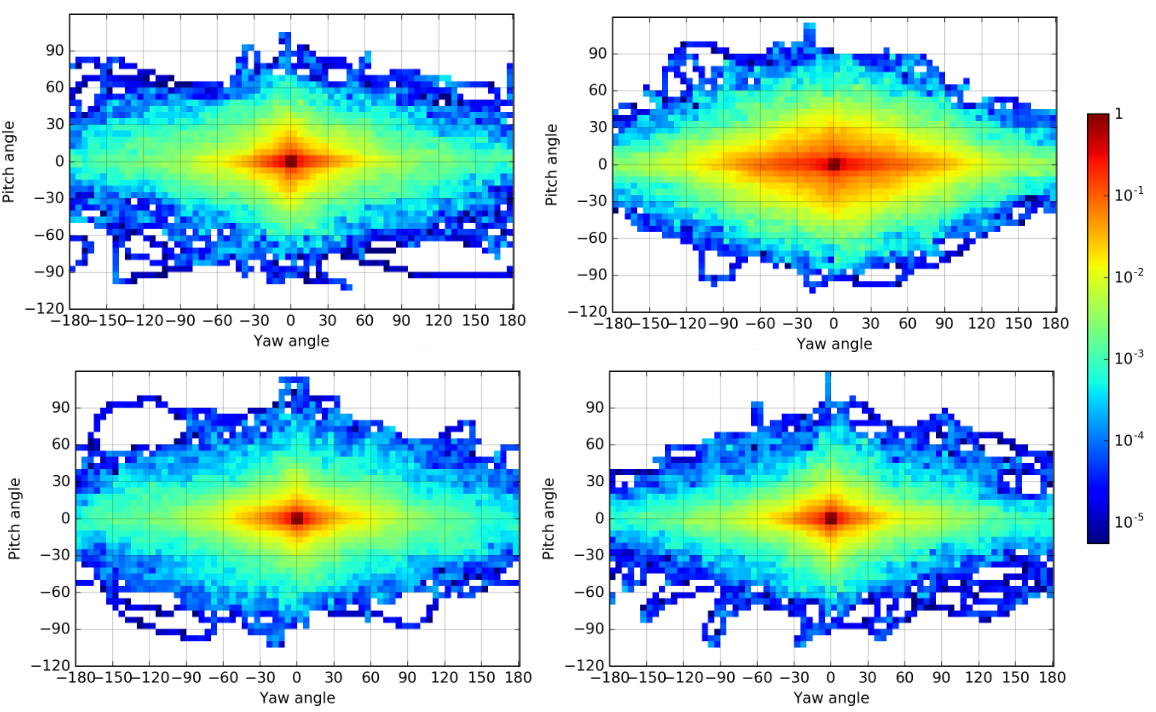}}}
  \hspace{-14pt}
  \subfigure[$T$=20s]{
    \includegraphics[trim = 0mm 2mm 0mm 2mm, clip, width=0.48\textwidth]{{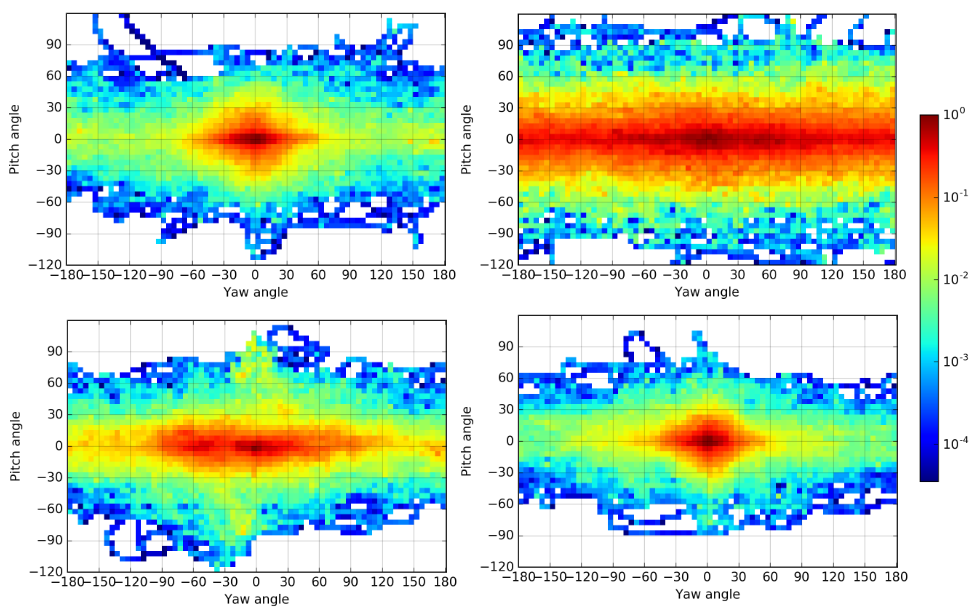}}}
  \hspace{-10pt}
  \vspace{-10pt}
  \caption{Example heat-maps for four categories of videos. Within each subfigure:  Rides (top-let), Exploration (top-right), Moving focus (botom-let), Static focus (botom-right).}
  \label{fig:example-heatmaps}
  \vspace{-12pt}
\end{figure*}

Figure~\ref{fig:example-heatmaps} illustrates
the dependency between uncertainty in the user viewing direction, and the time scale, 
using
\revtomm{two example categories}{four categories}
of videos
(each category defined in Section 3).
For
\revrev{both}{all}
categories, we show the relative probability distributions of the change
in the viewing direction of a set of example users
viewing the same subset of videos, conditioned
on the time duration $T$ (in seconds),
where $T$=2s (left) and $T$=20s (right).
\revtwo{Here, the origin (in the figure)}{The origin in the figure}
corresponds to no change and all changes are measured relative to the
viewing direction the user had $T$ seconds earlier.
It is clear from these results
that the playback quality selection for each potential viewing direction
can be best optimized
when done very close to the playback of a frame, suggesting the use of small buffer margins.
This shows that there
is an important tradeoff between
the goal of making good prefetching decisions with respect to the quality that should be prefetched
for each direction, and the goal of prefetching far ahead in time,
so to protect against future bandwidth variations or other randomness causing stalls.

This paper presents a measurement-driven characterization of the
prefetching aggressiveness
tradeoffs associated with different categories of 360$\degree$ videos,
providing both qualitative and quantitative insights
regarding how best to address these tradeoffs.
In particular, we present a characterization of the user
behavior, as well as an optimization framework that captures the prefetching aggressiveness tradeoffs
experienced by a content provider wanting to
optimize the quality selection in each viewing direction so to maximize the user's
expected quality of experience (QoE).
The optimization framework takes
into account both the probability of
the user having a particular viewing direction at a particular point in time,
and the buffer levels needed to avoid playback interruptions.

Our optimization framework assumes the use of chunk-based streaming (e.g., using DASH or alternative
HAS-based formats) and takes into account
\revtwo{the type of conditional probabilities that we observe from the viewing behaviors}{the viewing behaviors we observe}
for different categories of 360$\degree$ videos.  Using measurements from
\revtwo{}{our}
user study and
our optimization framework,
we provide insights into the best possible
tradeoffs when delivering 360$\degree$ videos over the internet.
Our main contributions are as follows:

{\em 1) User-driven head-movement characterization:}
  We record the orientation and rotation velocity of an Oculus Rift HMD
  during playback of different categories of 360$\degree$ videos.  More than 21 hours of
  viewing data (based on the viewing of 32 participants)
  is analyzed to characterize viewing patterns
  relevant for optimizing the wide-area delivery of 360$\degree$ video.
  Significant differences in the viewing patterns between the different
  categories are observed.  However, for most video categories the head movements
  (e.g., as in Figure~\ref{fig:example-heatmaps})
  are sufficiently confined
  that, although viewing direction prediction is most accurate at short time scales,
  prefetching with optimized quality selection can also be beneficial at larger time scales. 

  {\em 2) Optimized buffer-quality tradeoffs:}
  We present an optimization framework that captures the optimal tradeoffs between
  the goal of prefetching far ahead in time 
  (to protect against bandwidth variations and stalls) and
  the goal of making the best quality selections for each potential viewing
  direction when prefetching so as to maximize the expected playback utility
  (as is dependent on the probability distribution of the user's viewing direction when a particular
  play time is reached, and the qualities of the prefetched chunks for this play time).
  Using this framework we then
  evaluate the optimized tradeoffs for different categories of videos, utility functions,
  and bandwidth conditions.
  Our evaluation highlights differences across video categories
  and the possible impact of
  how bandwidth constrained clients are.
 
  {\em 3) Discussion of further design optimizations:}
  Finally, based on our findings, we present an adaptive policy framework and describe additional
  optimizations that could be done to improve
  client performance for
  the different video categories.
  The framework allows clients to simultaneously protect against bandwidth variations
  and provide personalized and content-based quality adaptation, best optimized at shorter time scales.
  Using our measurements, we then discuss additional optimizations that leverage
  biases in head movements related to the current rate of change in the viewing direction,
  the relative viewing direction compared to that at the start of the video,
  and if the viewer is in an initial exploration phase (that we have observed) or not.

The remainder of the paper is organized as follows.
Section~\ref{sec:background} provides a brief introduction to 360$\degree$ video.
Our measurement methodology and dataset are described in Section~\ref{sec:measurements}.
Section~\ref{sec:characterization} presents a characterization of the observed head movements,
before Section~\ref{sec:tradeoff} presents our optimization framework and
characterizes the prefetch aggressiveness tradeoffs.
Motivated by these tradeoffs and observations,
Section~\ref{sec:discussion} then presents an adaptive framework
that allows both substantial prefetching well ahead of playback deadlines and fine grained quality adaptation
based on predicted viewing directions, followed by a data-driven characterization
of some further considerations that such a system could account for.
Finally, Section~\ref{sec:related} presents related work and Section~\ref{sec:conclusions} concludes the paper.

\section{Background}\label{sec:background}

360$\degree$ videos provide users with an interactive video experience,
in which the users can freely select any viewing direction within a spherical virtual environment.
Users can experience 360$\degree$ videos in several ways.
On a PC, the user typically controls the view using either the W-A-S-D keyboard buttons
or by clicking and dragging using the mouse.
On a smartphone or tablet,
the user can change the view by swiping the screen or
by changing the orientation of the phone/tablet.
Finally, with
virtual reality (VR) head mounted displays (HMDs),
the users simply move their heads in the same way as if they were at the location
where the video was recorded, creating a more immersive experience.

In this project, we use the first consumer version (CV1) of the Oculus Rift headset, released March 2016.
The Oculus Rift hardware system consists of a sensor, a headset, and a remote.  The headset works as a display
using OLED panels with a resolution of 1080x1200 per eye, resulting in 2160x1200 across the entire field of view.
Oculus offers a 110$\degree$ horizontal field of view.
The VR applications run on a PC which transfers video and audio to the headset via an HDMI cable.
To allow users to move in the virtual space, the sensor offers position tracking by monitoring infrared LEDs
that are embedded in the headset.
The remote
can be
used
for navigation,
but was not needed for our experiments.

The headset features a set of Micro-Electro-Mechanical System (MEMS) sensors, namely a magnetometer, gyroscope and an accelerometer
which are combined so as to track the orientation of the headset~\cite{LYKA14}.
The headset orientation is interpreted according to an internal virtual coordinate system,
transmitted from the headset to the PC at a rate of 1000 Hz.
This allows applications to accurately track a user's head movements and update the view accordingly.

\section{Measurement Methodology}\label{sec:measurements}

We next describe our measurement setup.

{\bf Physical setup:}
We used a dedicated PC with Intel Xeon CPU E5-1620 V4 3.50GHz,
32GB RAM, and a NVIDIA GeForce GTX 1080 graphics card.  The PC ran Windows 10
and was connected to the Oculus Rift CV1 headset and sensor via USB 3.0.
To deliver audio and visuals,
an HDMI cable was connected directly between the headset and the HDMI port of the dedicated graphic card of the PC.
Once connected, the sensor was placed on a table facing
towards
an open area in the room where
the user wearing the headset was placed on a turning chair approximately 1.5 meters away.

\begin{figure}[t]
\centering
\includegraphics[trim = 0mm 40mm 0mm 4mm, clip, width=0.32\textwidth]{{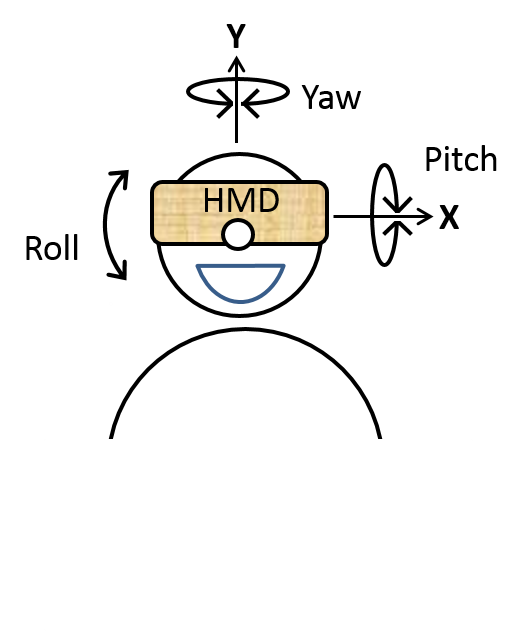}}
\vspace{-10pt}
\caption{Head-movement coordinates: Yaw, pitch, and roll.}
\label{fig:directions}
\vspace{-6pt}
\end{figure}

{\bf Sensor readings and viewing traces:}
Sensor readings were extracted using the Oculus Software Developer Kit (SDK) 1.8.0 and Oculus runtime 1.9.0 291603.
For this purpose, we developed a minimal application in C++ that
extracts the head orientation and rotation velocity from the MEMS sensors.
For easy interpretation,
we convert orientation readings (from quaternions) to yaw-pitch-roll format based on
the user's head orientation relative to that at
the start of each video and velocities to degrees per second.
The yaw-pitch-roll format is illustrated in Figure~\ref{fig:directions}.
Here, all directions are measured relative to the 0$\degree$ line.
Importantly,
to allow direct comparison of the viewing directions within a scene,
for yaw ($\pm 180\degree$), we set the 0$\degree$ line on a per-video basis.
In particular, rather than using the direction of the sensor (default) as zero line,
we instead use the head orientation that the user has at the start of the video.
The reason for this is that videos are played relative to this initial viewing direction.
This adjustment ensures that we record the same yaw angle
for two users looking at the same object at the same time instance within a video,
regardless of their original head positioning.
For pitch ($\pm 90\degree$), the 0$\degree$ line is parallel to the ground,
and for roll the zero value corresponds to holding the head straight.

{\bf Video playback and recording:}
Our application utilizes the Whirligig video player\footnote{\url{https://www.oculus.com/experiences/rift/1130182873666293/}}
to sequentially play a list of mp4 videos stored locally on the PC.
For each video viewing, the user's head orientation and movements are recorded and stored to trace files.

{\bf Video selection:}
We first identified
30 videos from YouTube's 360$\degree$ collection
that we found nicely represented different categories of 360$\degree$ videos.
We then used the trace collector application to download these videos in 4K resolution,
ensuring that we can offer good quality of experience during viewing.
The videos are 1-5 minutes long
(3 min on average) and divided into five categories.
\begin{itemize}

\item {\em Exploration:}
In these videos,
there is no particular object or direction of special interest
and the users are expected to explore the entire sphere throughout
the video duration.
Furthermore,
two
independent viewers
of the same video
are expected to have substantially different viewing angles at each point in time.
(Example: Camera positioned on top of a tall building overlooking a city.)

\item {\em Static focus:}
In these videos, the main focus of attention is deemed to always be at the same location in the video.
A static viewing behavior is expected since the
focus of attention
does not move.
With these videos,
most of the time a near-zero yaw angle is expected.
(Example: A theatre performance or a concert being displayed on a scene.)

\item {\em Moving focus:}
Story-driven videos where there is an object of special interest that is moving across
the 360$\degree$ sphere.  With these videos,
a high correlation
is expected between
the viewing angles of users over time,
since they typically would follow the objects of interest.
(Example: An action scene where the involved characters move around
the viewing sphere, causing the user to follow.)

\item {\em Rides:}
In these videos, the users take a virtual ride in which the camera
is moving forward at a high speed, making
users feel that they too are moving forward quickly.
In the majority of the video the user is expected to look forward,
as when taking a ride in real life.
(Example: Roller-coaster.)

\item {\em Miscellaneous:}
This
category includes videos that were
deemed to
have a mix of the characteristics of the other categories or
had a hard-to-classify ``unique feel'' to them.

\end{itemize}

The full set of videos and their categorization are summarized
\revtwo{}{in}
Table~\ref{tab:videos}.
We note that some of these categories are named in part based
on the expected viewing behavior of a user watching the video.
We believe that this allows for a natural categorization
that can be used on larger sets of videos.
Of course alternative
classifications are possible.
However, for the purpose of this study,
this categorization is sufficient
to characterize and evaluate how the differences in viewing behavior of
these diverse categories impact the best prefetching tradeoffs.

\begin{table*}
  \centering
  \caption{Summary of videos.  (To watch video, use URL of the form: \url{https://www.youtube.com/watch?v=VideoID}, where VideoID is replaced based on
    \revtwo{table.}{the table entries.}}
  \vspace{-10pt}
          {\small
            \begin{tabular}{|l|p{10.7cm}|}
              \hline
              Category & Video Name (Duration, VideoID) \\\hline
              Exploration &
              Zayed Road
              (3:00, \url{uZGrikvGen4}),
              Burj Khalifa
              (2:30, \url{bdq4H1CIehI}),
              Hadrain's Wall
              (3:36, \url{2zeKpeRZ8uA}),
              New York
              (1:59, \url{T3e-GqZ37uc}),
              White House
              (5:16, \url{98U2jdk8OGI}),
              Waldo
              (1:00, \url{hM9Tg_dQkxY}),
              Skyhub
              (4:00, \url{D9-i_F3xYhI})\\
              \hline
              Static &
              Christmas Scene
              (2:49, \url{4qLi-MnkxBY}),
              Boxing
              (3:29, \url{raKh0OIERew}),
              Elephants
              (2:49, \url{2bpICIClAIg}),
              Mongolia
              (1:52, \url{VuOfQzt2rI0}),
              Orange
              (2:43, \url{i29ITMfLVU0})\\
              \hline
              Moving &
              Christmas Story
              (4:14, \url{XiDRZfeL_hc}),
              Assassin's Creed
              (2:31, \url{a69EoIiYqoE}),
              Clash of Clans
              (1:23, \url{wczdECcwRw0}),
              Frog
              (3:13, \url{sk8hm7DXD5w}),
              Solar System
              (4:32, \url{ZnOTprOTHc8}),
              Invasion
              (4:04, \url{gPUDZPWhiiE})\\
              \hline
              Rides &
              F1
              (1:54, \url{2M0inetghnk}),
              Le Mans
              (3:00, \url{LD4XfM2TZ2k}),
              Roller Coaster
              (2:11, \url{LhfkK6nQSow}),
              Total War
              (1:49, \url{YSBWwnOHvM8}),
              Blue Angels
              (2:30, \url{H6SsB3JYqQg}),
              Ski
              (2:48, \url{kMCYo5rO6RY})\\
              \hline
              Misc. &
              Hockey
              (2:25, \url{8DKVvb17xsM}),
              Tennis
              (4:05, \url{U-_yX4e4Z_w}),
              Avenger
              (2:58, \url{3LSf6_ROCdY}),
              Trike Bike
              (3:14, \url{jU-pZSsYhDk}),
              Temple
              (4:36, \url{Lx14NDttRWo}),
              Cats
              (1:59, \url{0RtmVnD8_XM})\\
              \hline
          \end{tabular}}
          \label{tab:videos}
          \vspace{-6pt}
          \end{table*}

{\bf User study:}
An open invitation was sent out to different groups at the university,
allowing people to sign up for one-to-three 45-minute sessions (but at most one per day).
In total, 32 people signed up for a total of 45 sessions.
To avoid bias in the results and encourage viewers to follow their instincts,
at the start of each session,
users were not given any instructions on how to view a video,
but instead simply watched a four minute introduction video about
VR produced by Oculus.
This allows the user to get accustomed to the 360$\degree$ surrounding
and feel comfortable wearing the headset.
After the introduction,
the participants then view ten ``semi-random'' videos.  Videos for each session are
selected at random from the set of videos that the user has not viewed in the past
(so that users attending multiple sessions do not watch the same video twice)
and we make sure that all users watch one ``representative'' video from each category
(first video in each row of Table~\ref{tab:videos}).\footnote{By definition,
  it is impossible to choose a ``representative'' video for the Miscellaneous category.
  However, due to
  sports-related follow-up work,
  we selected a video (``Hockey'') in which viewers watch
  a hockey game from between the player benches.}
After the views to the representative videos had been accounted for,
the other videos got between 8-13 views each.
The additional views to the ``representative'' videos allows for more detailed analysis for these videos.
It should also be noted that some videos were avoided for users that
indicated that they had fear of heights or were prone to motion sickness
or dizziness (asked at the start of the session).
Finally, to avoid biases related to the order videos are played,
the order of the videos selected for a given session is randomized.

In total,
we recorded the head movements
from 439 unique viewings,
totaling 21 hours and 40 minutes. The age distribution of the 32 participants was:
20-29 (66\%), 30-39 (28\%), 40-49 (3\%), 50-59 (3\%).  56\%  of the participants
were male and 44\% female. Moreover, 25 participants had never tried VR
and only 3 had tried
it with Oculus.

No personal information is stored or included in our datasets.
For our analysis we only perform per-category and per-video analysis,
no per-user analysis, and only aggregate information is reported.
\revtomm{}{We present results for the four more well-defined categories
  {\em rides}, {\em exploration}, {\em moving focus}, and {\em static focus} (excluding the {\em miscellaneous} category).}
Finally, we note that we did not make any modifications to the videos
or otherwise try to effect the user experience.  The users simply
watched the videos as they otherwise would, while we used the API to
record their head movements.

\section{Category-based Characterization}\label{sec:characterization}

\subsection{Angular utilizations}

\begin{figure*}[t]
    \begin{minipage}[t]{0.54\textwidth}
    \vspace{1pt}
  \centering
  \includegraphics[trim = 0mm 2mm 0mm 0mm, clip, width=0.9\textwidth]{{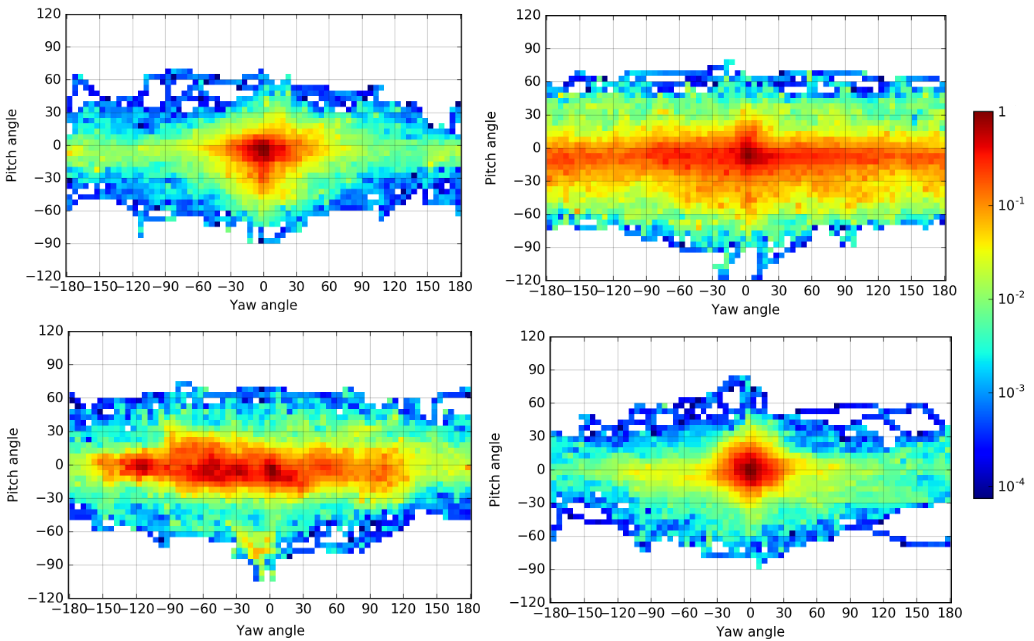}}
  \vspace{-10pt}
  \caption{Heatmap of most utilized yaw and pitch angles: Rides (top-left), Exploration (top-right),
    Moving focus (bottom-left), Static focus (bottom-right).}
  \label{fig:heat-duration}
  \vspace{-6pt}
\end{minipage}
\hfill
\begin{minipage}[t]{0.44\textwidth}   
  \centering
  \subfigure[Rides]{
    \includegraphics[trim = 8mm 2mm 8mm 0mm, width=0.48\textwidth]{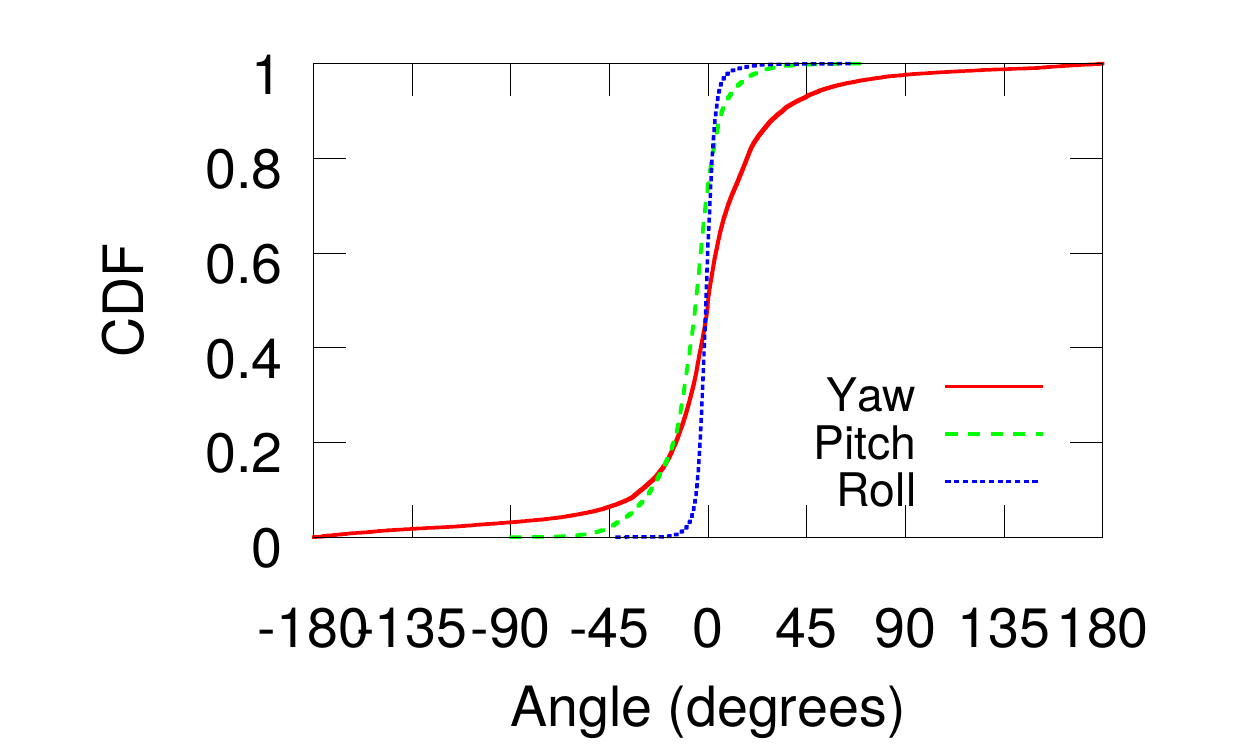}}
  \hspace{-10pt}
  \subfigure[Exploration]{
    \includegraphics[trim = 8mm 2mm 8mm 0mm, width=0.48\textwidth]{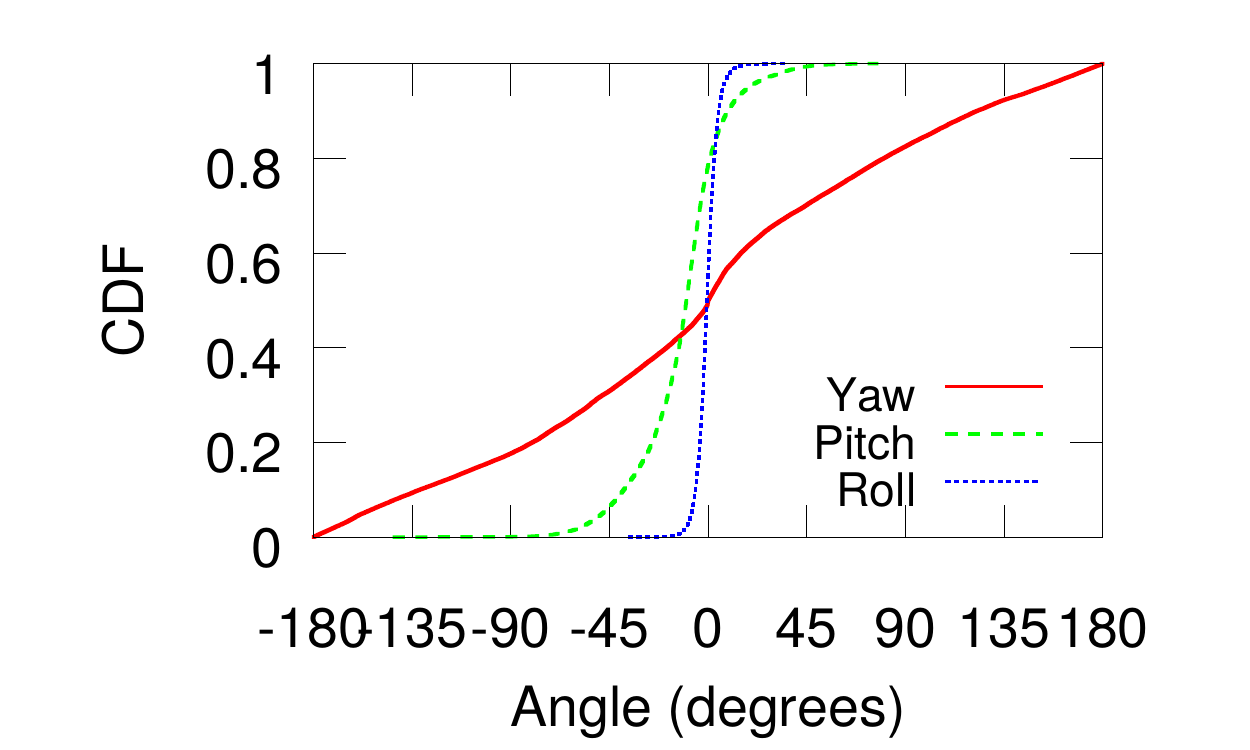}}\\
  \vspace{-10pt}
  \subfigure[Moving focus]{
    \includegraphics[trim = 8mm 2mm 8mm 0mm, width=0.48\textwidth]{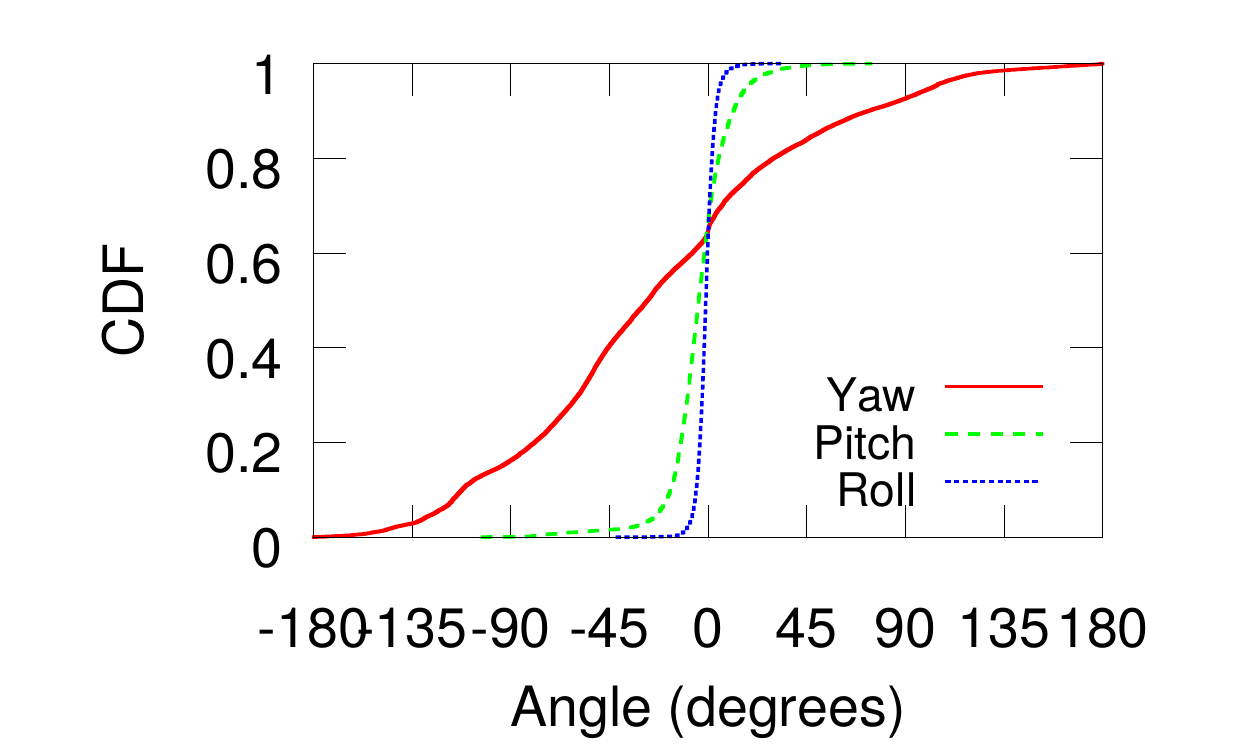}}
  \hspace{-10pt}
  \subfigure[Static focus]{
    \includegraphics[trim = 8mm 2mm 8mm 0mm, width=0.48\textwidth]{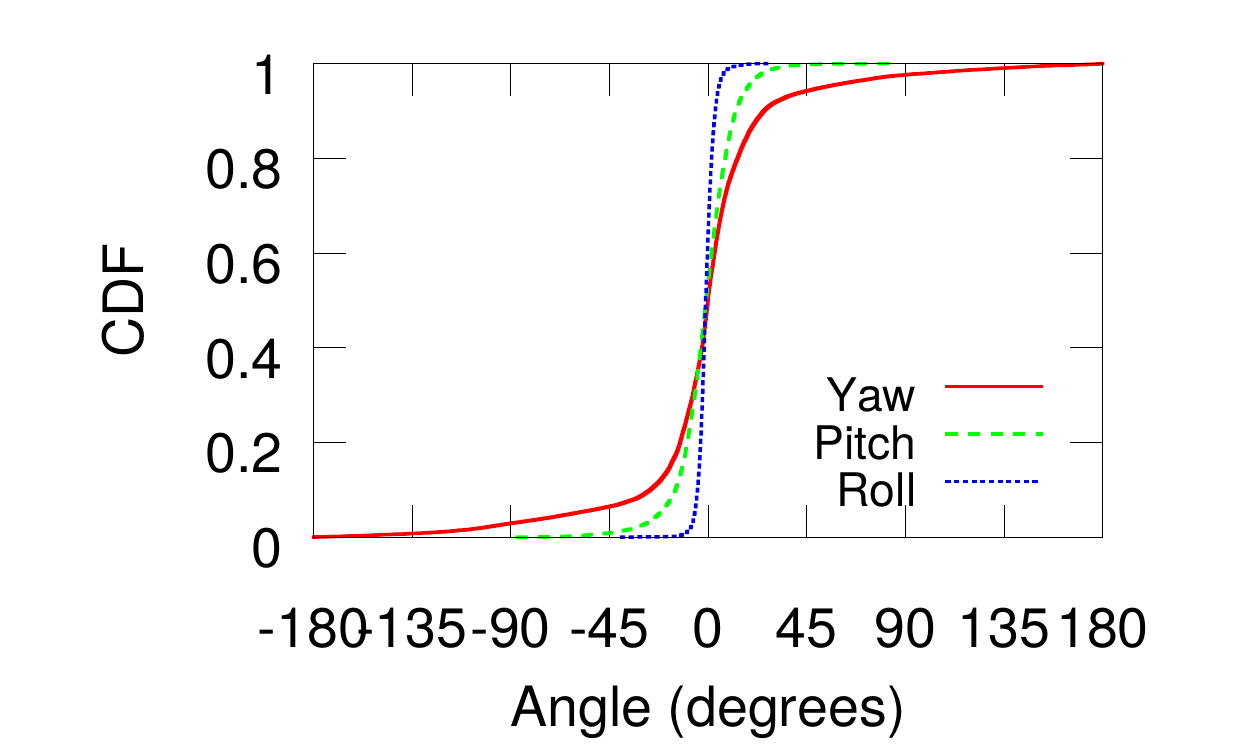}}
    \vspace{-12pt}
    \caption{CDFs of angle utilization.}
  \label{fig:angle-duration}
  \vspace{-6pt}
\end{minipage}
\end{figure*}

We begin by looking at how the viewing angles have been utilized for the videos
of the different categories.
Figure~\ref{fig:heat-duration} shows a heatmap of the most utilized yaw and pitch angles
over the full duration of the videos.  A quantification of the angular utilizations
(also including the roll) is provided in Figure~\ref{fig:angle-duration}.
Here, for each category,
we show the cumulative distribution function (CDF)
of the observed angles for
yaw (red), pitch (green) and roll (blue).

We note that yaw is the most dominant orientation movement across all categories,
with angular utilizations much more widely distributed than
for pitch and (especially) roll.
This suggests that predicting and accounting for changes in yaw,
is most important when trying to adapt the prefetch quality based on
\revtwo{users'}{the}
expected viewing direction by the time a prefetched frame is played.

In general,
for the videos that we selected,
the distributions are relatively symmetric.
Except for the moving focus category (bottom-left),
for which there is a slight
bias towards the left,
we do not see any major biases between
leftwards or rightwards utilizations.
The slight bias observed for the moving focus videos
is likely due to the particular choice of videos.
Furthermore, except for the exploration videos,
there is only a very small bias to look downwards rather than upwards.
For the exploration videos this bias is
reinforced
by videos (taken in Dubai) where the viewer is positioned at the top of a very large
building or when flying above the city.
For roll,
we observe only small non-biased changes over the video playback durations
(e.g., 98\% within $\pm 10\degree$).

Finally, and perhaps most importantly,
both Figures~\ref{fig:heat-duration} and~\ref{fig:angle-duration}
clearly show that there are significant differences
in the angular
\revtwo{utilizations themselves}{utilizations}
between the different categories.
For example, if we focus on the (dominant) yaw-angle distributions,
we observe substantially more evenly spread angular utilization with the exploration (top-right) videos
than with the rides (top-left) and static focus (bottom-right) videos.
With both these latter categories,
users spend most of the time
in the original direction
(e.g., 80\% of the time within $\pm 30\degree$ and 90\% within $\pm 60\degree$)
in which the video playback was initiated.
These results show that for these categories of videos,
the original (or intended primary) viewing direction
can be used as a good predictor of what directions to prioritize
aggressively prefetching data for future playback.

\begin{figure}[t]
  \centering
  \subfigure[Rides]{
    \includegraphics[trim = 8mm 3mm 8mm 0mm, width=0.235\textwidth]{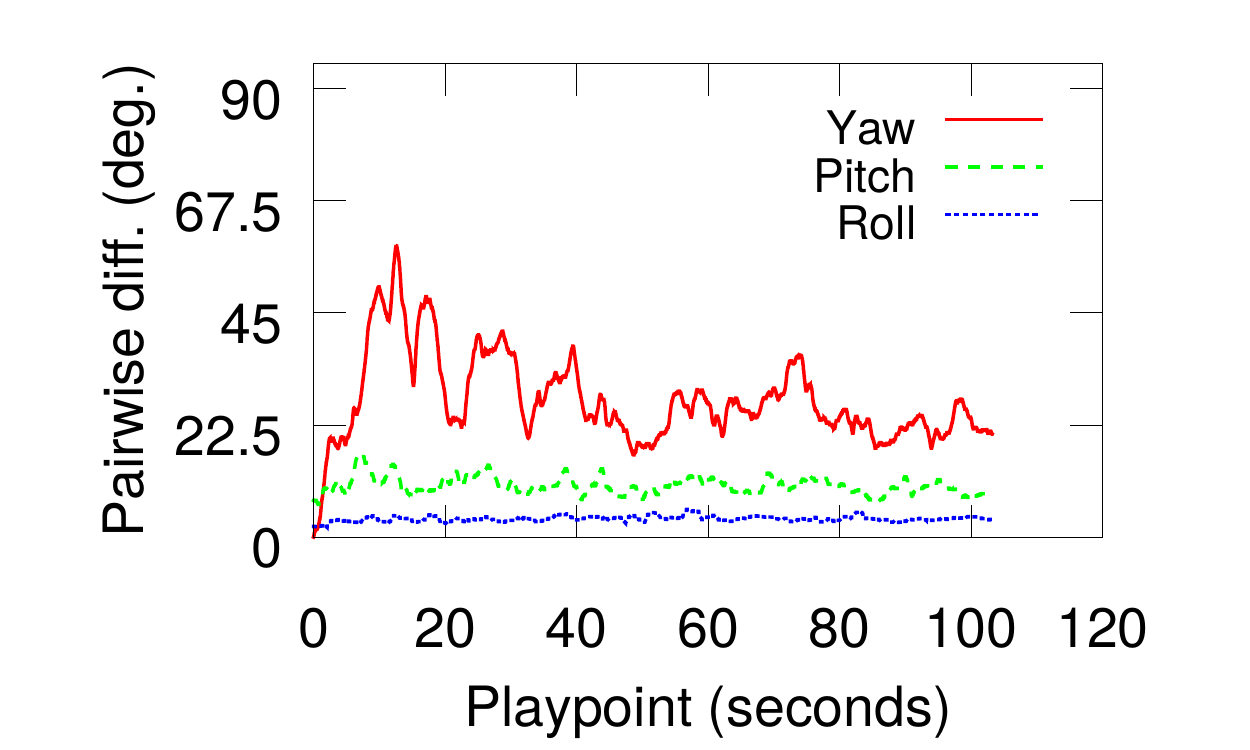}}
  \hspace{-10pt}
  \subfigure[Exploration]{
    \includegraphics[trim = 8mm 3mm 8mm 0mm, width=0.235\textwidth]{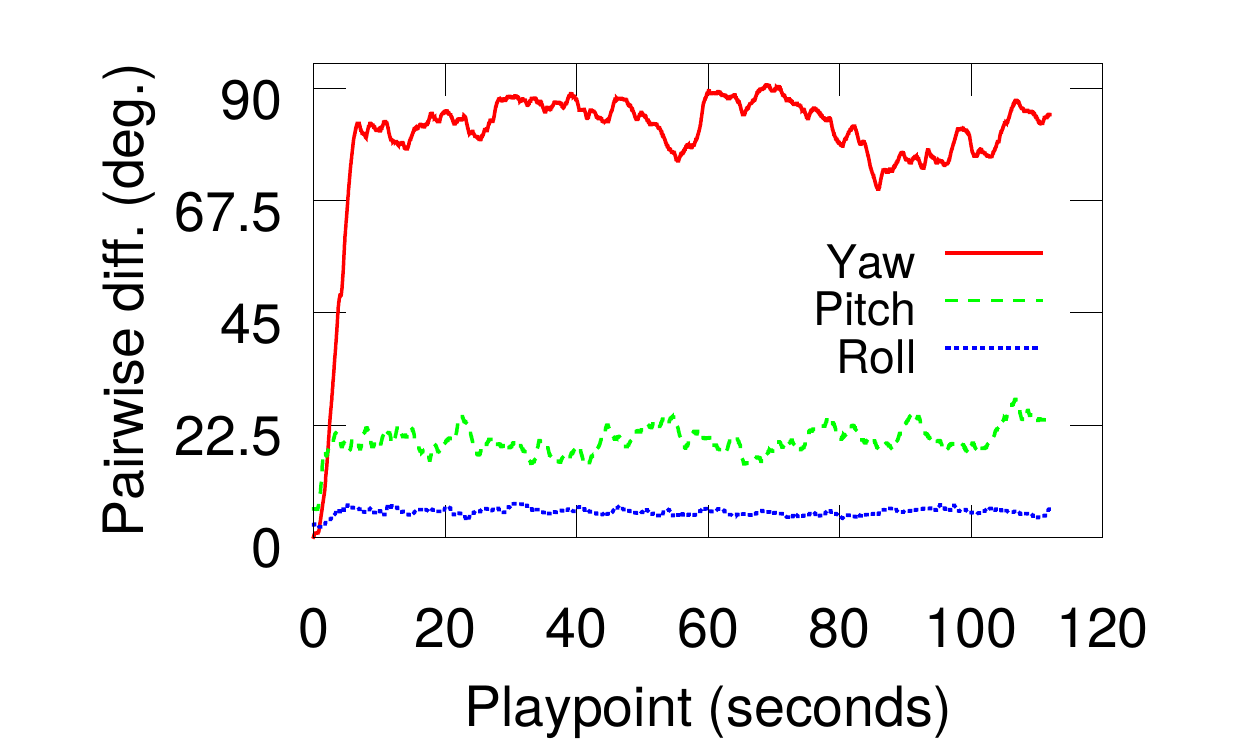}}
  \hspace{-10pt}
  \subfigure[Moving focus]{
    \includegraphics[trim = 8mm 3mm 8mm 0mm, width=0.235\textwidth]{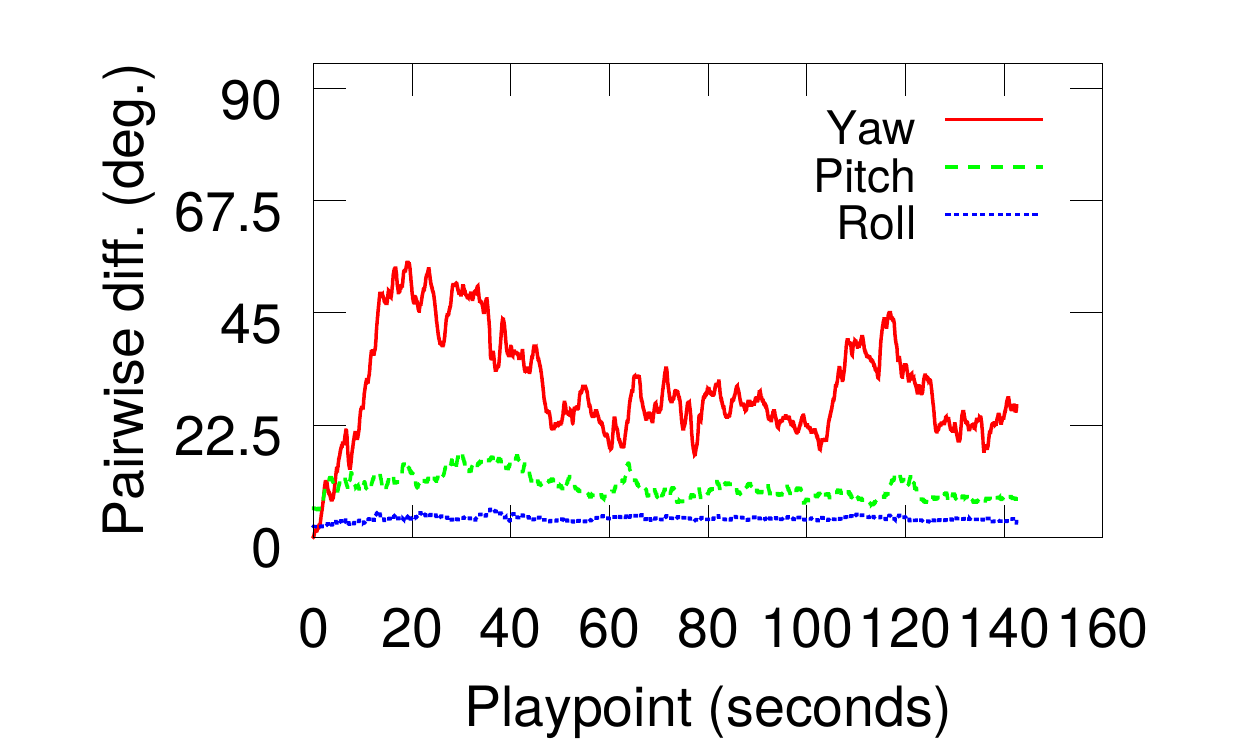}}
  \hspace{-10pt}
  \subfigure[Static focus]{
    \includegraphics[trim = 8mm 3mm 8mm 0mm, width=0.235\textwidth]{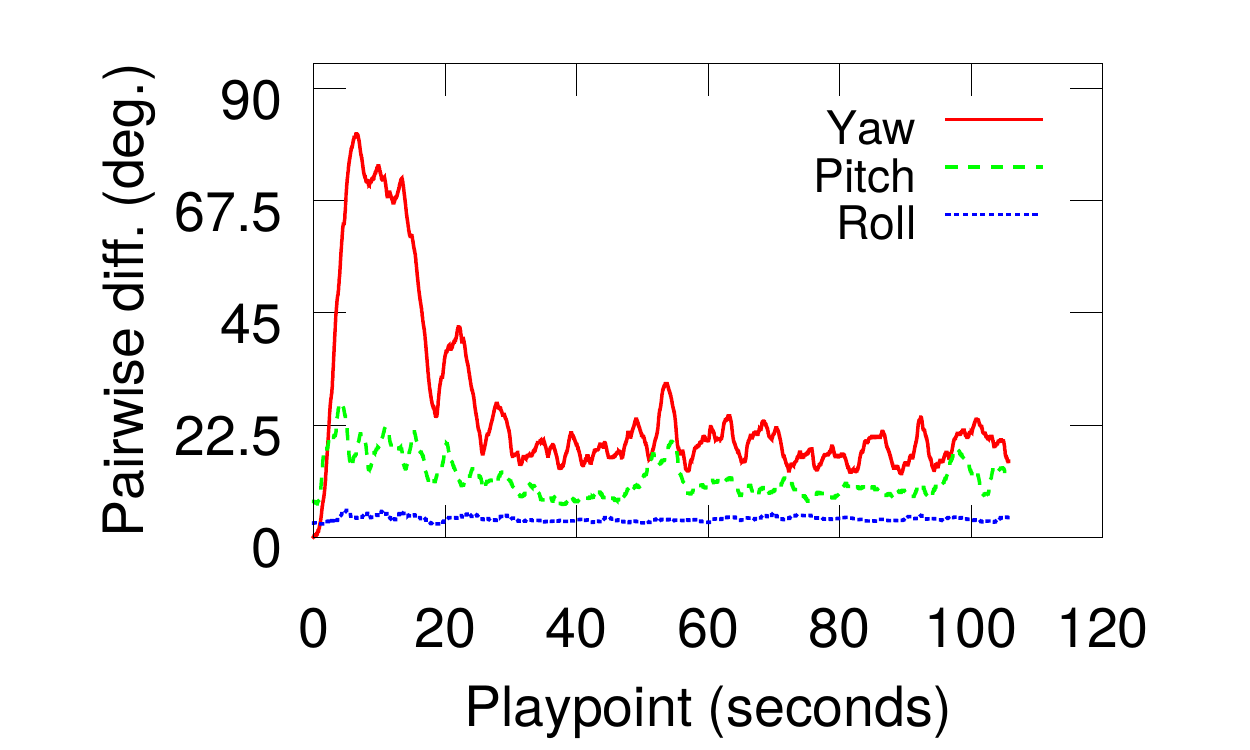}}
    \vspace{-14pt}
    \caption{Average pairwise angular difference between users.}
  \label{fig:angular-diffs}
  \vspace{-6pt}
\end{figure}

While the angular utilizations (Figure~\ref{fig:angle-duration}) of the moving focus (bottom-left) videos
are more similar to those of the exploration videos (top-right), we have found that
these videos differ substantially in how predictable the movements are.
This is illustrated in Figure~\ref{fig:angular-diffs}.
Here, we show the average angle difference between each pair of users
watching the same video, as a function of the playtime
(with time-axis constrained by the duration of the shortest video).

\begin{figure}[t]
  \centering
  \includegraphics[trim = 0mm 2mm 0mm 0mm, clip, width=0.46\textwidth]{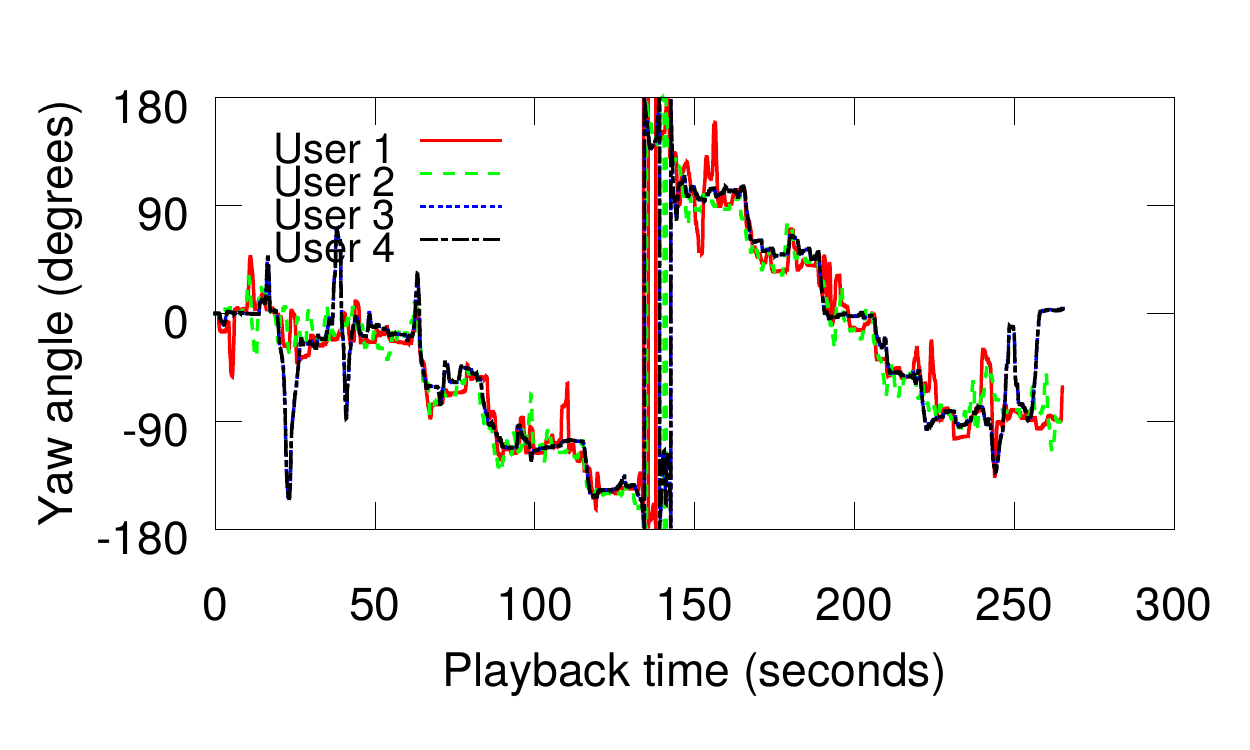}
  \vspace{-14pt}
  \caption{The yaw angle over time for four example users watching the ``Solar system'' video.}
  \label{fig:example-moving}
  \vspace{-6pt}
    \end{figure}

In contrast to the results shown
in Figure~\ref{fig:angle-duration},
Figure~\ref{fig:angular-diffs} shows that the average angle difference between each pair of users watching the same moving focus
(bottom-left) video
is more similar to that for rides (top-left) and static focus (bottom-right) videos
than for exploration (top-right) videos.
These results suggest
that the viewing patterns of other users who have watched
moving focus, rides, and static focus
videos
are valuable when predicting the viewing directions of future viewers;
hence, allowing better prefetching when differentiating the quality for different directions.

Figure~\ref{fig:example-moving} provides a more concrete example of how
different viewers may have similar viewing patterns despite using the full
spectrum of yaw angles.  Here, we show the yaw angle for four example
viewers as they watch the same moving focus video (in which the viewer
is taken on a narrated journey through the solar system).
The above results suggest that we use a reasonable categorization of videos,
in which each category has distinct properties visible in the observed viewing
characteristics.  Note in particular, from Figure~\ref{fig:angular-diffs},
the considerably different behavior for the exploration (top-right) videos.
For this category of videos,
the average pairwise difference is only slightly within 90$\degree$
(where 90$\degree$ corresponds to
the average pairwise difference that would be observed if viewing directions were random).
Clearly, users watching these videos tend to move completely
independently,
focus on highly different things, and
have mostly uncorrelated viewing angles
for the full video duration.

Finally, we note that
there often appears to be some exploratory behavior when viewing the
videos of the other categories as well, in particular at the beginning of a video.
This is particularly visible when considering the static focus
(bottom-right) videos in Figure~\ref{fig:angular-diffs}.
Here, we see a substantial initial spike in the pairwise
angular differences between users during the first 20 seconds of the playback.
This suggests that users tend to explore once they are put in a new environment,
as at the start of a new video.  We will examine this further in Section~\ref{sec:discussion}.

\begin{figure}[t]
  \centering
  \subfigure[Rides]{
    \includegraphics[trim = 2mm 2mm 2mm 0mm, width=0.235\textwidth]{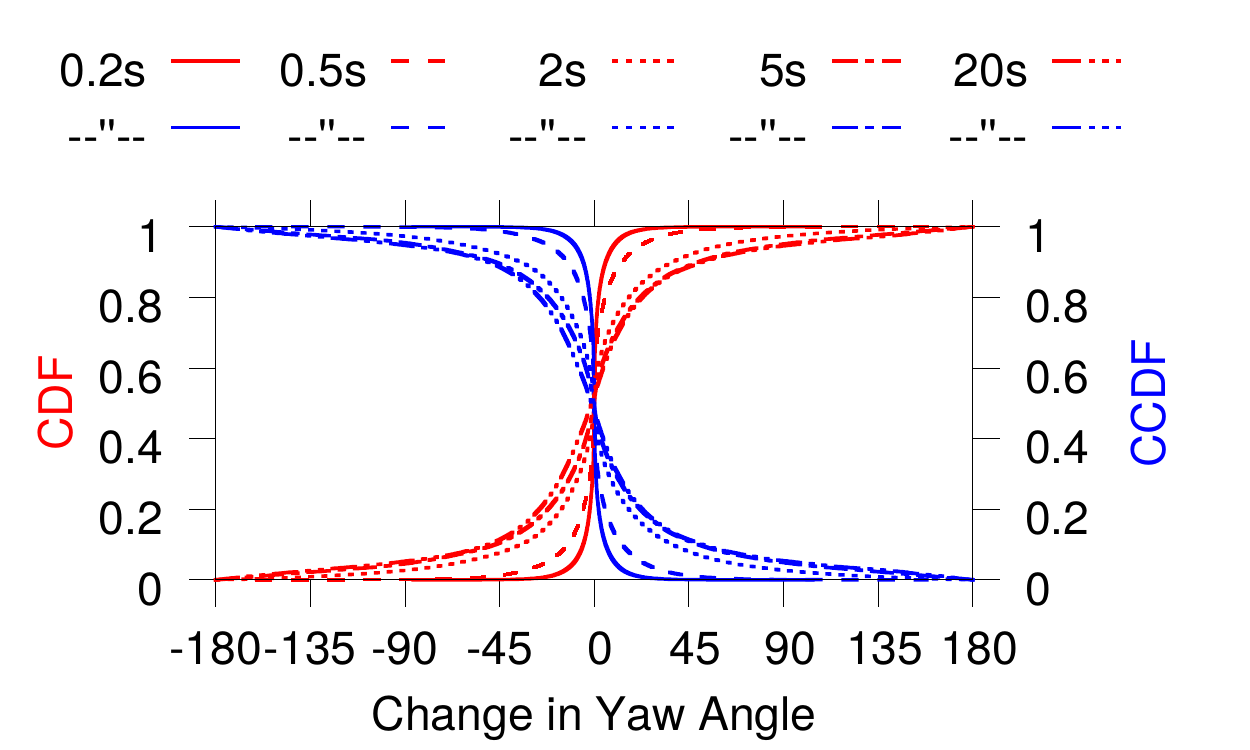}}
  \hspace{-12pt}
  \subfigure[Exploration]{
    \includegraphics[trim = 2mm 2mm 2mm 0mm, width=0.235\textwidth]{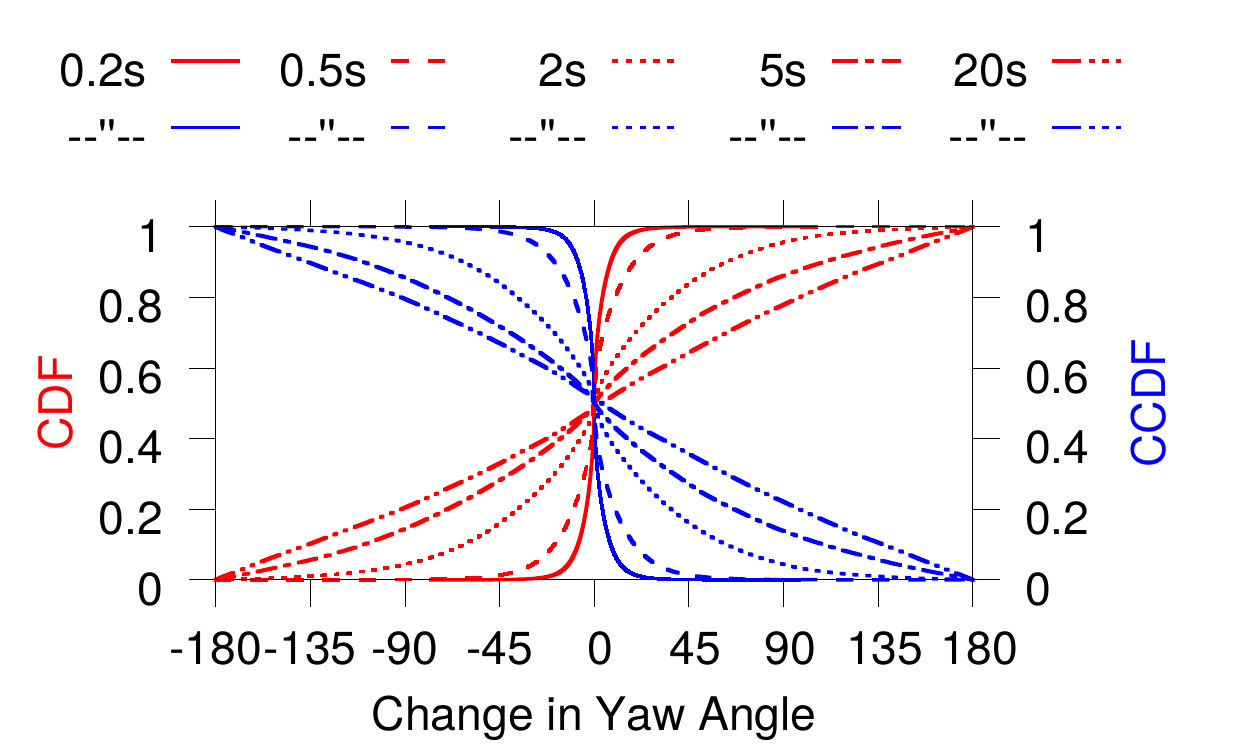}}
  \hspace{-12pt}
  \subfigure[Moving focus]{
    \includegraphics[trim = 2mm 2mm 2mm 0mm, width=0.235\textwidth]{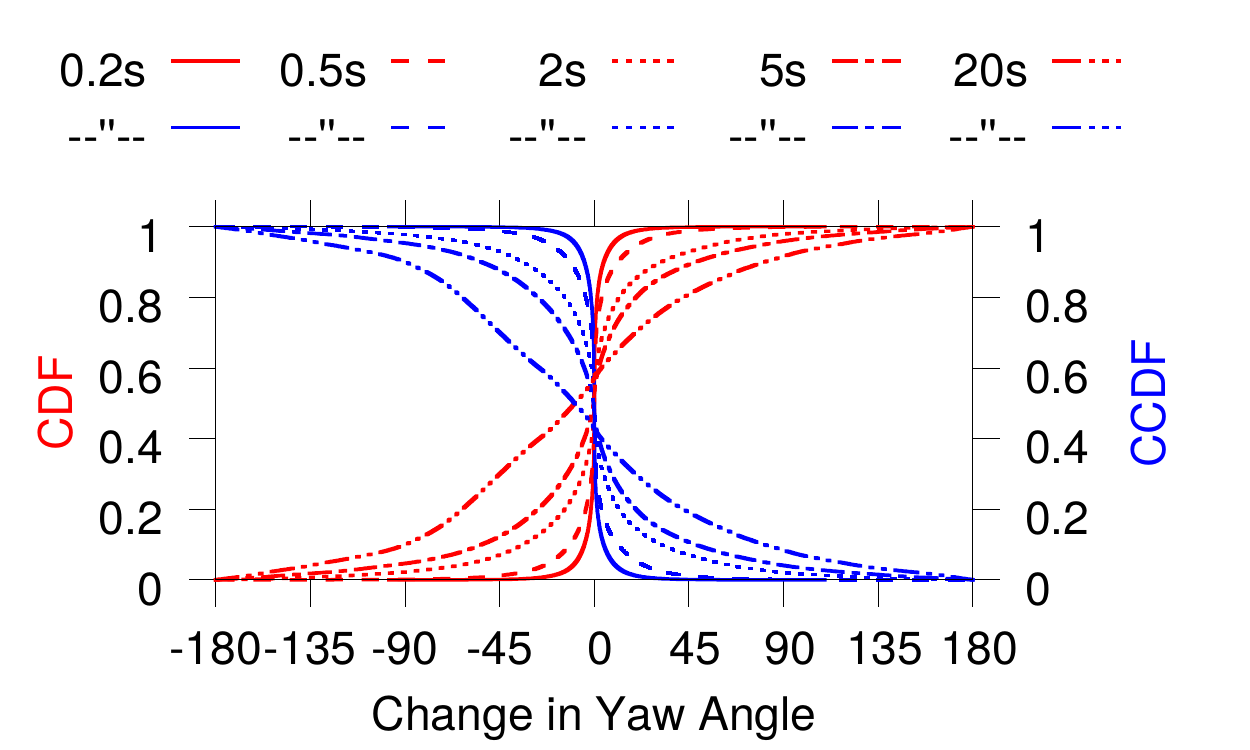}}
  \hspace{-12pt}
  \subfigure[Static focus]{
    \includegraphics[trim = 2mm 2mm 2mm 0mm, width=0.235\textwidth]{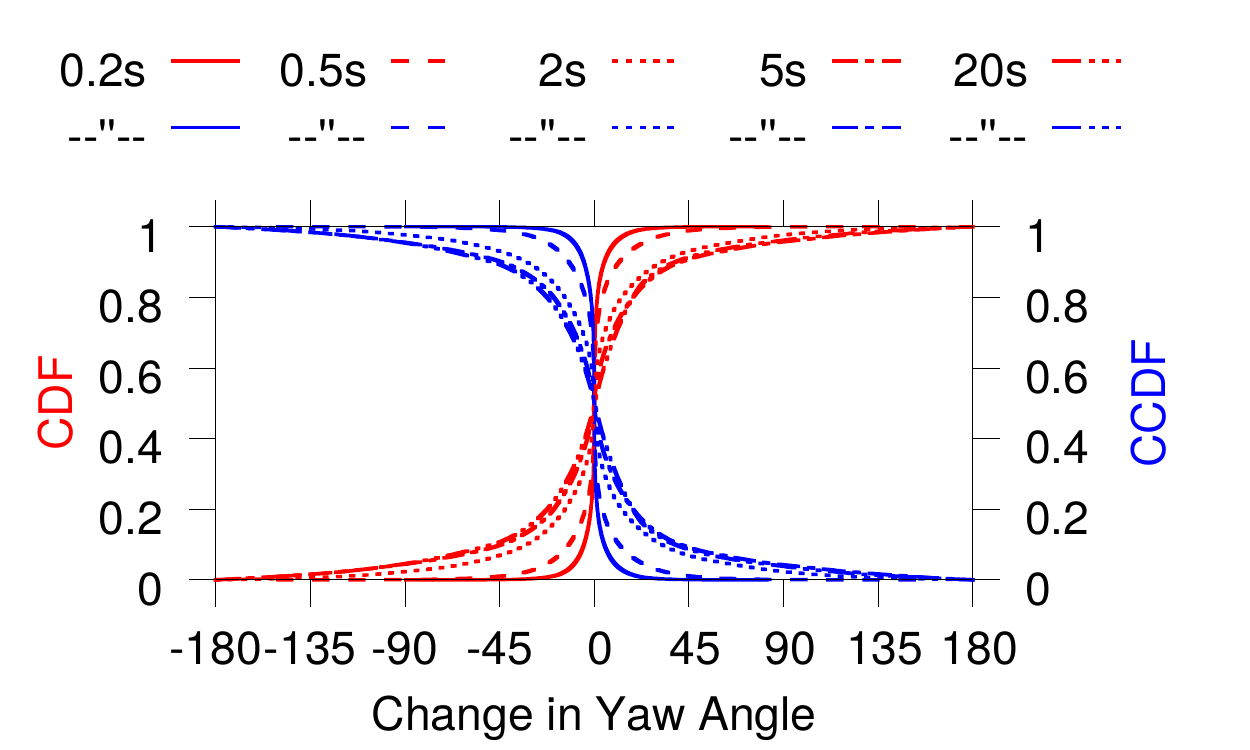}}
  \vspace{-12pt}
  \caption{CDFs and CCDFs of the yaw change over different time intervals $T$.}
  \label{fig:yaw-CDFs}
  \vspace{-6pt}
  \end{figure}

\begin{figure}[t]
  \centering
  \subfigure[Rides]{
    \includegraphics[trim = 2mm 2mm 2mm 0mm, clip, width=0.235\textwidth]{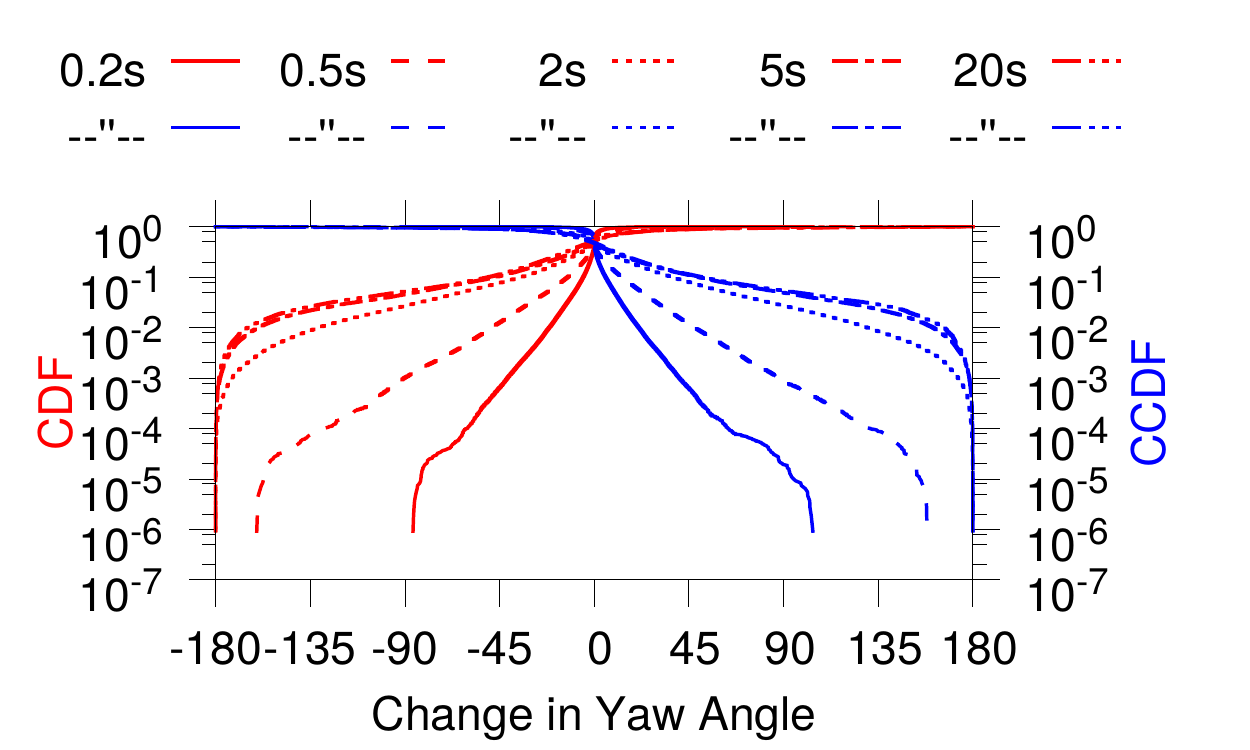}}
  \hspace{-12pt}
  \subfigure[Exploration]{
    \includegraphics[trim = 2mm 2mm 2mm 0mm, clip, width=0.235\textwidth]{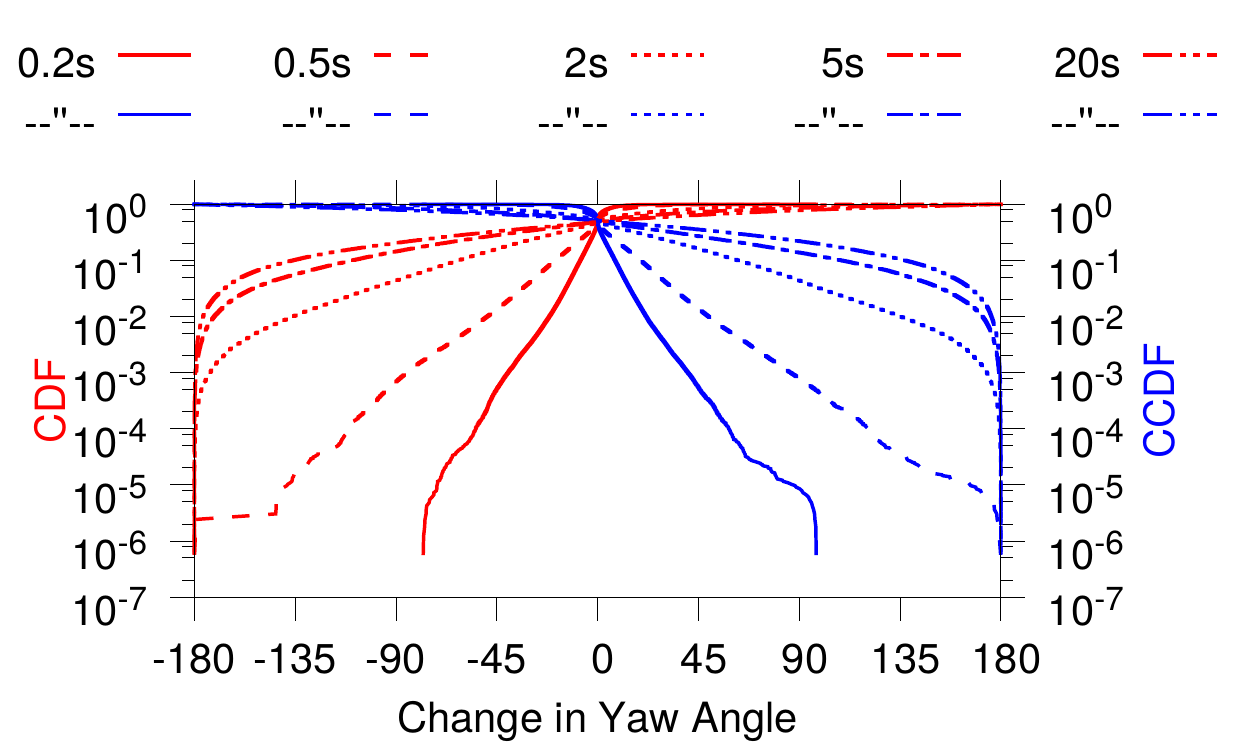}}
  \vspace{-8pt}
  \subfigure[Moving focus]{
    \includegraphics[trim = 2mm 2mm 4mm 0mm, clip, width=0.235\textwidth]{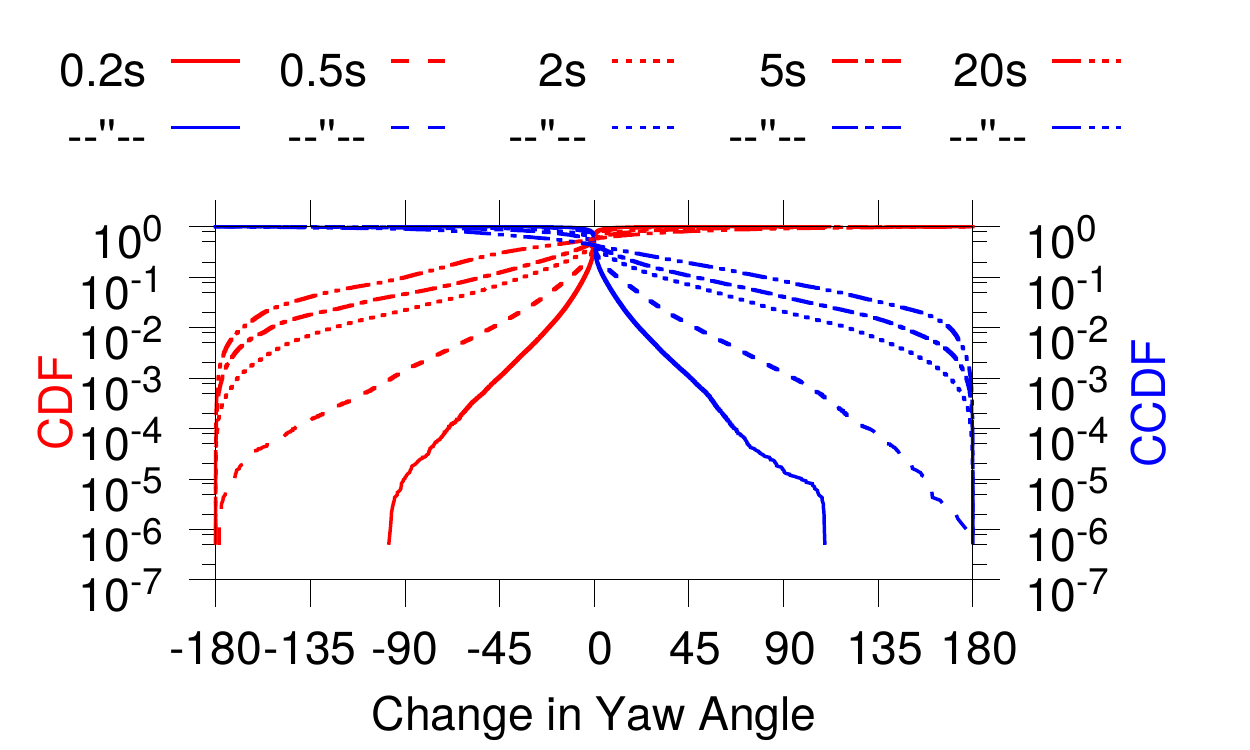}}
  \hspace{-10pt}
  \subfigure[Static focus]{
    \includegraphics[trim = 2mm 2mm 2mm 0mm, clip, width=0.235\textwidth]{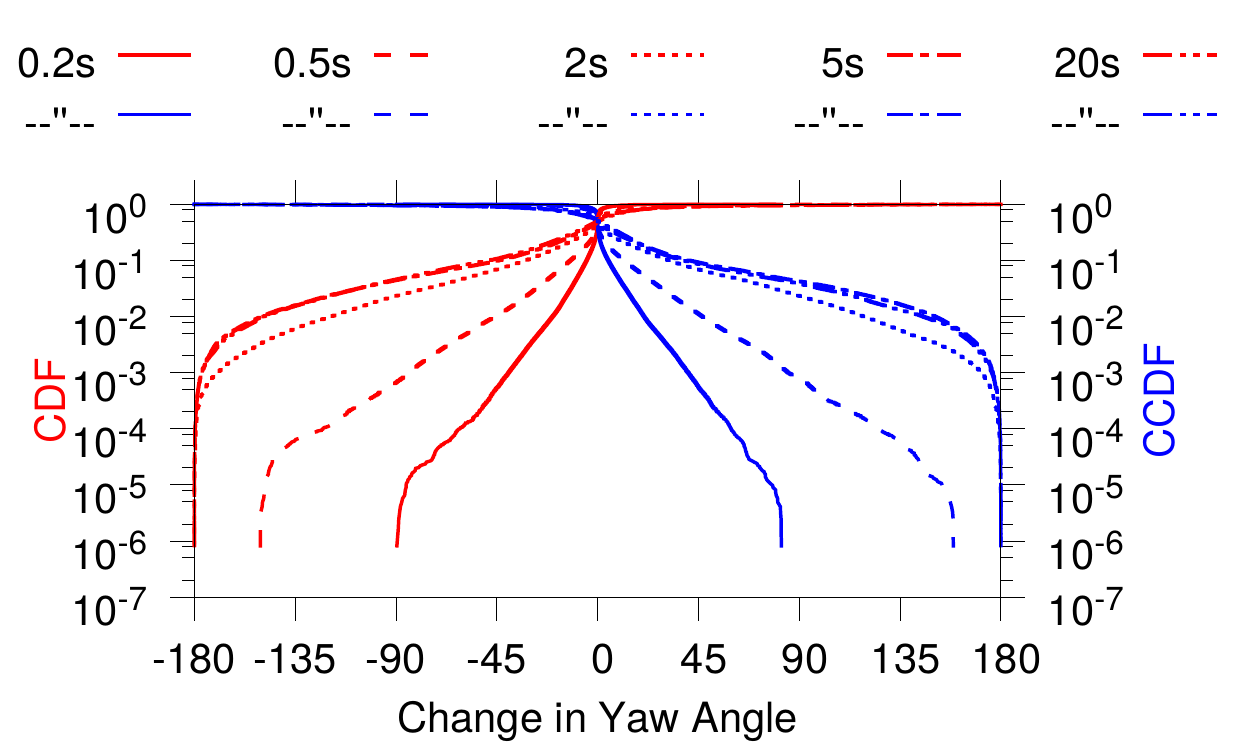}}
  \vspace{-6pt}
  \caption{Log-scale version of Figure~\ref{fig:yaw-CDFs}.}
  \label{fig:yaw-CDFs-log}
  \vspace{-6pt}
\end{figure}

\subsection{Changes in Viewpoint}

For shorter time scales, one of the most natural and commonly used predictors
of a user's future viewpoint is the current viewpoint.  However,
since users can quickly turn their head or body,
when using such a predictor,
it is important to understand how such viewpoint changes vary across different categories
of videos and how the absolute changes depend on the time $T$ over which the change is measured.
We next take a closer look at how much the viewpoint differs between two time points separated by $T$ seconds.

To cover a wide range of 360$\degree$ delivery technologies
we consider time intervals $T$ between 0.2 seconds to 20 seconds.
Here, the low-end intervals (e.g., 0.2-1 second) may be most applicable for low-latency scenarios
where edge servers are used to render frames that the users may view
and the multi-second range (e.g., 2-20 seconds) may be most applicable for HAS-based designs,
which typically use 2-10 second chunks and typically must account for significant wide-area
bandwidth variations.

In both scenarios, we foresee either the client adapting the qualities it requests for each direction
based on view change probabilities or the servers adapting the qualities for each direction based on the
\revtwo{client's}{user's}
current viewing direction and statistics about view change probabilities. 
\revtwo{We also note}{Note}
that some buffering is necessary in all scenarios as the human tolerance threshold
for differences in the displayed content and the actual viewing direction is approximately 50 ms~\cite{Abra14,AHJ+01,BLM+04}
\revtwo{and we at the minimum}{and at the minimum we}
must account for network jitter and other network effects,
which may be on the order of
\revtwo{magnitude of at least 100ms}{100 ms or more}
in modern LTE networks, for example~\cite{TLL+18}.
Further delays beyond
\revtwo{the 50 ms threshold}{50 ms}
\revtwo{can even}{can}
result in user discomfort and motion sickness.

Figure~\ref{fig:yaw-CDFs} shows the CDFs and Complementary CDFs (CCDFs) of the change
in the yaw angle for each of four video categories.
For the short time scales (e.g., 200ms) we have only observed
very minor differences between the video categories.  The extreme values (i.e., the tail values)
of these distributions are bounded by the speed with which the participants moved their heads
when watching the videos.
For example, for all categories 99\% of the head movements are
within $\pm 28\degree$ yaw angle and $\pm 13\degree$ pitch angle.
The corresponding numbers for 99.9\% are: $\pm 46\degree$ and $\pm 19\degree$, respectively.
Across all experiments,
the
largest changes that we observe over 200ms are $95\degree$ for yaw and $35\degree$ for pitch.
Again, the maximum is almost the same for all four video categories.

The above
200 ms results suggest that a significant portion of the potential viewing field
\revtwo{potentially would}{may}
not need to be pre-rendered in edge-based rendering systems
operating at the extremely low-latency (i.e., sub-200 ms) time scale.
However, modern networks do not provide performance guarantees
sufficient to operate at such time scales yet~\cite{TLL+18}.
Instead, most video streaming services today therefore require substantial buffers to
protect against bandwidth variations and other interruptions,
and HAS-based solutions with larger buffers are likely to
continue to dominate the markets for the foreseeable future.

Considering head movements over longer time scales,
for all categories the full range (i.e., $\pm180\degree$) of head movements is covered already after a second.
In fact, for the explore and moving focus categories,
we observe such full range yaw rotations already at the 0.5 second time scale.
For the rides and static focus videos the most extreme movements
reach at most $100\degree$ and $90\degree$, respectively, over this 0.5 second time window.
To better quantify these extreme changes in viewing direction,
Figure~\ref{fig:yaw-CDFs-log} presents log-scale versions of
the Figure~\ref{fig:yaw-CDFs}
\revtomm{plots for the rides and moving focus categories.
  The (omitted) plot for exploration is similar to moving focus and the
(omitted) plot for static focus is similar to that of rides.}{plots.}

Overall, we observe diminishing increases in the variations as the time interval $T$ increases
and the distributions become more
similar to the life-time distributions (shown in Figure~\ref{fig:angle-duration}).
This is most clear for rides and static focus.
For these categories,
the differences between the 5-second and 20-second curves
are much smaller than for the other neighboring curves
and the 20-second curves provide similar accuracy
as when using the life-time distribution,
suggesting that for these two categories it is
equally good to use the current viewing direction as the
zero-degree line (used for the life-time distributions) as predictors when maintaining a 20-second buffer.

In general, as the buffer needed becomes greater, the more value
should be placed on the angular distribution and the viewpoint directions
of other users that have watched the same video, for example.
This is particularly apparent when considering the moving focus category
(e.g., as exemplified in Figure~\ref{fig:example-moving}).
For this category we also see that the  current direction is
a slightly better predictor than the zero-degree line.
Interestingly, for the exploration category, on the other hand,
the zero-degree line is a (slightly) better predictor for the 20-second time scale.
To see this,
note that the 20-second distribution for the exploration category
closely resembles that of a uniform distribution
(showing that yaw-based angular prioritization has no benefit
for the exploration category when maintaining a 20+ second buffer),
whereas the life-time distribution
is somewhat more concentrated around the zero-degree line.

In summary, our results show clear differences in the viewpoint distributions
and how predictable the head movements are
(e.g., as measured by how concentrated
\revtwo{the distributions}{distributions}
are),
both across video categories and with regards to the time scale over which
\revtwo{the prediction}{prediction}
is done.
We next determine the
(optimized) tradeoffs between the buffer used
and the expected achievable playback utility.

\section{Optimized Prefetching Tradeoff}\label{sec:tradeoff}

Let us now consider the basic prefetch aggressiveness tradeoff between clients trying to
maintain a buffer $T$ (to protect against stalls due to bandwidth variations or other service interruptions)
and the expected playback utility $E[u|T]$ experienced by the client,
where we use playback utility to measure the client's perceived playback quality.
To better understand this tradeoff we consider the optimized prefetch schedule
of a client that operates in steady-state, maintaining a fixed buffer of $T$ seconds,
and
\revtwo{}{that}
prefetches data at a known average download rate
\revtwo{$D$ (over the prefetch duration of a chunk).}{$D$.}
For this client, we then determine the optimized prefetch schedule
for different buffer levels $T$, allowing us to investigate the
tradeoff between the optimized $E[u|T]$ and $T$.

\subsection{High-level optimization model}

Without loss of generality, consider the optimized download schedule of an arbitrary chunk
with playback duration $\Delta$.
For each
viewing direction $\theta_n$
of such a chunk
we assume that the player can
choose a
\revtwo{}{tile $n$ of}
quality level $l$
($0 < l \le L$)
with encoding rate $q_{n,l}$ or to not download any data for that direction.
We use $l=0$ to indicate this last choice.
Furthermore,
we assume that each
\revtwo{view direction}{tile}
can be delivered independently
(or together with other tiles, as a combined chunk)
and that these tiles can be stitched together either at the client
or at the server (when creating a combined chunk).
 
For simplicity,
since the majority of the head movements are yaw rotations
(with small pitch and negligible roll),
we will focus our analysis here on the yaw angle.
We also discretize the possible angles (possible viewing directions).  In particular,
\revtwo{we let
$\theta_0 \le \theta_1 \le ... \le \theta_{N-1}$
correspond to increasing angles from the initial viewing direction $\theta_0$,
with wrap-around when $n=N$,
such that $\theta_N = \theta_0$,
where $N$ is the total number of viewing directions considered.}{we let directions
  $n = 0, 1, ..., N-1$ correspond to increasing yaw angles from the initial viewing direction,
  with wrap-around when $n=N$ (i.e., direction $N$ is the same as the initial viewing direction,
  direction 0).  Each of these viewing directions corresponds to a single tile.
  Note that the entire view field at any point in time may encompass a number of these tiles.}
Generalizations that also take into account pitch are relatively straightforward,
whereas models that also take into account roll require significantly more geometry and notation.

At a high level,
given the available bandwidth budget in the next $\Delta$ seconds,
an average buffer of $T$, and a known conditional probability distribution
$p_n(T)$ that the user looks in direction
\revtwo{$\theta_n$}{$n$}
another $T$ seconds after the chunk was requested by the client,
we maximize the expected playback utility:
\begin{align}
  E[u|T] & = \sum_{n=0}^{N-1} p_n(T) u(n | q_0,q_1, ...,q_{N-1}),
\end{align}
where
\revtwo{$u(\theta_n | q_0,q_1, ..., q_{N-1})$}{$u(n | q_0,q_1, ..., q_{N-1})$}
is the playback utility
experienced by the client when looking in direction
\revtwo{$\theta_n$}{$n$}
after having prefetched tile qualities $q_0,q_1, ...,q_{N-1}$ for
the $N$ directions.
For the purpose of our numeric evaluation,
we calculate the probabilities $p_n(T)$  as follows:
\begin{align}
  p_n(T) & = \int_{\theta_{n}}^{\theta_{n+1}} p(\theta|T)\textrm{d}\theta,
\end{align}
\revtwo{using}{where $\theta_n$ is the yaw angle corresponding to direction $n$, using }
CDFs such as those studied in
\revtwo{Section 5.}{Section 4.}

To model the expected utility
\revtwo{$u(\theta_n | q_0,q_1, ...,q_{N-1})$,}{$u(n | q_0,q_1, ...,q_{N-1})$,}
we use a utility model that weights
(i) the playback utility $u(q_n)$ of the encoding rate $q_n$ in the current viewing direction
\revtwo{$\theta_n$,}{$n$,}
(ii) the playback utility of neighboring viewing directions (i.e., $u(q_{n-1})$, $u(q_{n+1})$), and
(iii) the relative utility differences compared to neighboring directions (i.e., $-|u(q_n)-u(q_{n-1})|$ and $-|u(q_n)-u(q_{n+1})|$),
the last of which potentially may cause negative effects.
Since we are not aware of any research that has provided relative weights to these factors
for the context of 360$\degree$ videos, we use variable-sized
\revtwo{weights for the different factors}{weights}
and evaluate their relative importance.
It turns out that by careful selection of constants, only a single parameter $\beta$ is needed.  To see this,
let us give each of the above factors the relative weights
(i) $(1-\alpha-\beta)$, (ii) $\frac{\alpha}{2}$ times their
\revtwo{relative weights}{probability ratios}
($\frac{p_{n-1}}{p_n}$ and $\frac{p_{n+1}}{p_n}$),
and (iii) $\frac{\beta}{2}$, respectively.
Summing over all directions,
the objective function can now be simplified as follows:
\begin{align}
  E[u|T] & = \sum_{n=0}^{N-1} p_n \biggl( (1-\alpha-\beta)u(q_n) + \frac{\alpha}{2}(\frac{p_{n-1}}{p_n} u(q_{n-1}) + \frac{p_{n+1}}{p_n} u(q_{n+1})) \nonumber\\
  & ~~~~ - \frac{\beta}{2} (|u(q_{n})-u(q_{n-1})|+|u(q_{n})-u(q_{n+1})|\biggr) \nonumber\\
  & = (1-\beta) \sum_{n=0}^{N-1} p_n u(q_n) - \beta  \sum_{n=0}^{N-1} \frac{p_n+p_{n+1}}{2}|u(q_{n})-u(q_{n+1})|.
\end{align}
Here, we have dropped the argument $T$ from $p_n(T)$ and used $q_n$ to represent the selected
encoding rate for direction $n$.  Note that this expression is independent of $\alpha$.

\begin{table*}[t]
  \centering
  \caption{Notation for streaming model for a single chunk.}
  \label{tab:notation}
  \vspace{-12pt}
         {\small
           \begin{tabular}{|c|p{10.7cm}|}\hline
             {\bf Symbol} & {\bf Definition} \\ \hline
             $L$          & Number of non-zero quality levels of the video \\\hline
             $N$          & Number of tiles or discrete viewing directions \\\hline
             $x_{n,l}$     & Binary variable indicating that the client will prefetch tile $n$ of the chunk
             at quality level $l$\\\hline
             $q_{n,l}$    & Playback encoding of tile $n$ of the chunk with quality level $l$ \\\hline
             $b_{n,l}=b(q_{n,l})$    & Size of tile $n$ of the chunk with quality level $l$ \\\hline
             $u_{n,l}=u(q_{n,l})$    & Playback utility of playing tile $n$ (direction $n$) of the chunk with quality level $l$ \\\hline
             $\Delta$   & Playback duration of the chunk \\\hline
             $D$       & Download rate \\\hline
             $T$          & Average maintained buffer size in seconds \\\hline
             $E[u|T]$     & Expected playback utility, conditioned on $T$ \\\hline
             $p_n(T)$     & Probability looking in direction $n$, $T$ time later \\\hline
             $\beta$      & Parameter used to weight factors ($0 \le \beta \le 1$) \\\hline
         \end{tabular}}
         \vspace{-10pt}
         \end{table*}

{\bf Discussion of utility model:}
Similar HAS/DASH models
have been used to capture the quality of experience (QoE)
of ``regular'' non-interactive streaming video, rather than 360$\degree$ video.
For example, Yin et al.~\cite{YJSS15} try to capture the QoE by the following metrics:
(i) the average video quality,
(ii) the average quality variations,
(iii) number and duration of rebufferings, and
(iv) the startup delay.
Looking at a single chunk during steady-state streaming,
the startup delay and average quality variations are not applicable to our model.
Instead, we use the expected buffer ($T$) to capture the protection against stalls/rebufferings
(i.e., factor (iii)) and consider the quality variations relative to nearby
viewing angles, in the case that a user changes viewing direction and also
to take into account that the
\revtwo{peripheral view}{viewfield, especially the peripheral view,}
typically will be made up by tiles
from neighboring directions.  It is therefore good if the quality differences
between neighboring tiles are not too obvious.  We note that this objective
will somewhat offset greedy maximization of the average (expected) video quality,
and weight the two factors
using a
variable $\beta$ ($0 \le \beta \le 1$).

\subsection{Detailed optimization model}

Let $q_{n,l}$ denote the video
encoding rate of tile $n$ with quality level $l$ (ordered from lowest to highest quality),
let $b_{n,l} = b(q_{n,l})$ denote the size of the tile (proportional to its encoding rate),
let $u_{n,l} = u(q_{n,l})$ denote the estimated playback utility of this tile, and
let $x_{n,l}$ be a binary decision variable (indicating that tile $n$ of the chunk is downloaded
at quality level $l$ whenever $x_{n,l} = 1$).
(See Table~\ref{tab:notation}.)
We can now formulate the
problem of choosing optimal quality encodings
as a packing problem:
\begin{align}
  \textrm{maximize~~} & ~~E[u|T],
\end{align}
where
\begin{align}
  E[u|T] & = (1-\beta) \biggl(\sum_{n=0}^{N-1} p_n(T) \sum_{l=0}^L x_{n,l} u_{n,l}\biggr)
  \nonumber\\
  & ~~~~
  - \beta \biggl( \sum_{n=0}^{N-1} \frac{p_n(T)+p_{n+1}(T)}{2}
  \nonumber\\
  & ~~~~~~~~ \times
  \sum_{l=0}^L \sum_{l'=0}^L  x_{n,l} x_{n+1,l'} |u_{n,l}-u_{n+1,l'}| \biggr),  \label{eqn:objectiveC}
\end{align}
such that
\begin{eqnarray}
  \sum_{l=0}^L x_{n,l} = 1, & 0 \le n < N,  \label{eqn:constraintC1}\\
  \sum_{n=0}^{N-1} \sum_{l=0}^L x_{n,l} b_{n,l} \le D\Delta,  & \label{eqn:constraintC2}\\
  x_{n,l} \in \{0,1\}, & 0 \le n < N, 0 \le l \le L.    \label{eqn:constraintC3}
\end{eqnarray}

\begin{algorithm}[t]
  {\small
    \KwIn{Number of directions $N$, total bytes to download $\Delta D$, probabilities $p_n$, tile sizes $b_{n,l}$, and utilities $u(q_{n,l})$.}
    \KwOut{Optimal set of qualities (equivalent to $x_{n,l}$) and value of the corresponding
      (optimal) objective function.}
    \ForEach{ $0 \le l_0 \le L$ }{
      \ForEach{ $0 \le C \le \Delta D$ }{
        \ForEach{ $0 \le n \le N-1$ }{
          \ForEach{ $0 \le l \le L$ }{
            $DP(l_0,l,n,C) \leftarrow$ equation (\ref{eqn:dpA})
          }
        }
      }
    }
    Return $\max_{0 \le l_0 \le L} DP(l_0,l_0,N-1,\Delta D)$ and corresp. parent pointers\;
    \caption{Calculate optimal single-slot prefetch schedule.}
    \label{alg:dpA}
  }
  \end{algorithm}

As is typically the case,
for our evaluation,
we assume that
$u(q_{n,l})$
is a concave function of $q_{n,l}$
and that $b_{n,l}$ is a linear function of $q_{n,l}$.
Furthermore, we model the case when there is missing data
(i.e., when $x_{n,0}=1$)
using $b_n,0 = 0$ and
a negative utility $u_{n,0} = -f u_{n,L}$,
where $f$ is a penalty factor.

For the special case that $\beta=0$,
the above problem simplifies to the standard 0-1 knapsack problem.
It is therefore trivial to formulate any 0-1 knapsack problem
using our formulation, and our problem is therefore
at least as hard as the 0-1 knapsack problem,
which is NP-complete.
As with the standard 0-1 knapsack problem,
under some circumstances this problem can be solved using dynamic progrmaming.
We next describe one such formulation and then discuss variations thereof.

\subsection{Dynamic programming solution}

For simplicity,
in the following,
we assume that the size $b_{n,l}$ of each tile,
as well as the amount of data that can be downloaded during the interval $\Delta$ (i.e., $D\Delta$),
can be represented using integers (e.g., measured in kilobytes).\footnote{The development of
  $(1+\epsilon)$ FPTAS approximations appear possible when these quantities are non-integers.}
Under these assumptions,
given a list of tiles,
we can formulate the sub-problem of determining the maximum utility for directions $0$:$n$,
given a total prefetching capacity $C$ for these directions.
To allow us to take into account both
the utility differences between neighboring tiles
and
\revtwo{that we operate}{operation}
over a circular space (with modulus $N$),
we also condition each sub-problem on the encoding selections for tiles 0 and  $n+1$.
\revtwo{Now, assuming that the}{Denoting the}
quality selections for tiles $n+1$ and $0$
\revtwo{are}{by}
$l$ and $l_0$, respectively,
\revtwo{and that the remaining prefetching capacity to schedule the remaining directions \revtwo{are}{is} $C$, we}{we}
can write out the following optimization recursion:
\begin{align}
  DP(l_0, l, n, C)
  & = \left\{ \begin{array}{l}
    (1-\beta) p_0 u(q_{0,l_0}) - \beta \frac{p_0+p_1}{2}|u(q_{0,l_0})-u(q_{1,l})|,
    \\
    \qquad \qquad \qquad \textrm{if}~ n=0, b_{0,l_0} \le C\\
    \max_{\{l'|b_{n,l'}+b_{0,l_0} \le C\}}\biggl[(1-\beta) p_{n} u(q_{n,l'})
      \\\qquad
      - \beta \frac{p_{n}+p_{n+1}}{2} |u(q_{n,l})-u(q_{n+1,l'})| \\
      \qquad \qquad + DP(l_0,l',n-1, C-b_{n,l'}) \biggr],
    \\
    \qquad \qquad \qquad \qquad \textrm{if}~ 1 \le n \le N-1\\
  \end{array}\right.\label{eqn:dpA}
\end{align}
with the boundary cases that $DP(l_0,l,n,C) = -\infty$ whenever $b_{n,l}+b_{0,l_0}>C$.
Given this recursion,
the optimal solution can be calculated by considering all choices for direction 0;
i.e., $\max_{0 \le l \le L} DP(l,l,N-1,D\Delta)$.
Algorithm~\ref{alg:dpA} summarizes the calculations,
from which the optimal solution
\revtwo{can be easily obtained through the addition of parent pointers.}{is obtained through parent pointers.}

{\bf Runtime analysis:}
There are $\Theta(CNL^2)$ sub-problems.
Each takes $\Theta(L)$ to calculate,
resulting in a total run-time of $\Theta(CNL^3)$.

\subsection{Example characterization}

Using the above optimization formulation,
we next characterize the optimized prefetch aggressiveness tradeoff.
Throughout the experiments presented in this section we
assume encoding rates proportional to those found in
an example YouTube video (equal to: 144, 268, 625, 1124, 2217, 4198 kbps)
and present results for different prefetch capacities $C=D\Delta$
(the number of bytes that can be downloaded during a timeslot),
the four different video categories considered in the paper,
and four different utility functions:
(i) a linear model, (ii) a square-root model, (iii) a logarithmic model,
and (iv) a large-screen model proposed by Vleeschauwer et al.~\cite{VVB+13}:
\begin{equation}\label{eqn:utility}
      u(q) = b \cdot \frac{(q / \theta)^{1-a}-1}{1-a},
\end{equation}
where $a > 1$, $b> 0$ and $\theta > 0$ are screen dependent parameters,
in our case set to $a=2$, $b=10$, $\theta=0.2 Mbps$.
Finally, a negative utility $u_{n,0}$ is used.
In the experiments,
we vary the stall penalty $\frac{u_{n,0}}{u_{n,L}}$ between -0.1 (small) to -100 (large).

To allow easier comparison,
all utilization values reported were normalized such that the maximum
utility (when playing at the highest available quality encoding) was always 1.
Furthermore, each tile could be selected in one of the discrete sizes:
144, 268, 625, 1124, 2217, 4198 ``units'' and the capacity $C$ was measured
in the same units.  (These units allow us to
avoid
having to pick a chunk
playback duration $\Delta$.)

\begin{figure}[t]
  \centering
  \subfigure[Linear utility]{
    \includegraphics[trim = 8mm 2mm 8mm 0mm, width=0.23\textwidth]{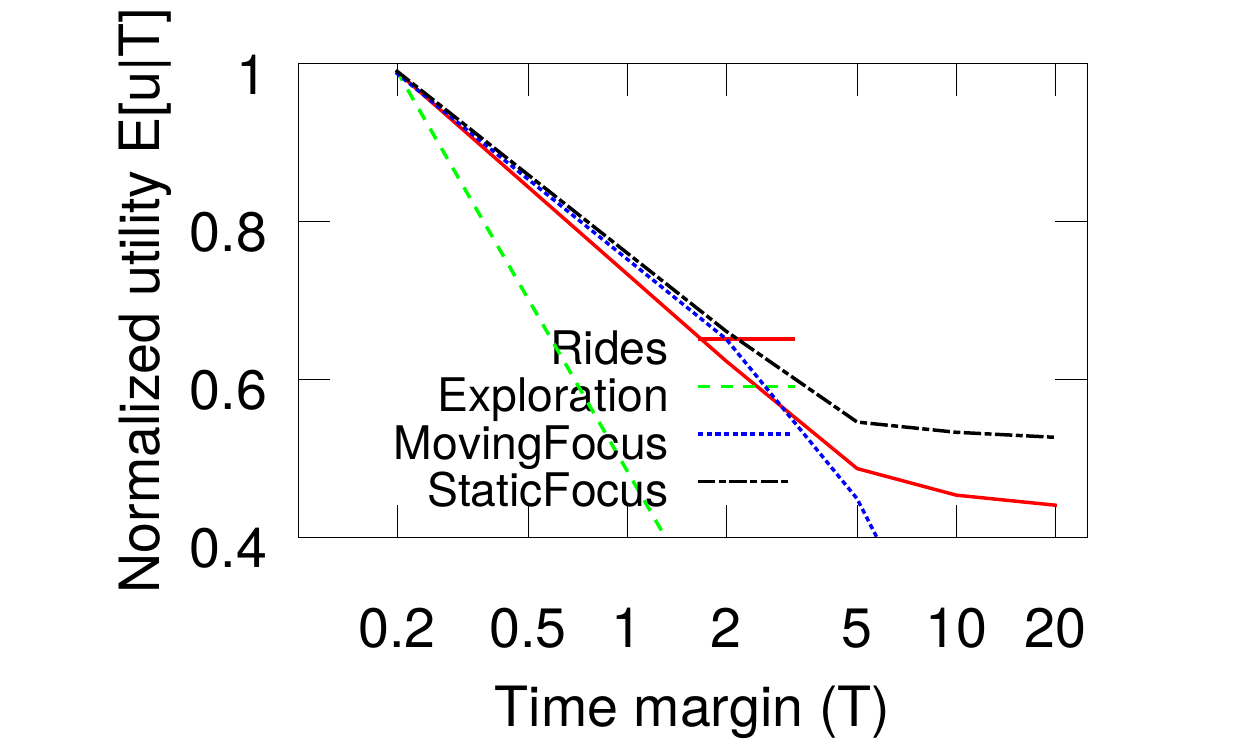}}
  \hspace{-12pt}
  \subfigure[Square-root utility]{
    \includegraphics[trim = 8mm 2mm 8mm 0mm, width=0.23\textwidth]{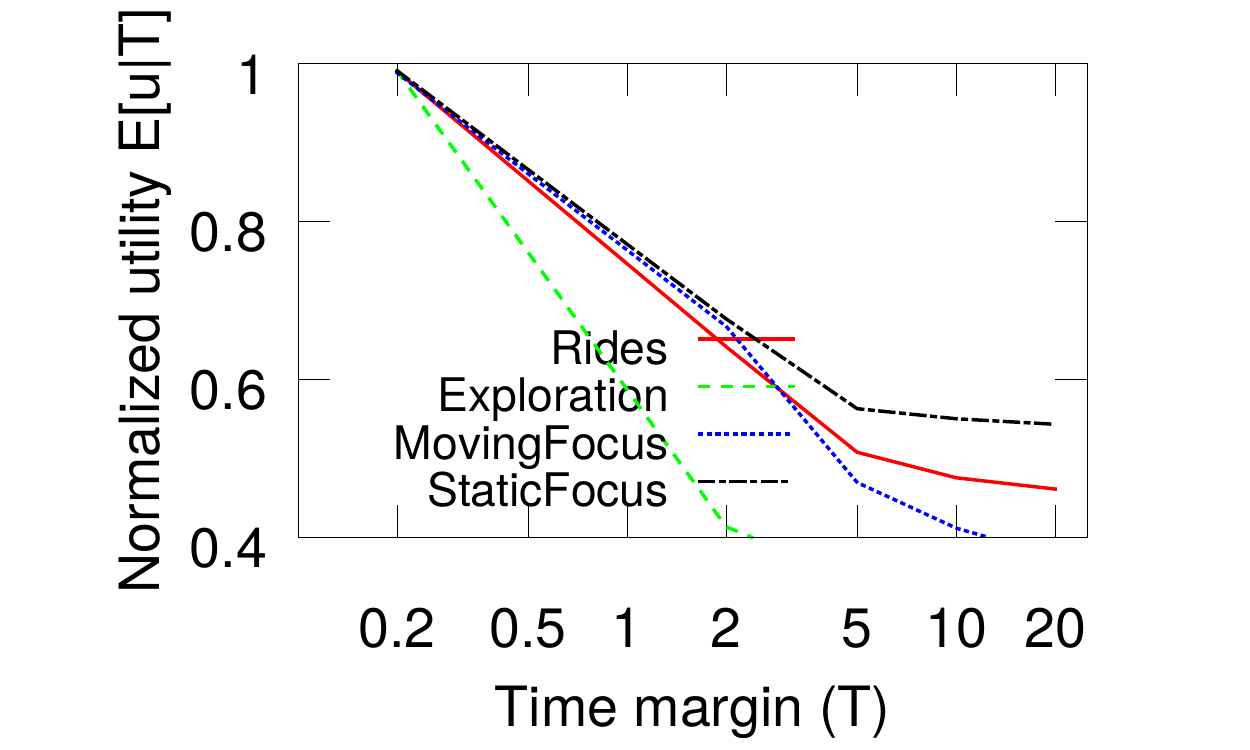}}
  \hspace{-12pt}
  \subfigure[Logarithmic utility]{
    \includegraphics[trim = 8mm 2mm 8mm 0mm, width=0.23\textwidth]{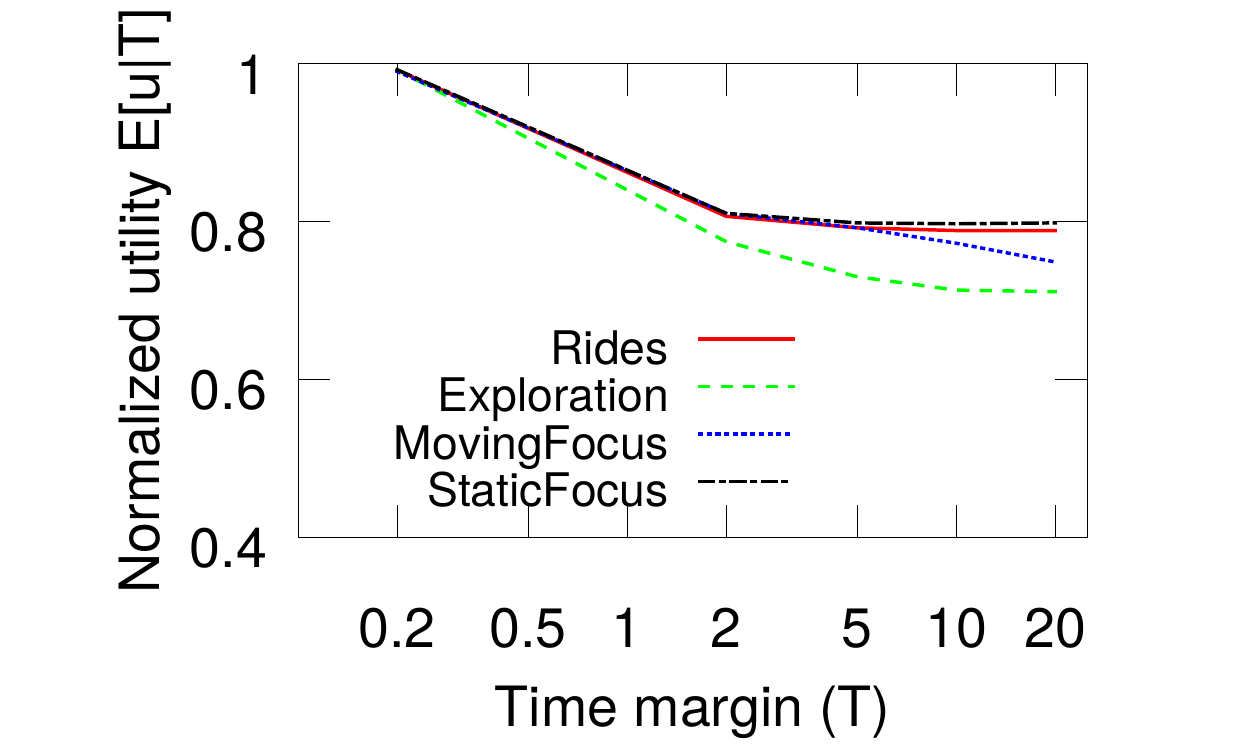}}
  \hspace{-12pt}
  \subfigure[Large-screen utility]{
    \includegraphics[trim = 8mm 2mm 8mm 0mm, width=0.23\textwidth]{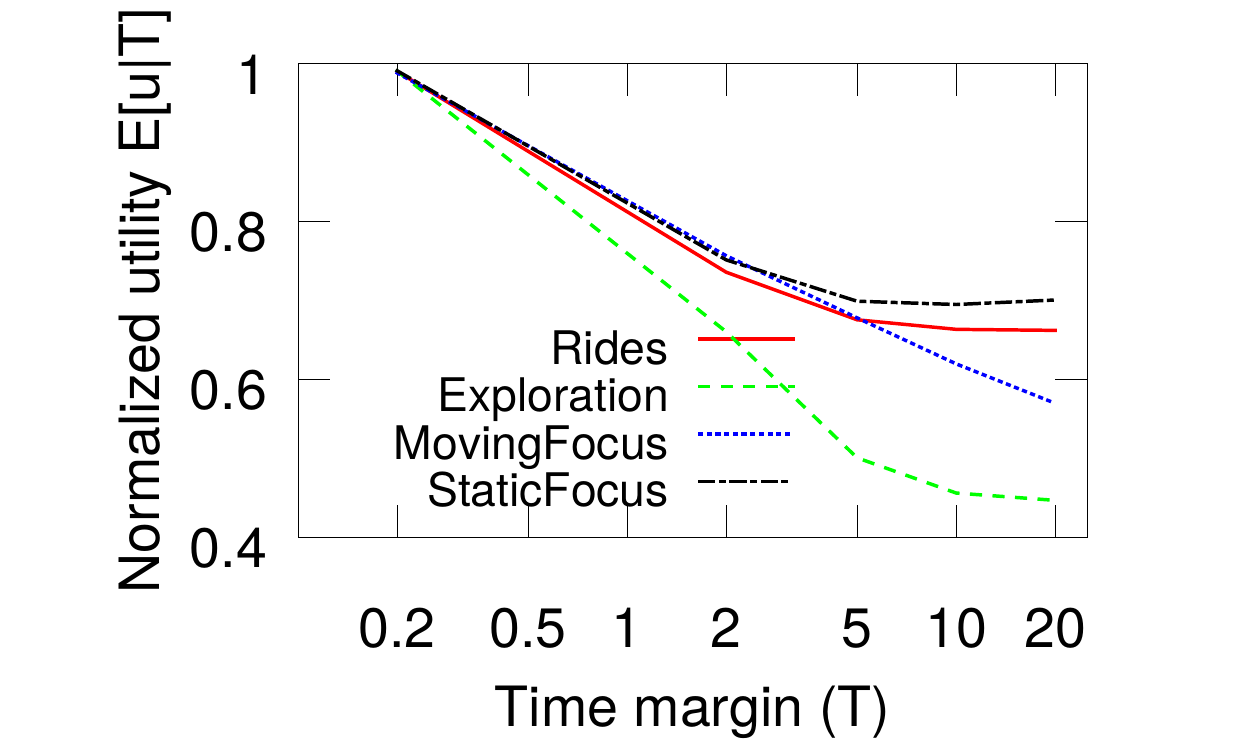}}
  \vspace{-12pt}
  \caption{Example tradeoffs for different 360$\degree$ video categories. ($C=2500$ and $\frac{u_{n,0}}{u_{n,L}}=-1$ in all sub-arxiv-figs.)}
  \label{fig:tradeoff-videotype}
  \vspace{-6pt}
\end{figure}

Figure~\ref{fig:tradeoff-videotype} compares the tradeoff curves
for the different video categories, with each sub-figure using a
different utility function but the same prefetch capacity $C=2500$,
stall penalty $\frac{u_{n,0}}{u_{n,L}}=-1$, and $\beta=0.001$.
We note that in all cases the static focus and exploration categories provide the
two extremes and the other two categories fall in between.
Interestingly, all the static focus and rides curves flatten out after 5 seconds.
This suggests that these categories can use prefetching to build larger buffers
at little expense in expected utility.
In contrast, moving focus and especially exploration typically have a more gradual tradeoff curve.
For these categories, there are noticeable benefits to the expected utility (assuming stalls do not occur)
when using more short-term prefetching;
e.g., using $T=5$ compared to $T=20$, for example.
However, since smaller buffers come at high risks,
we note that these categories may benefit from more incremental prefetching algorithms
in which each tile is gradually prefetched using layered encoding.
In the next section we describe and discuss one such candidate solution.

\begin{figure}[t]
  \centering
  \subfigure[Rides]{
    \includegraphics[trim = 8mm 2mm 8mm 0mm, width=0.24\textwidth]{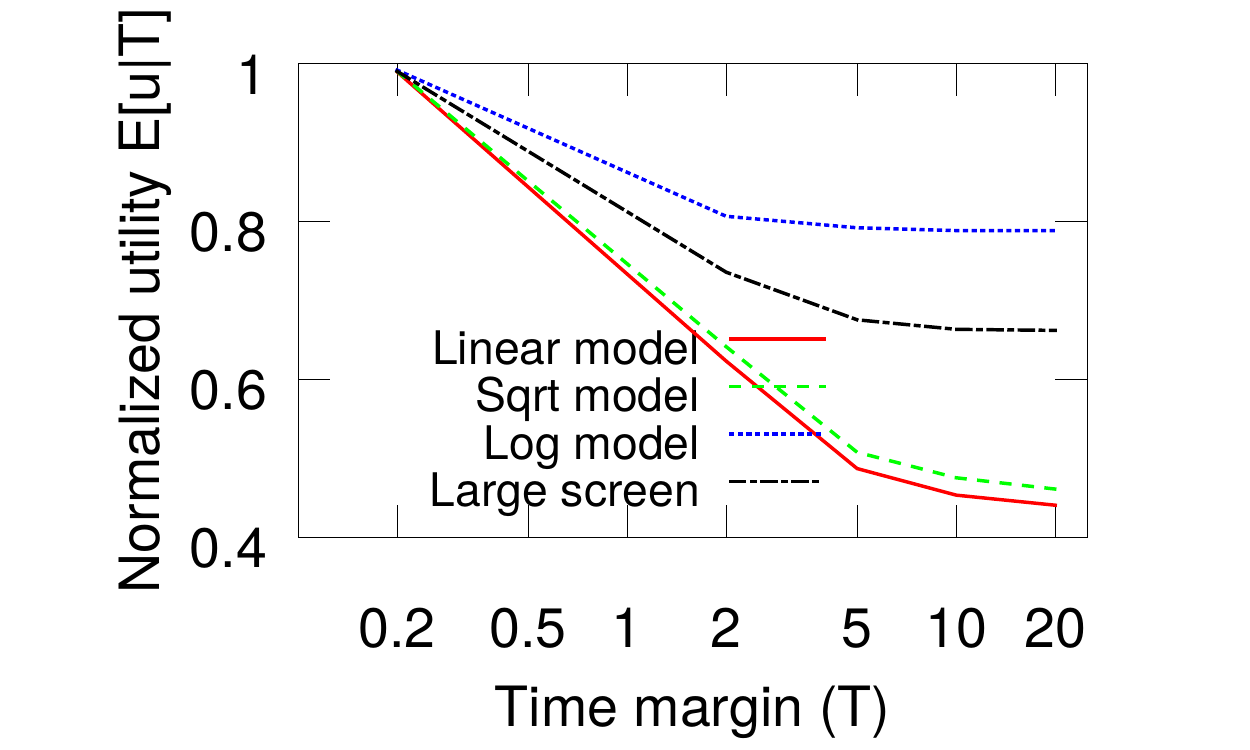}}
  \hspace{-12pt}
  \subfigure[Exploration]{
    \includegraphics[trim = 8mm 2mm 8mm 0mm, width=0.24\textwidth]{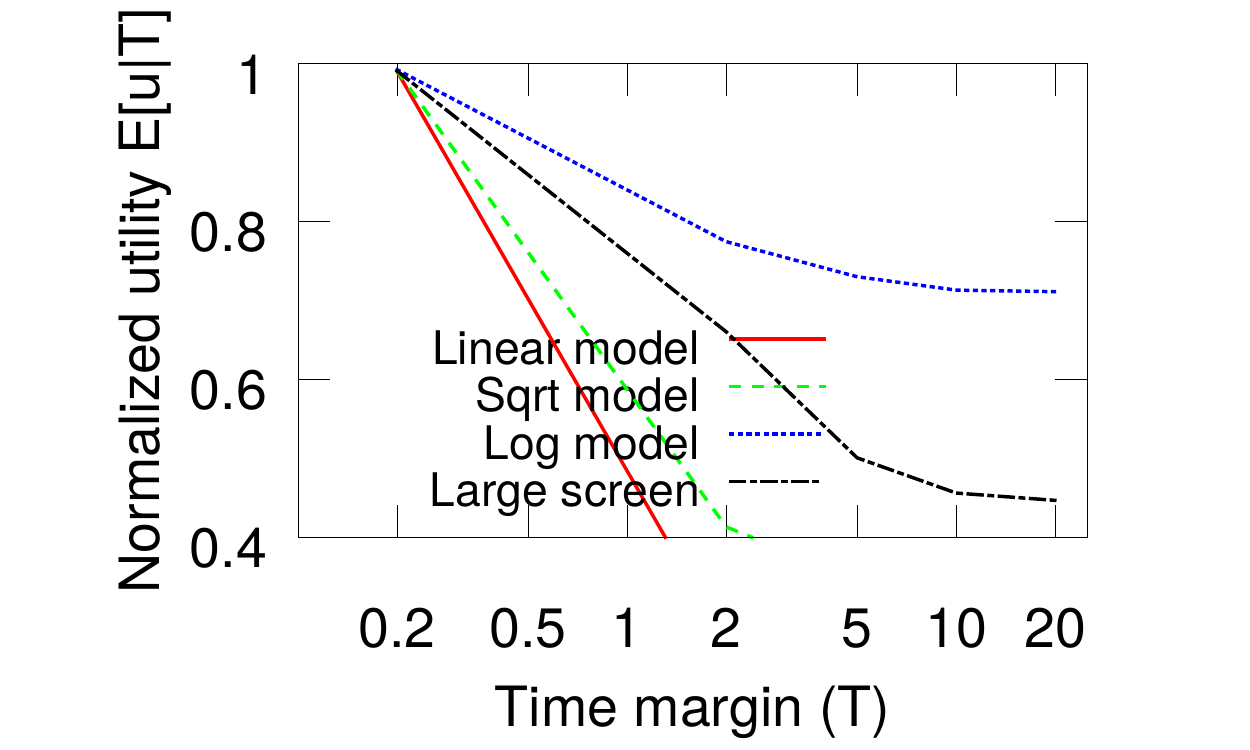}}
  \hspace{-12pt}
  \subfigure[Moving focus]{
    \includegraphics[trim = 8mm 2mm 8mm 0mm, width=0.24\textwidth]{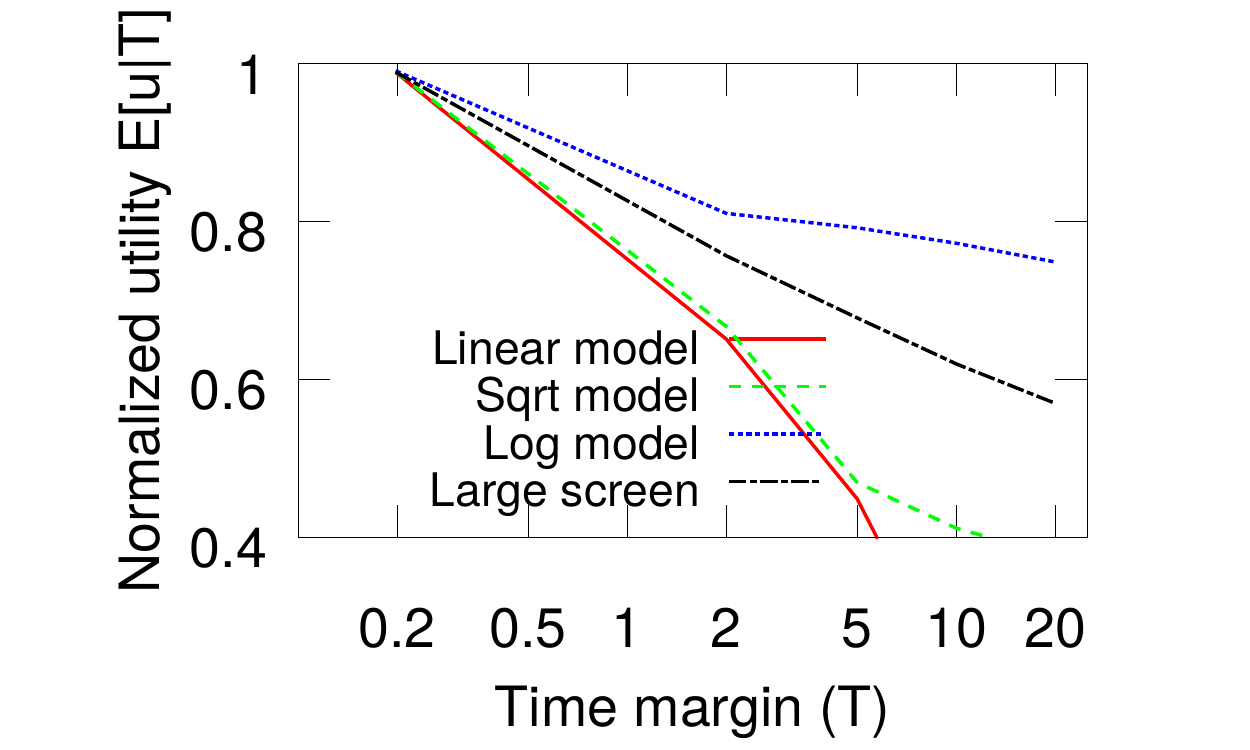}}
  \hspace{-12pt}
  \subfigure[Static Focus]{
    \includegraphics[trim = 8mm 2mm 8mm 0mm, width=0.24\textwidth]{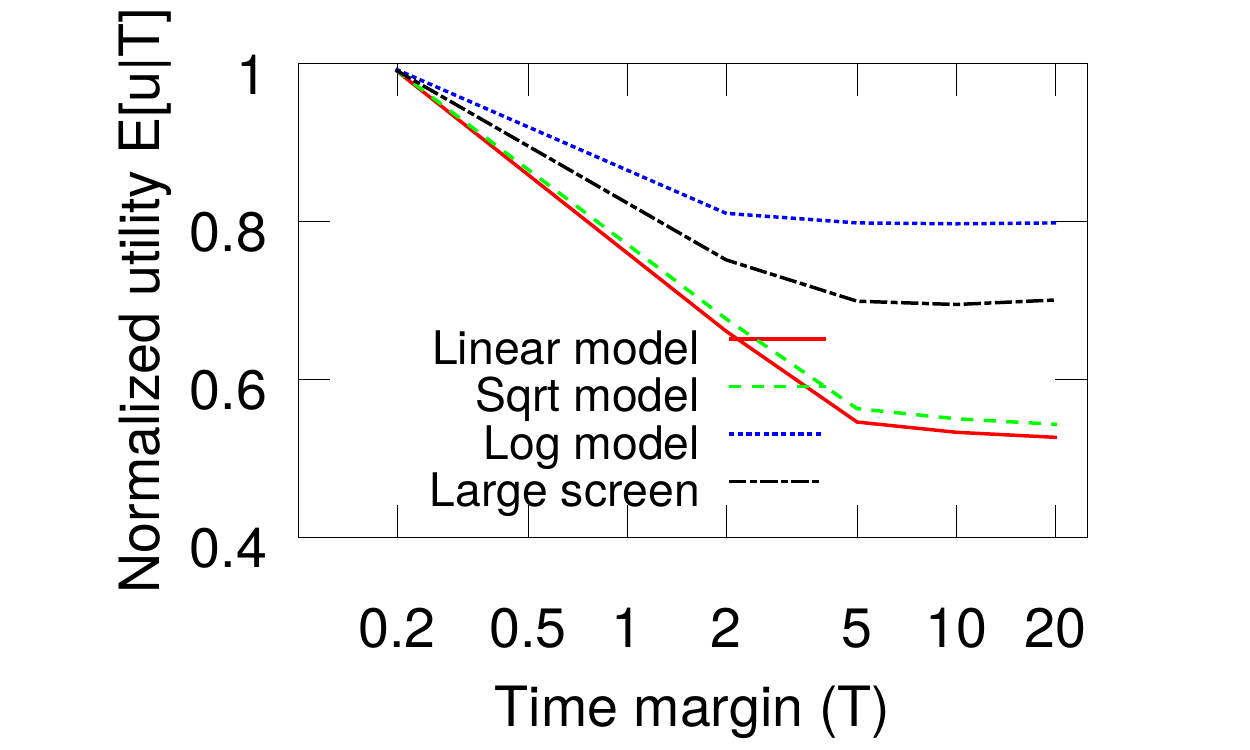}}
  \vspace{-12pt}
  \caption{Example tradeoffs for different utility functions. ($C=2500$, $\frac{u_{n,0}}{u_{n,L}}=-1$, $\beta=0.001$.)}
  \label{fig:tradeoff-utility}
  \vspace{-6pt}
\end{figure}

Comparing the sub-arxiv-figs themselves, we note a general ordering between the utility functions.
This is
\revtwo{perhaps more}{more}
clearly illustrated in Figure~\ref{fig:tradeoff-utility},
where we have extracted the corresponding
\revrev{curves for two example categories: static focus and exploration.}{curves for each of the example categories.}
These results show that a user's sensitivity to temporary quality degradations after sudden head movements
(captured by the different utility functions) significantly
\revtwo{impact}{impacts}
the importance of more accurate
prediction, or as we will see next use of additional prefetch capacity.
For example, the linear (pessimistic) assumption consistently results in the lowest utility,
while the logarithmic (optimistic) assumption consistently results in the highest utility.
For the reminder of this section,
\revtwo{we will focus}{we focus}
on the large-screen model,
which is motivated by existing work~\cite{VVB+13} and provides intermediate results.

To study the impact of various system parameters and conditions,
we next show example results for
\revtomm{the two extreme categories: static focus and exploration.
The results for the other two categories typically fall in between,
with results for rides being more similar to those of static focus and
the results for moving focus being more similar to the exploration results.}{each of the four categories,
  but note that static focus and exploration typically reprsent the two extreme categories,
  with the results for the other two categories typically falling in between,
  the results for rides being more similar to those of static focus and
  the results for moving focus being more similar to the exploration results.}
Throughout this analysis we
use the large-screen utility function,
consider one parameter at a time,
and use the following default values:
penalty factor $\frac{u_{n,0}}{u_{n,L}}=-1$, $\beta=0.001$, and capacity $C=2500$.

\begin{figure}[t]
  \centering
  \subfigure[Rides]{
    \includegraphics[trim = 8mm 2mm 8mm 0mm, width=0.24\textwidth]{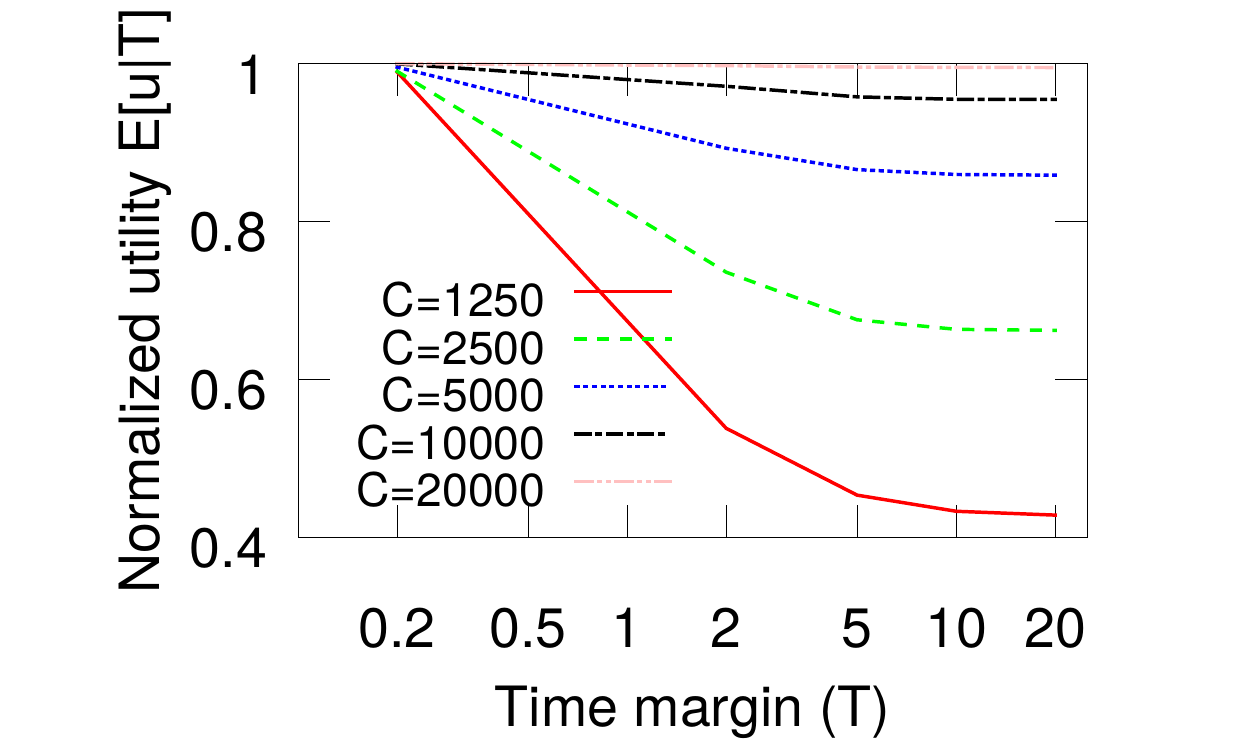}}
  \hspace{-12pt}
  \subfigure[Exploration]{
    \includegraphics[trim = 8mm 2mm 8mm 0mm, width=0.24\textwidth]{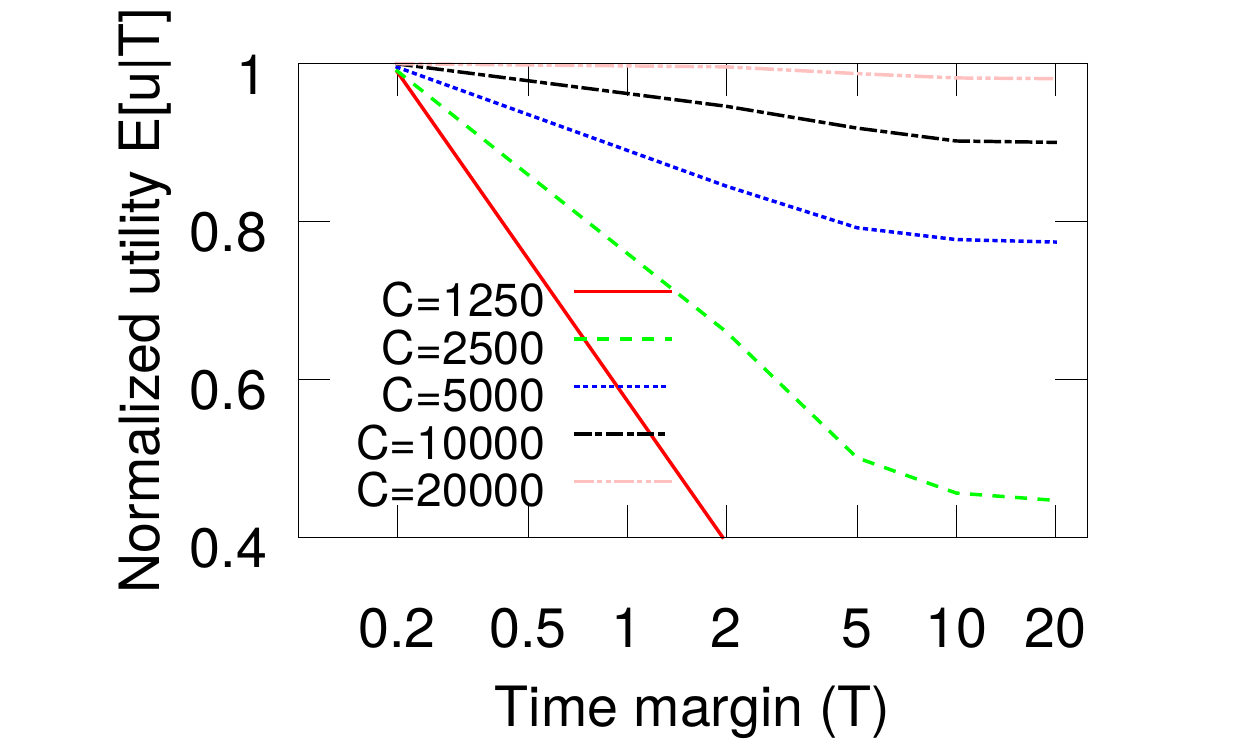}}
  \hspace{-12pt}
  \subfigure[Moving focus]{
    \includegraphics[trim = 8mm 2mm 8mm 0mm, width=0.24\textwidth]{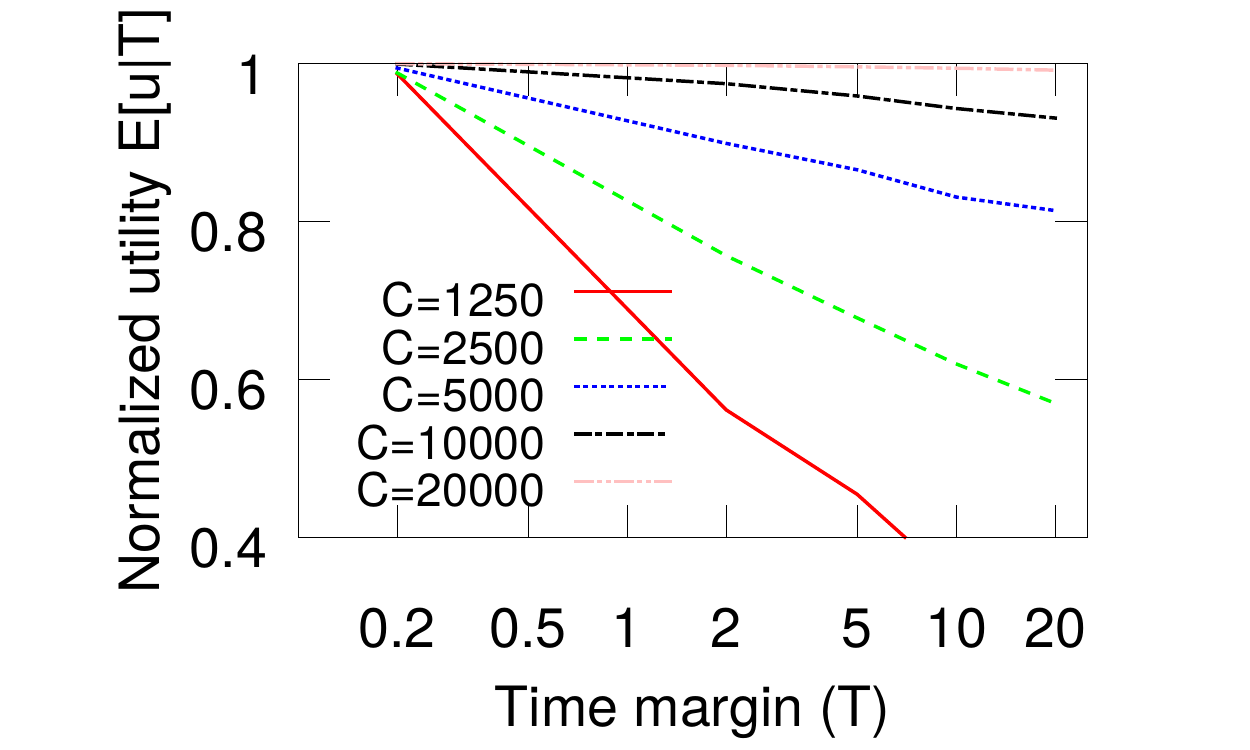}}
  \hspace{-12pt}
  \subfigure[Static focus]{
    \includegraphics[trim = 8mm 2mm 8mm 0mm, width=0.24\textwidth]{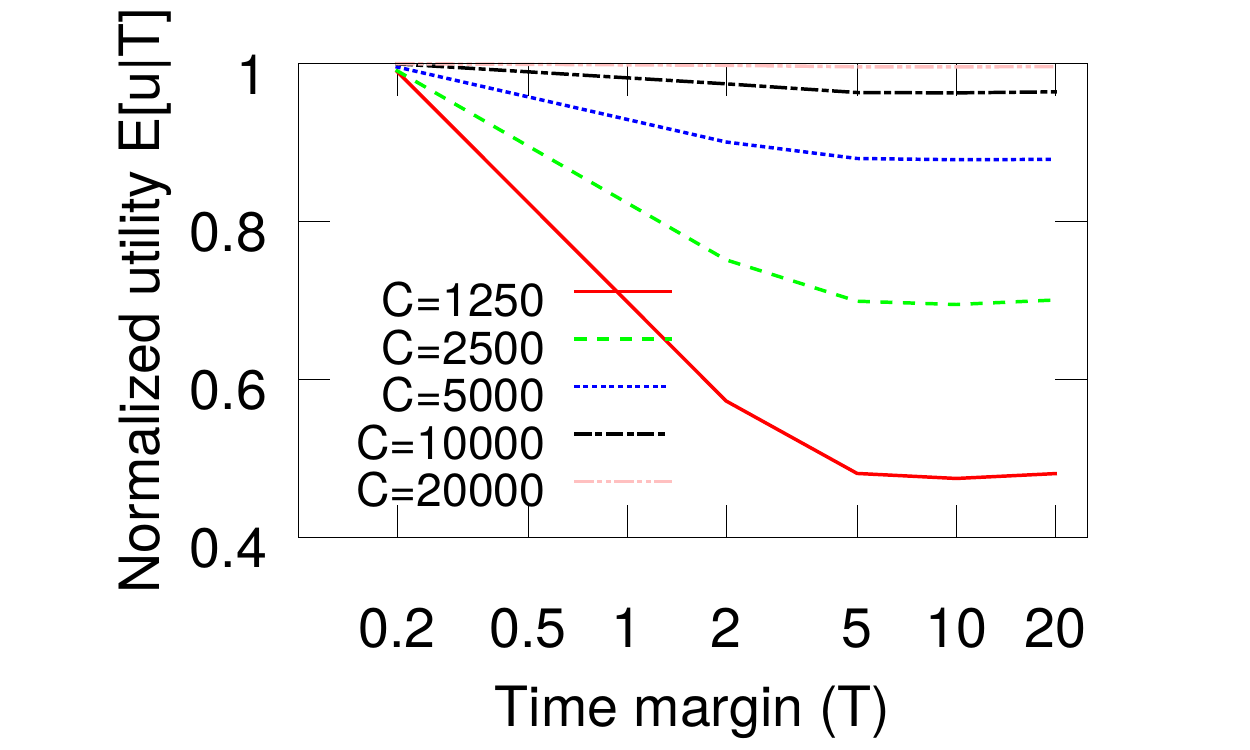}}
  \vspace{-12pt}
  \caption{Example tradeoffs for different capacities. (Large-screen utility function with $\frac{u_{n,0}}{u_{n,L}}=-1$ and $\beta=0.001$.)}
  \label{fig:tradeoff-capacity}
  \vspace{-6pt}
\end{figure}

{\bf Diminishing prefetch capacity returns:}
Figure~\ref{fig:tradeoff-capacity} illustrates the impact of the capacity $C$.
We observe diminishing returns from doubling the capacity (from 1,250 all the way up to 20,000)
and note that even a capacity of 5000 is able to achieve a utility of 0.837 and 0.737 even with $T=20$ seconds.
For static focus this is achieved by selecting rates somewhat more
aggressive rates towards the front (with tile qualities: 1$\times$2217, 4$\times$625, 1$\times$268),
than for the exploration category (2$\times$1124, 4$\times$625).

\begin{figure}[t]
    \centering
    \subfigure[Rides]{
      \includegraphics[trim = 8mm 2mm 8mm 0mm, width=0.24\textwidth]{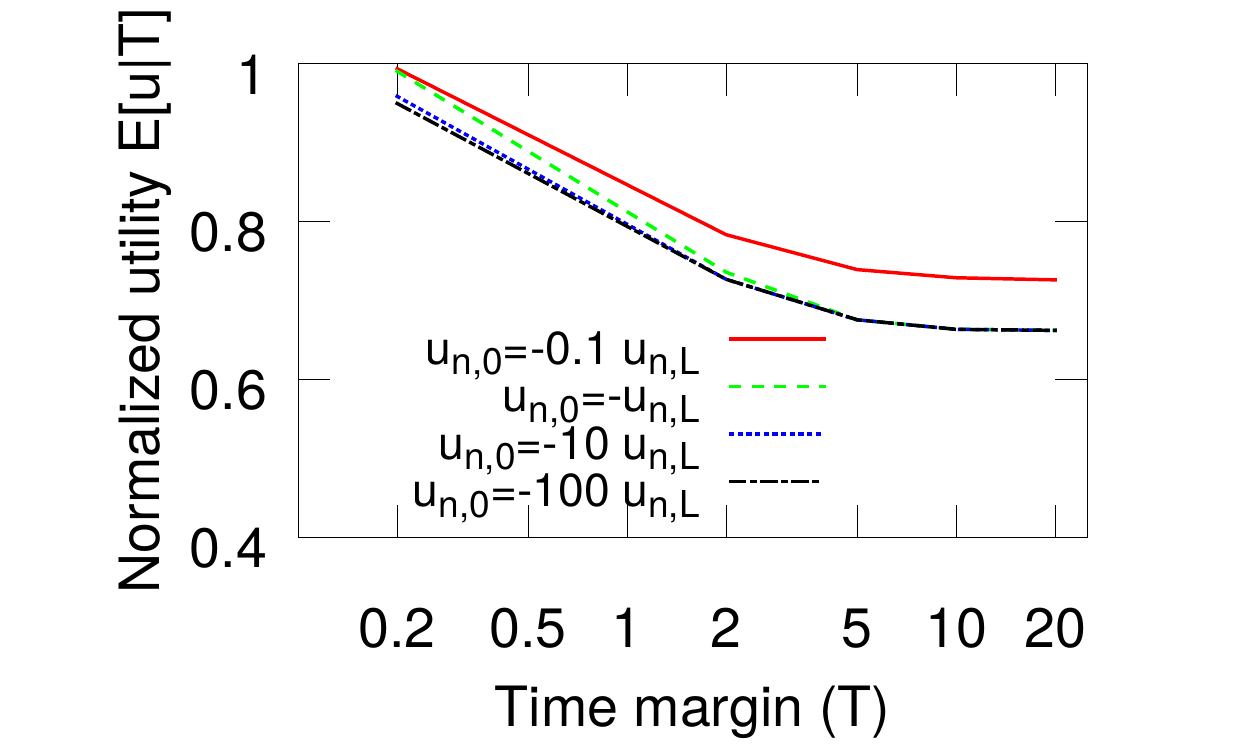}}
    \hspace{-12pt}
    \subfigure[Exploration]{
      \includegraphics[trim = 8mm 2mm 8mm 0mm, width=0.24\textwidth]{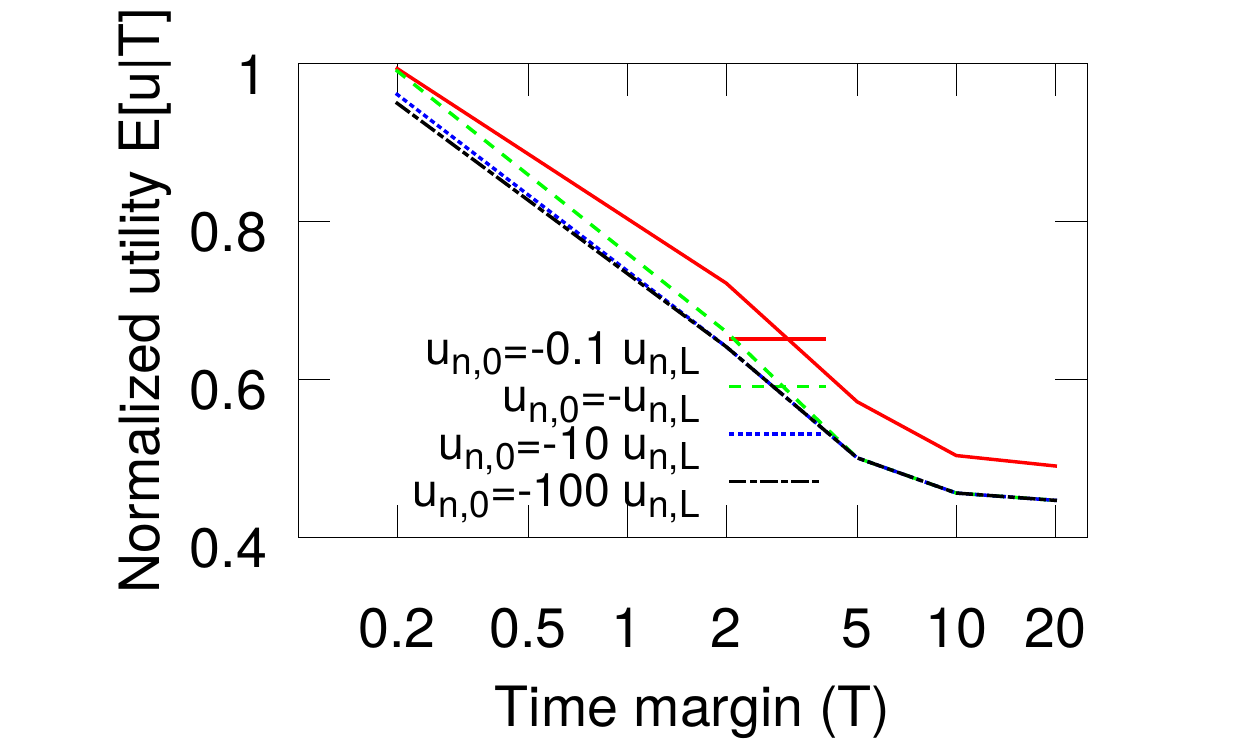}}
    \hspace{-12pt}
    \subfigure[Moving focus]{
      \includegraphics[trim = 8mm 2mm 8mm 0mm, width=0.24\textwidth]{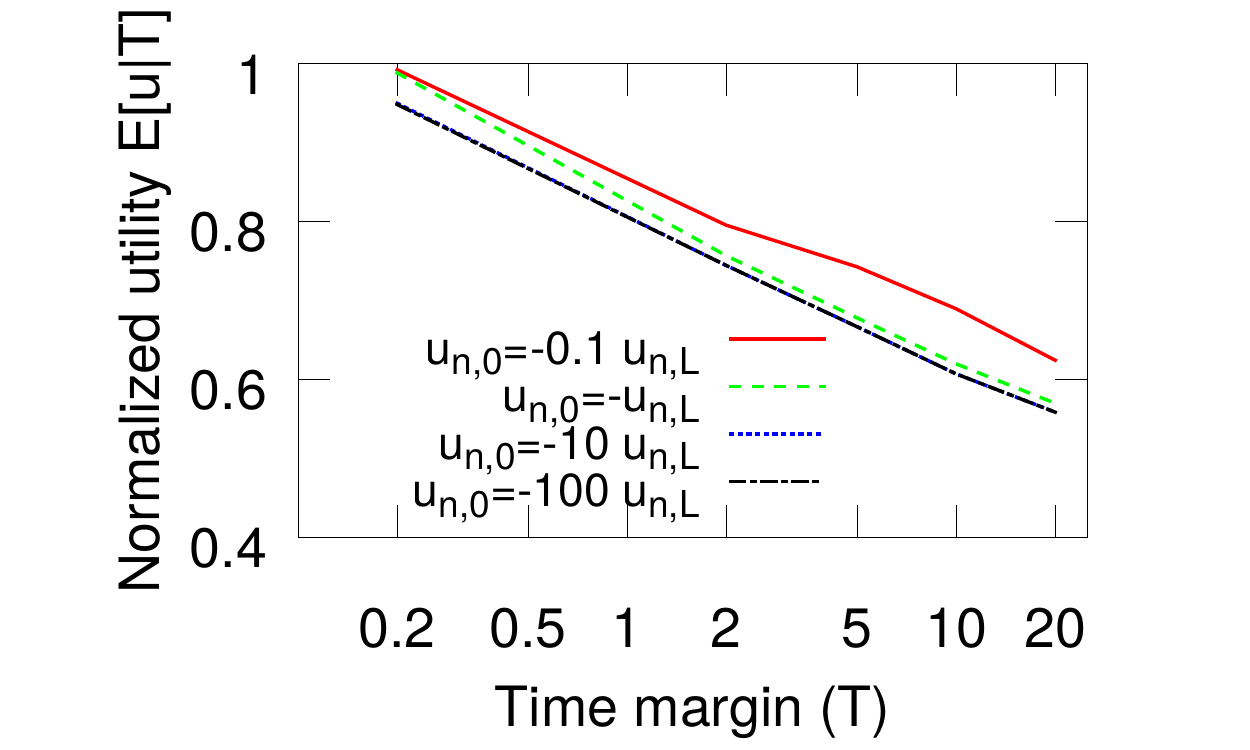}}
    \hspace{-12pt}
    \subfigure[Static focus]{
      \includegraphics[trim = 8mm 2mm 8mm 0mm, width=0.24\textwidth]{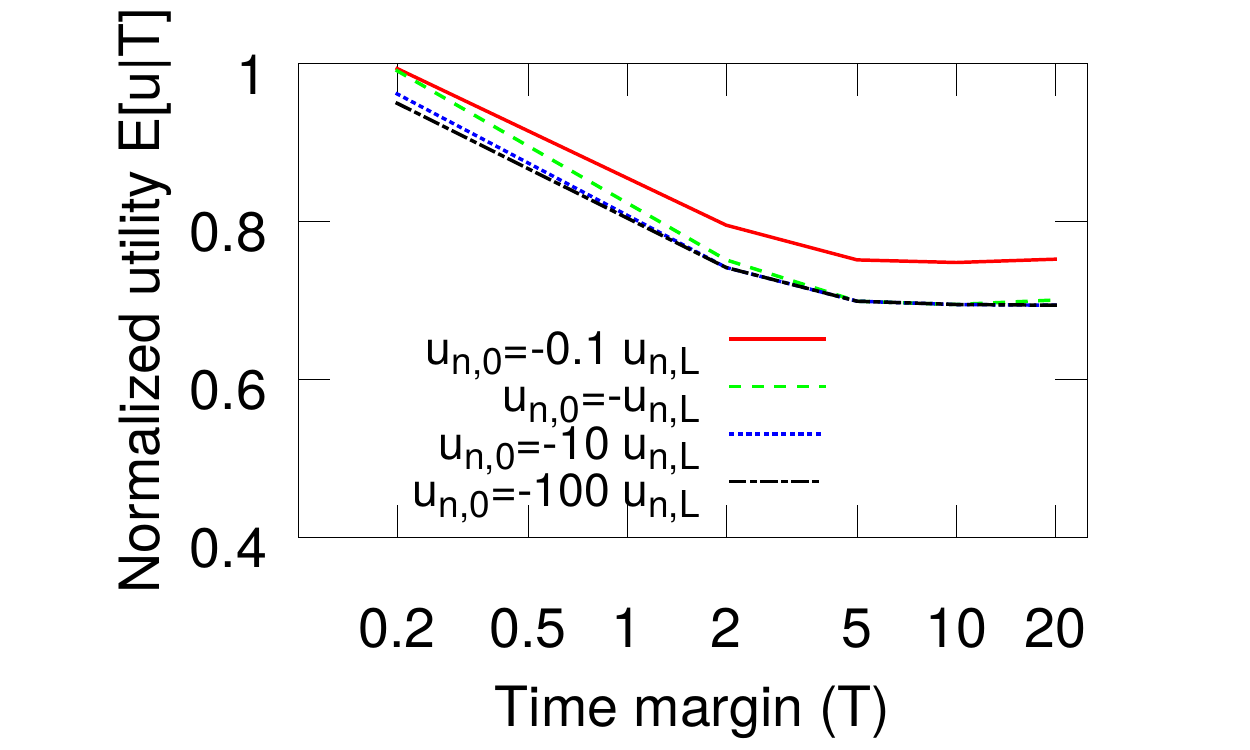}}
    \vspace{-12pt}
    \caption{Example tradeoffs for different penalties $\frac{u_{n,0}}{u_{n,L}}$. (Large-screen utility function with $C=2500$ and $\beta=0.001$.)}
    \label{fig:tradeoff-penalty}
    \vspace{-6pt}
  \end{figure}

{\bf Limited impact of stall penalty:}
Figure~\ref{fig:tradeoff-penalty} shows example tradeoffs for different stall penalty factors $\frac{u_{n,0}}{u_{n,L}}$,
ranging from small (-0.1) to large (-100).
We note that there only are very small differences observed here.
In fact, for capacities $C=5000$ and larger,
the results are independent of the stall penalty,
since the optimal solutions with these capacities always involved obtaining at least
the lowest encoding rate for each tile.  This highlights the importance of always protecting
against stalls.

\begin{figure}[t]
  \centering
  \subfigure[Rides]{
    \includegraphics[trim = 8mm 2mm 8mm 0mm, width=0.24\textwidth]{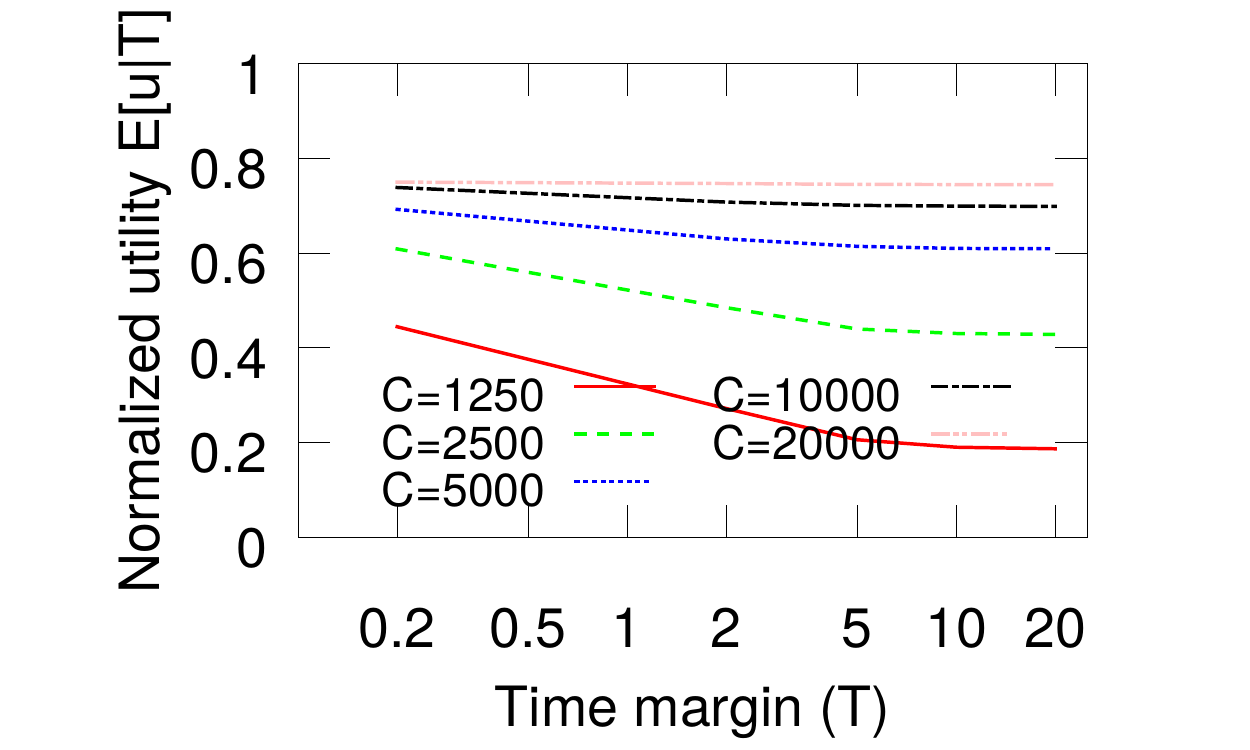}}
  \hspace{-12pt}
  \subfigure[Exploration]{
    \includegraphics[trim = 8mm 2mm 8mm 0mm, width=0.24\textwidth]{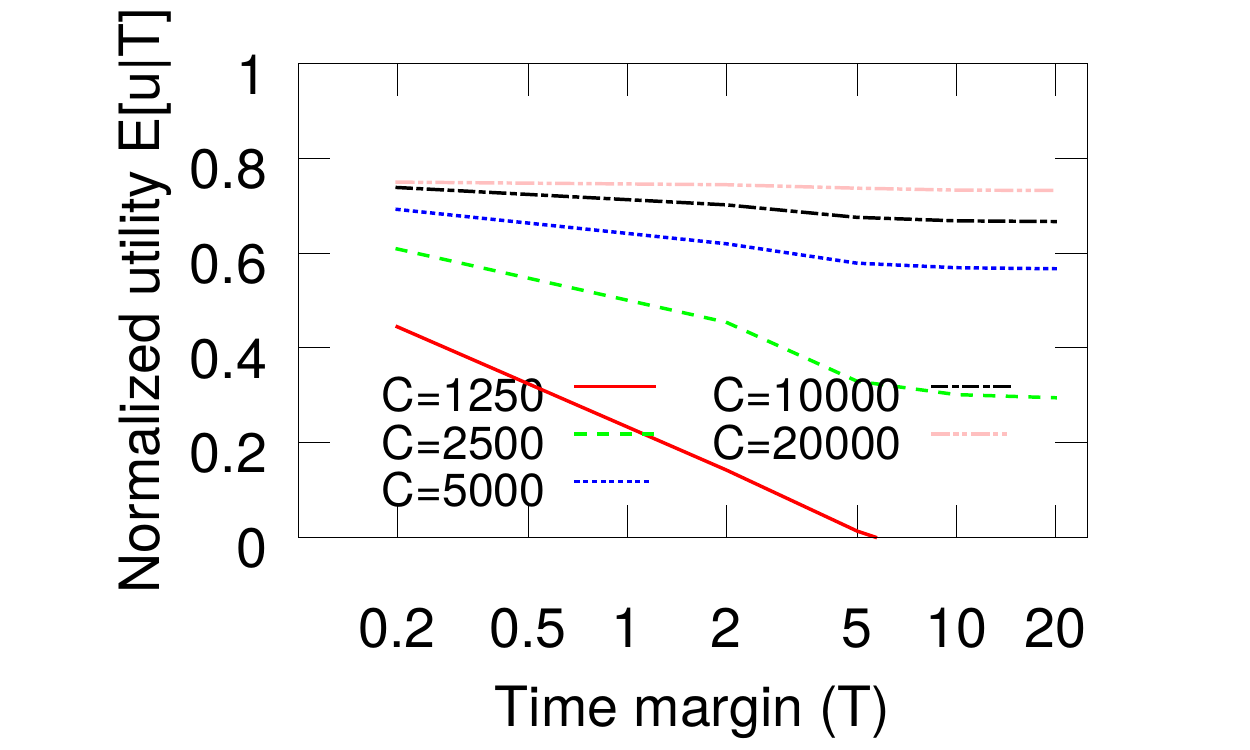}}
   \hspace{-12pt}
  \subfigure[Moving focus]{
    \includegraphics[trim = 8mm 2mm 8mm 0mm, width=0.24\textwidth]{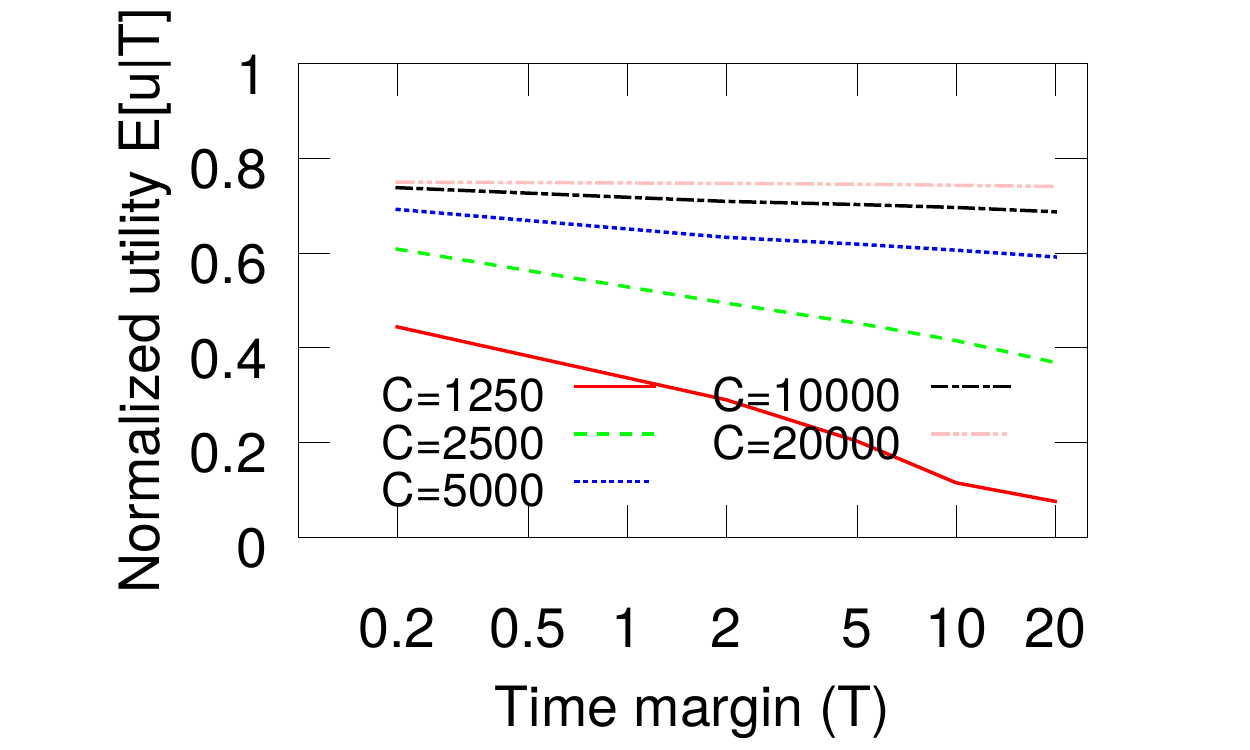}}
  \hspace{-12pt}
  \subfigure[Static focus]{
    \includegraphics[trim = 8mm 2mm 8mm 0mm, width=0.24\textwidth]{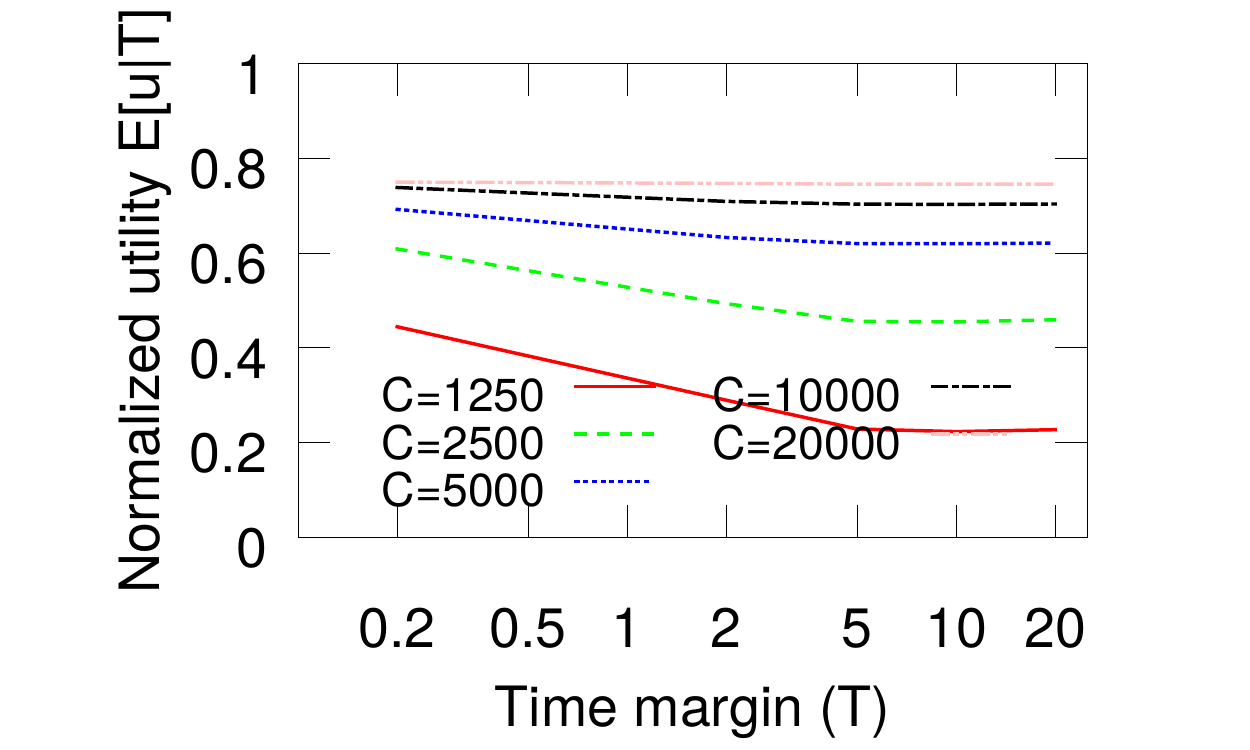}}
  \vspace{-12pt}
  \caption{Capacity tradeoffs with large $\beta=0.25$. (Large-screen utility function with $\frac{u_{n,0}}{u_{n,L}}=-1$.)}
  \label{fig:tradeoff-capacity-alpha025}
  \vspace{-6pt}
  \end{figure}

{\bf Impact of weight given to quality differences between neighboring tiles:}
While we expect that the most realistic $\beta$ typically would be small,
we have experimented with different $\beta$ values.  In general,
a larger (negative) $\beta$ factor does not impact the quality choices substantially,
but does shift the weighted utility curves downwards.
This is illustrated by comparing Figure~\ref{fig:tradeoff-capacity-alpha025}
(with $\beta=0.25$) and Figure~\ref{fig:tradeoff-capacity} (with $\beta=0.001$).
As expected, we have found that there are somewhat more evenly distributed quality selections
with $\beta=0.25$.  For example, the right-most points for the $C=5000$ curves for
static focus (which we discussed above) now use (3$\times$1124, 2$\times$625, 1$\times$268),
compared to (1$\times$2217, 4$\times$625, 1$\times$268) with $\beta=0.001$.
For exploration, the two optimal solutions are identical.
For similar reasons, we see somewhat flatter tradeoff curves with static focus
when $\beta$ increases, whereas they remain similar for the exploration category.
Again, remember that the absolute utilities in the two different plots
should not be compared against
each other,
only the relative shapes, since they represent substantially
different utility models.  Yet, the high $\beta$ results show that
the insights obtained for smaller $\beta$ also hold for larger $\beta$.

\begin{figure*}[t]
  \centering
  \subfigure[Linear utility]{
    \includegraphics[trim = 8mm 2mm 8mm 0mm, width=0.23\textwidth]{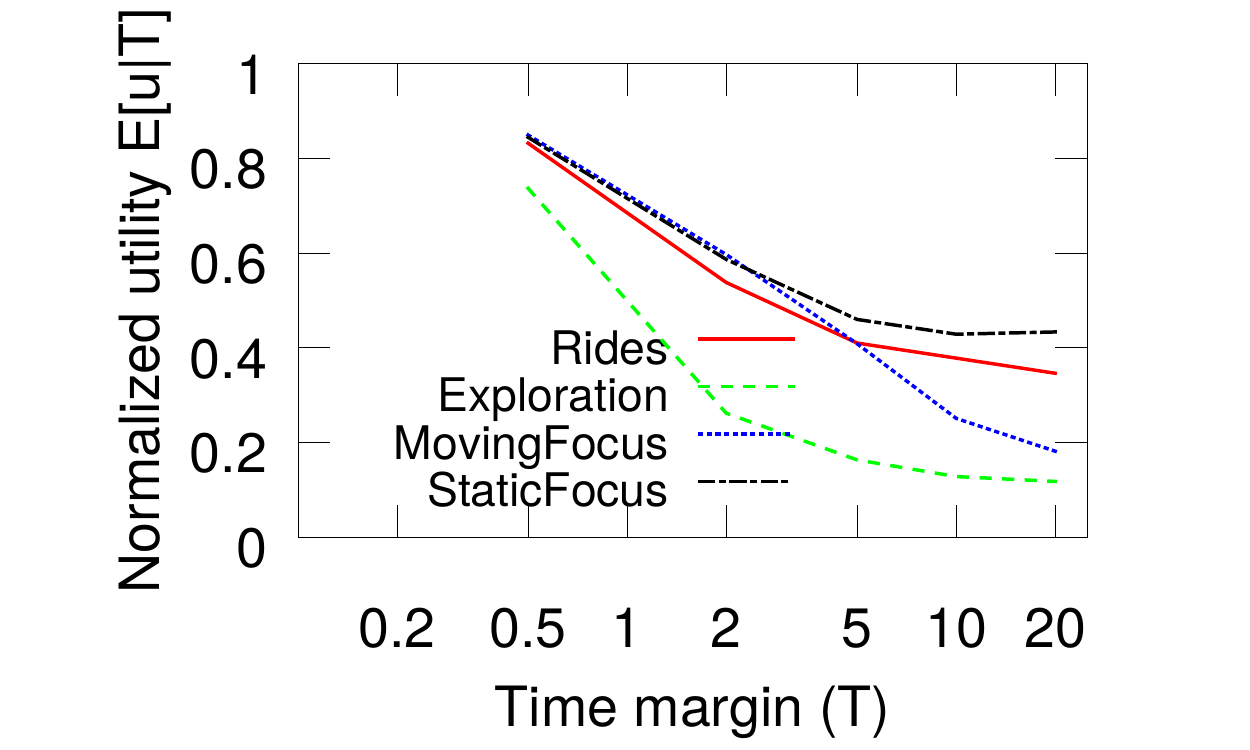}}
  \hspace{-12pt}
  \subfigure[Square-root utility]{
    \includegraphics[trim = 8mm 2mm 8mm 0mm, width=0.23\textwidth]{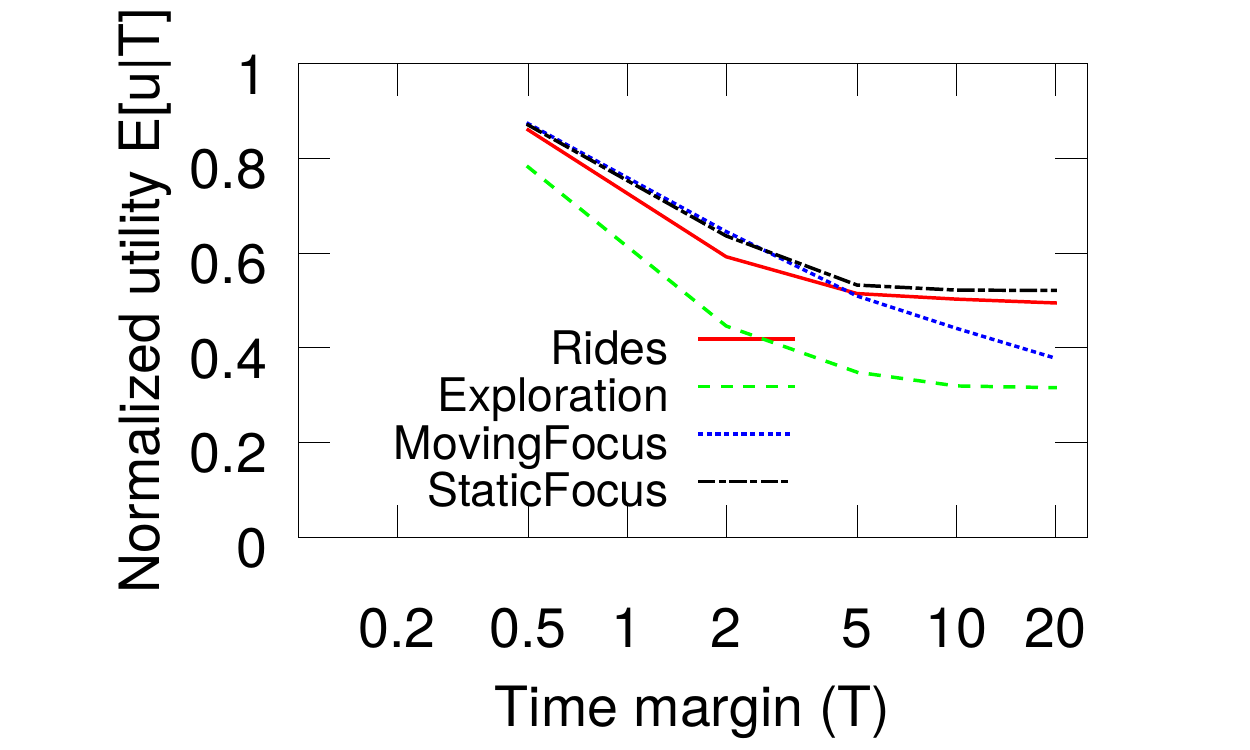}}
  \hspace{-12pt}
  \subfigure[Logarithmic utility]{
    \includegraphics[trim = 8mm 2mm 8mm 0mm, width=0.23\textwidth]{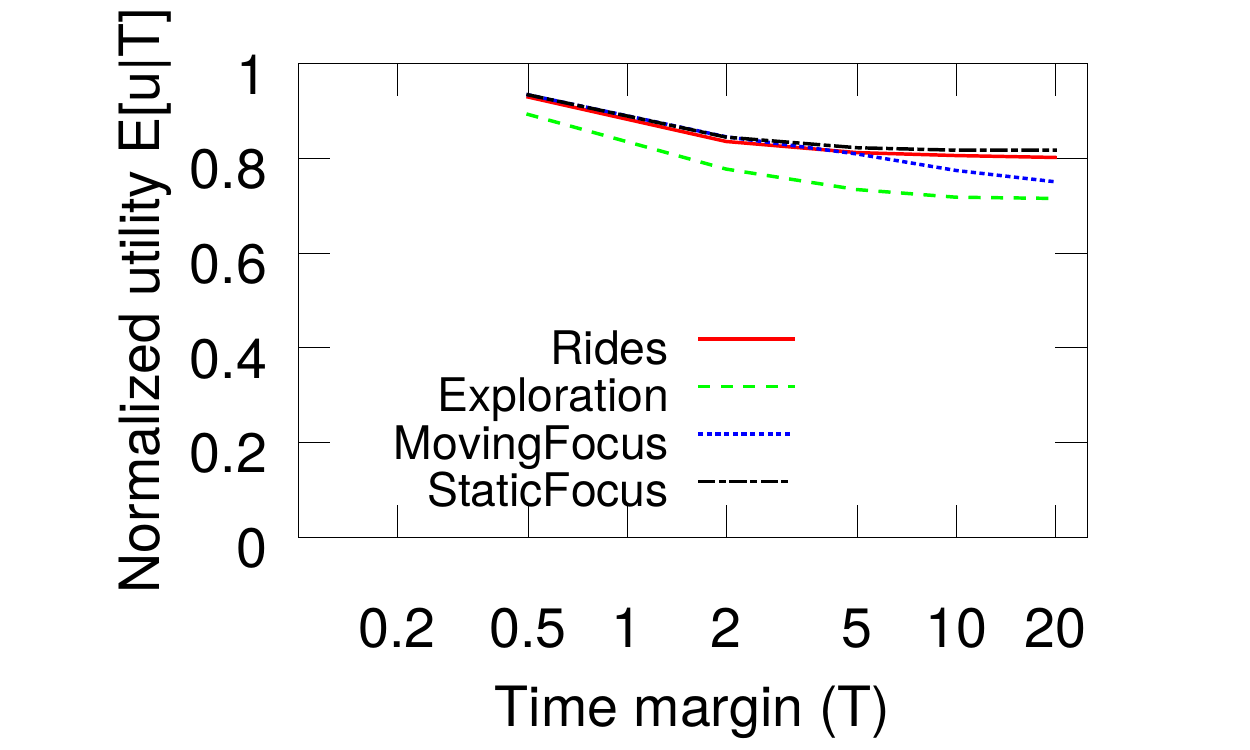}}
  \hspace{-12pt}
  \subfigure[Large-screen utility]{
    \includegraphics[trim = 8mm 2mm 8mm 0mm, width=0.23\textwidth]{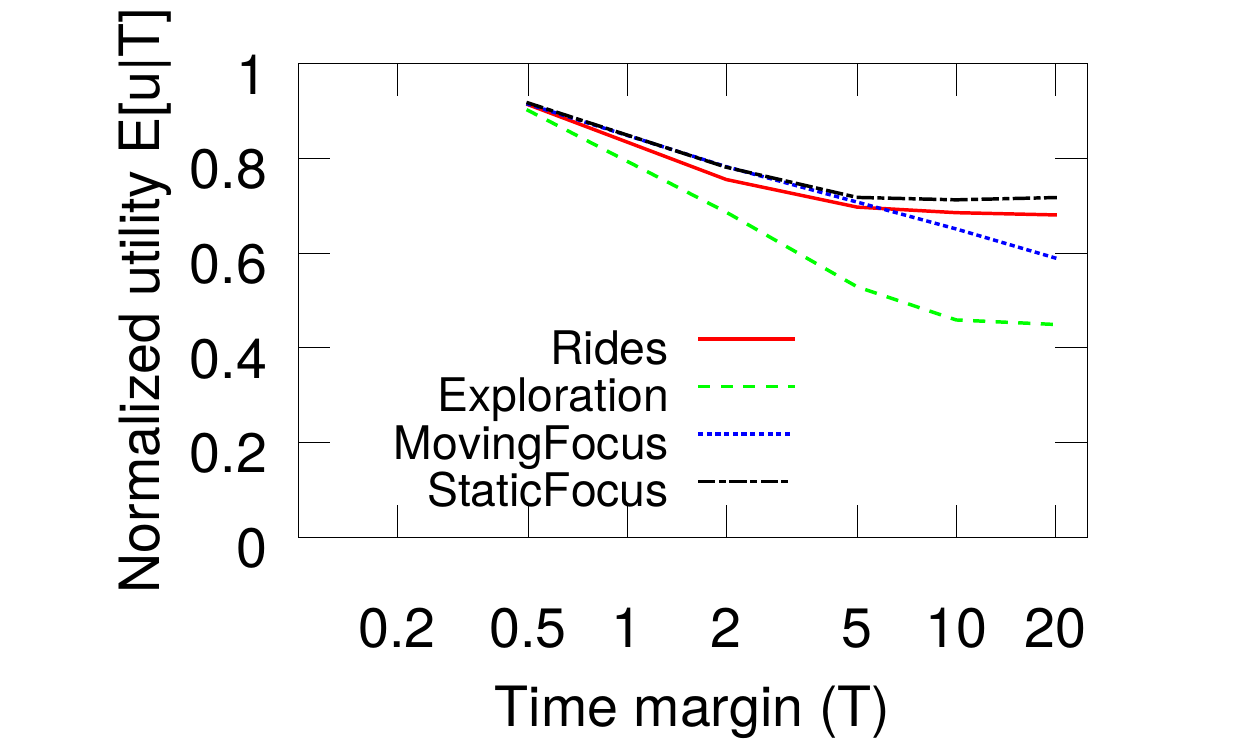}}\\
  \vspace{-12pt}
  \subfigure[Rides]{
    \includegraphics[trim = 8mm 2mm 8mm 0mm, width=0.24\textwidth]{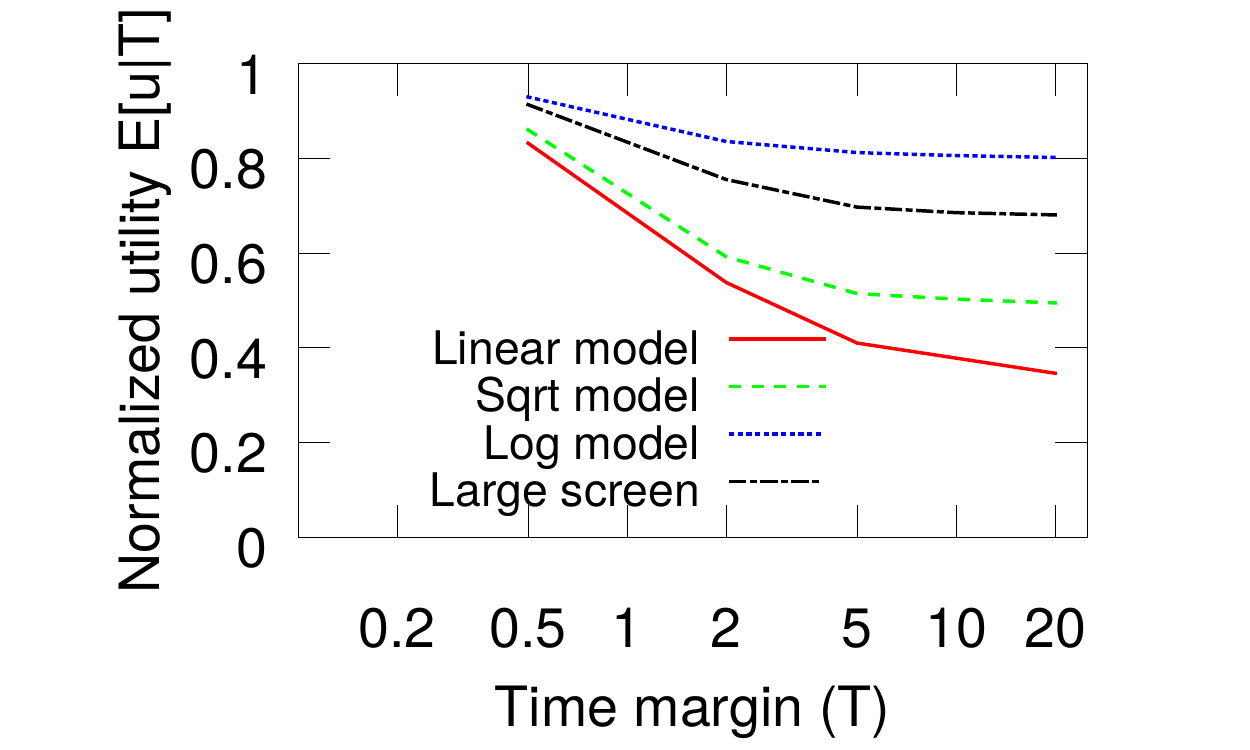}}
  \hspace{-12pt}
  \subfigure[Exploration]{
    \includegraphics[trim = 8mm 2mm 8mm 0mm, width=0.24\textwidth]{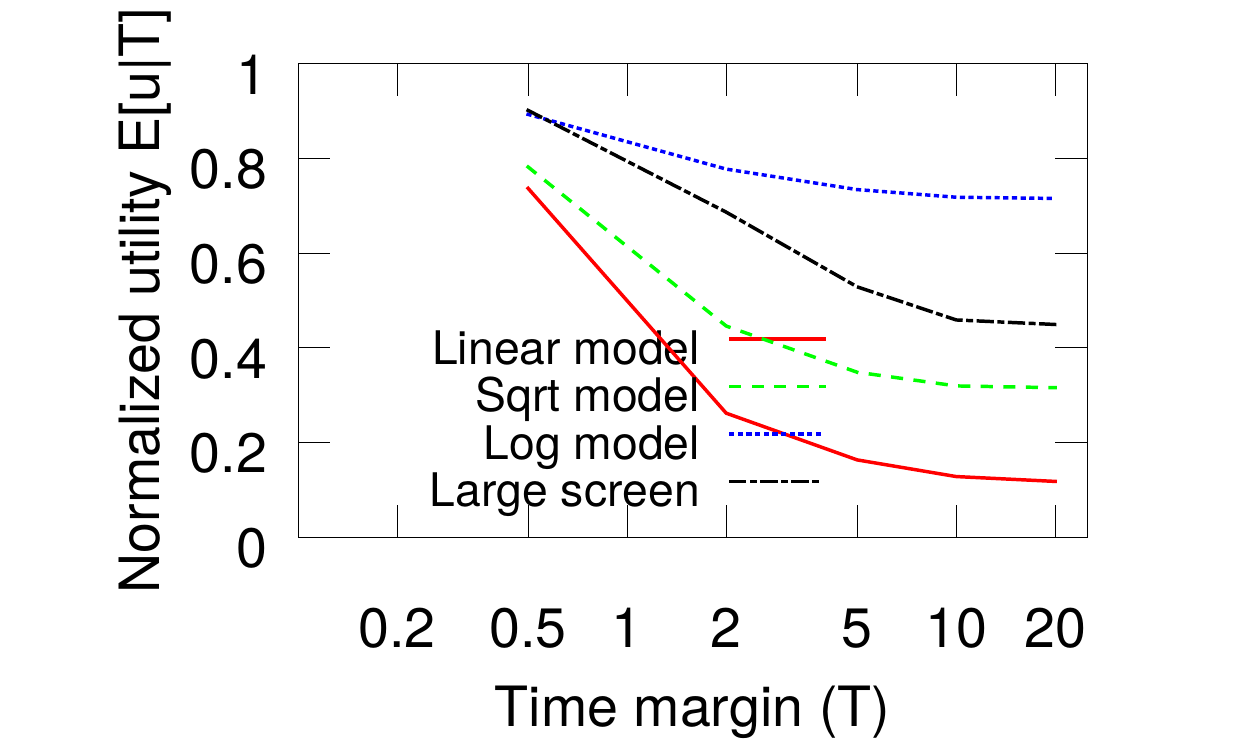}}
  \hspace{-12pt}
  \subfigure[Moving focus]{
    \includegraphics[trim = 8mm 2mm 8mm 0mm, width=0.24\textwidth]{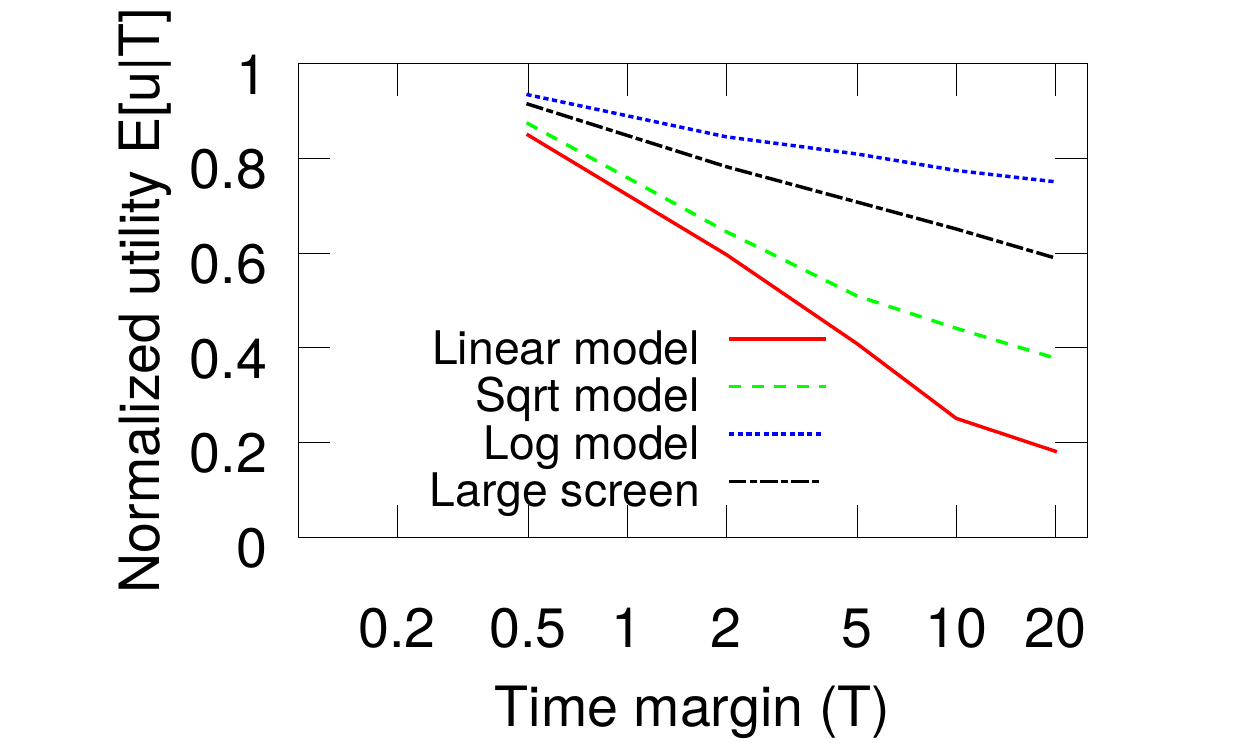}}
  \hspace{-12pt}
  \subfigure[Static Focus]{
    \includegraphics[trim = 8mm 2mm 8mm 0mm, width=0.24\textwidth]{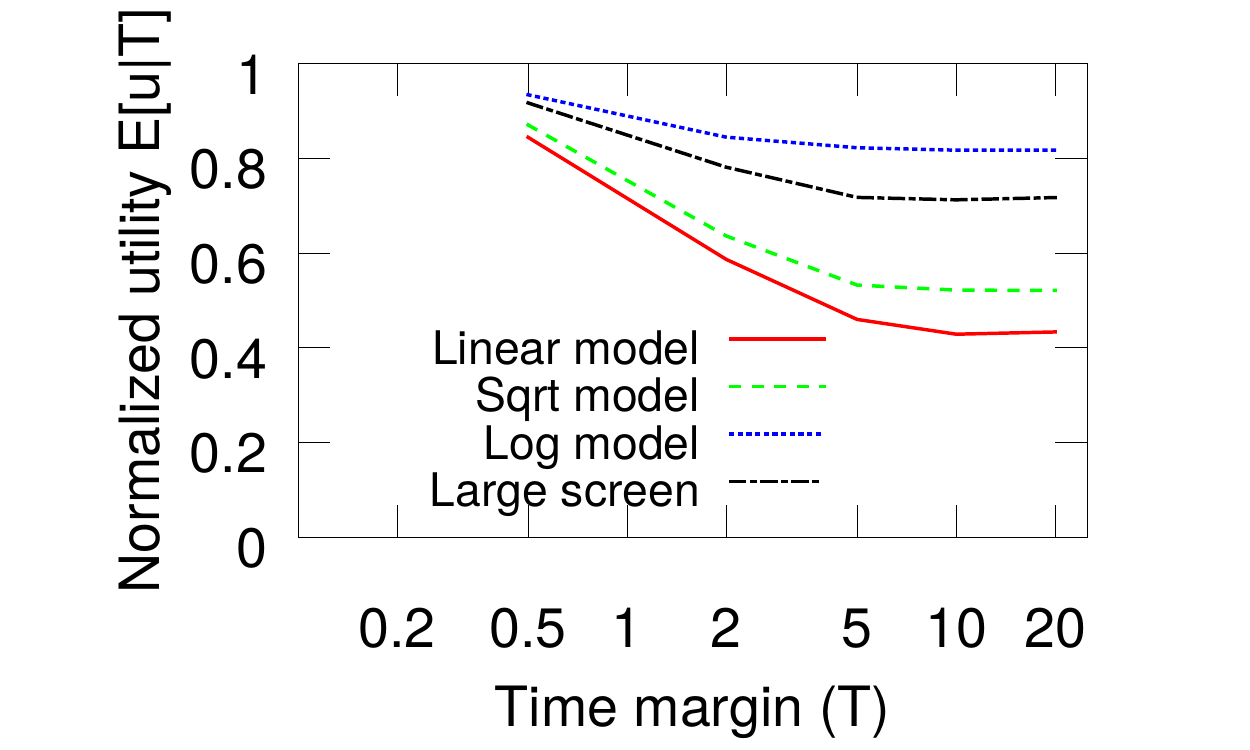}}\\
  \vspace{-12pt}
  \subfigure[Rides]{
    \includegraphics[trim = 8mm 2mm 8mm 0mm, width=0.24\textwidth]{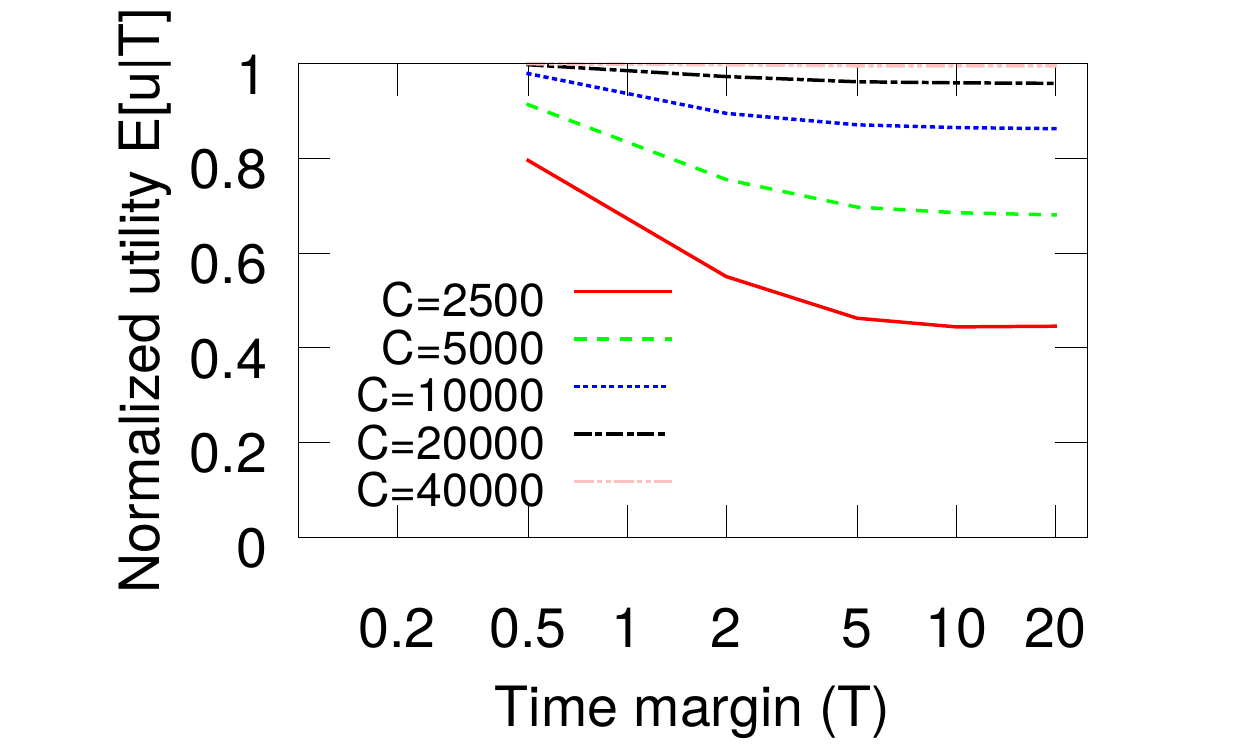}}
  \hspace{-12pt}
  \subfigure[Exploration]{
    \includegraphics[trim = 8mm 2mm 8mm 0mm, width=0.24\textwidth]{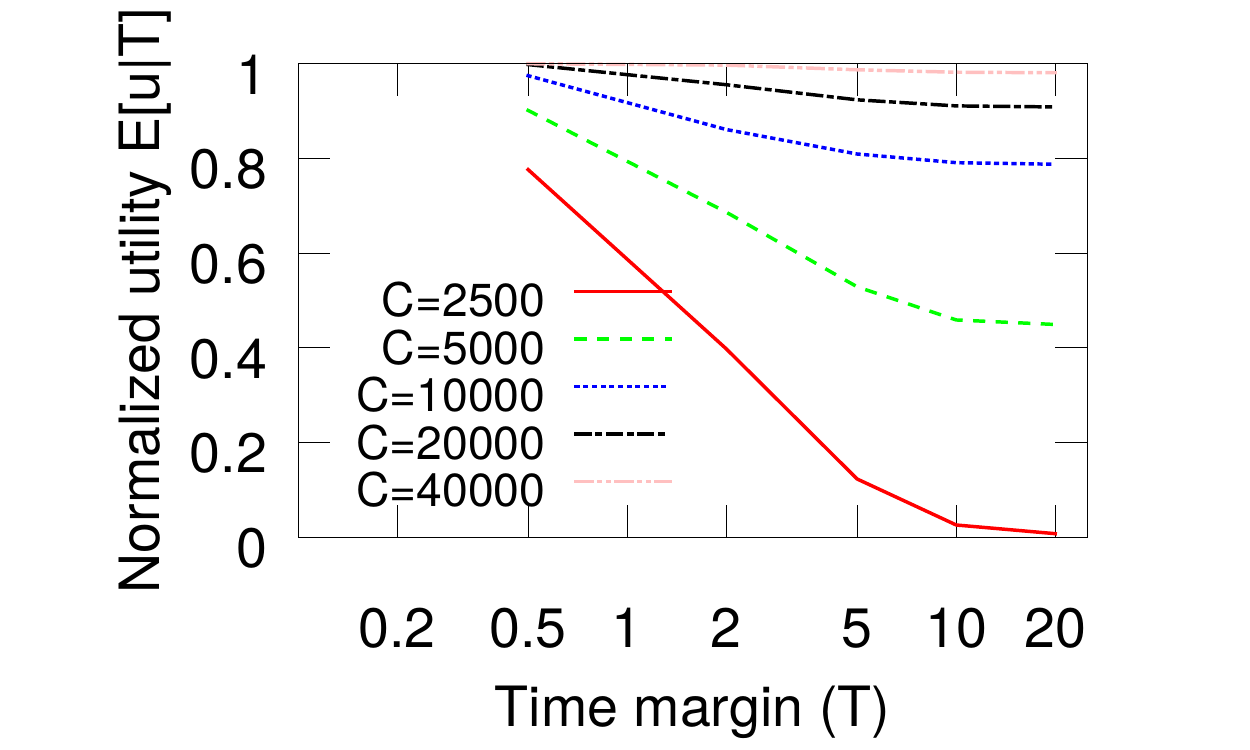}}
  \hspace{-12pt}
  \subfigure[Moving focus]{
    \includegraphics[trim = 8mm 2mm 8mm 0mm, width=0.24\textwidth]{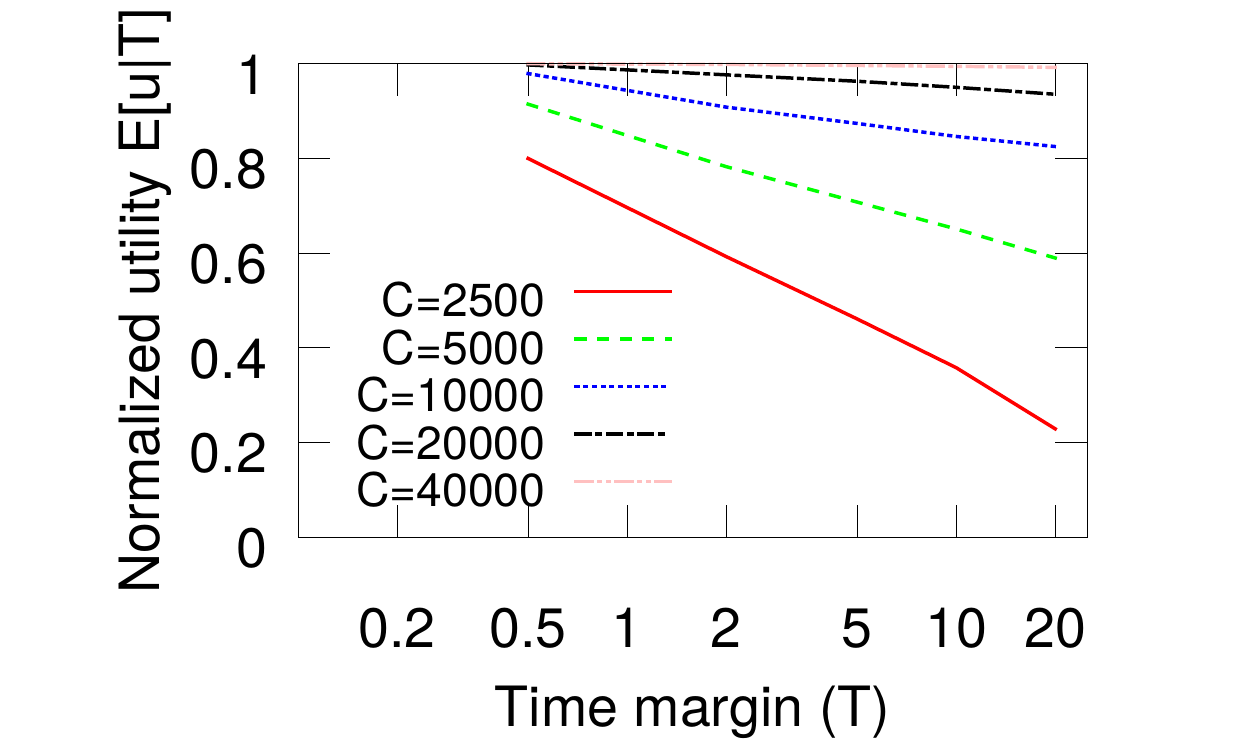}}
  \hspace{-12pt}
  \subfigure[Static focus]{
    \includegraphics[trim = 8mm 2mm 8mm 0mm, width=0.24\textwidth]{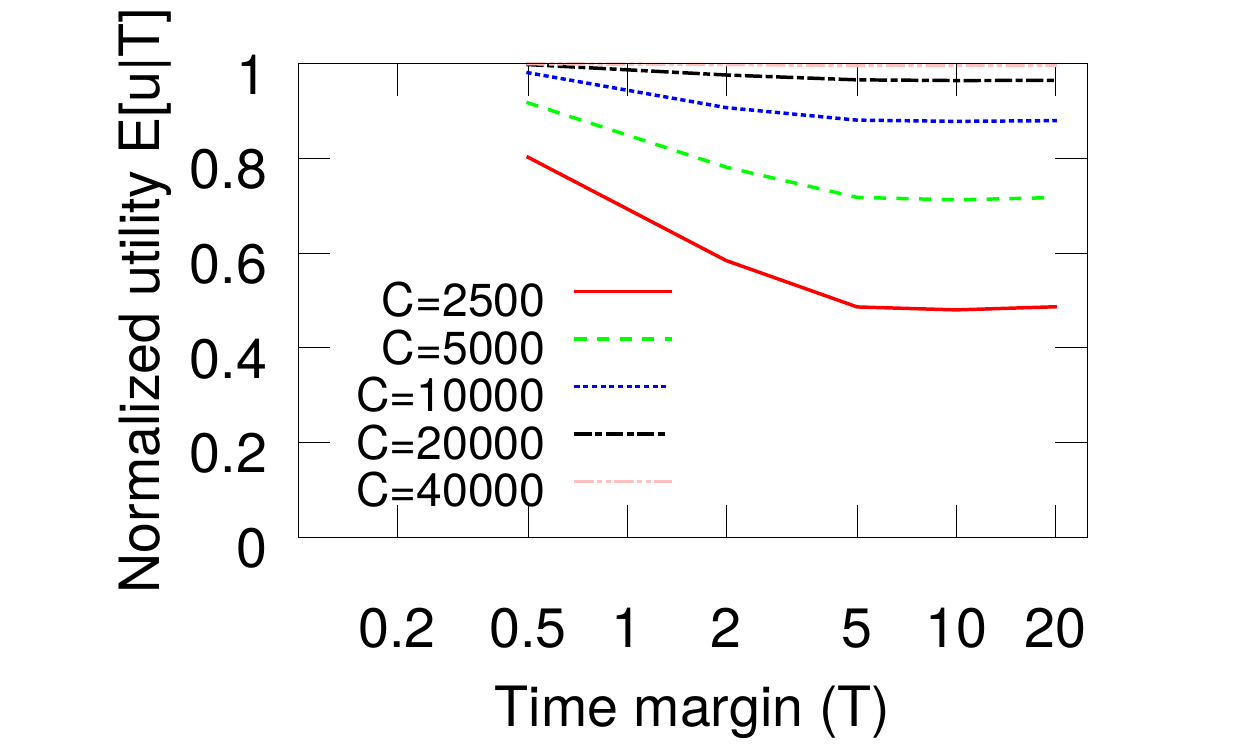}}
  \vspace{-12pt}
  \caption{Example tradeoffs with twice as many tiles $N=12$ as in Figures~\ref{fig:tradeoff-videotype},~\ref{fig:tradeoff-utility} and~\ref{fig:tradeoff-capacity}.
    (Default settings: $C=5000$, $\frac{u_{n,0}}{u_{n,L}}=-1$, $\beta=0.001$ using large-screen model.)}
  \label{fig:tradeoff-N12}
  \vspace{-6pt}
  \end{figure*}

{\bf Impact of number of tiles:}
The qualitative tradeoff results and general differences between the different categories presented here
also hold when using other number of tiles.
This is examplified in Figure~\ref{fig:tradeoff-N12},
which shows example results when using $N=12$ tiles instead of $N=6$ tiles.
In particular, the three rows in Figure~\ref{fig:tradeoff-N12} shows the
$N=12$ results corresponding to the $N=6$ results in Figures~\ref{fig:tradeoff-videotype},~\ref{fig:tradeoff-utility} and~\ref{fig:tradeoff-capacity}, respectively.
To allow a fair comparison, using our normalized per-tile units, we use double the per-chunk capacity $C$.

\section{Discussion of further design optimizations}\label{sec:discussion}  

When considering the optimal solutions, at both short and long time scales $T$,
it is typically optimal to download some minimum quality in each viewing direction
so as to protect against stalls (or missing tiles).
Motivated by this observation we argue
that prefetching can be split over multiple time scales.

{\bf Framework to split prefetching across time scales:}
In the following we describe a simple framework
that allows us to simultaneously perform both
(i) long-term prefetching so to protect against bandwidth variations and other unforeseen service outages, for example, and
(ii) fine grained optimized prefetching based on the current viewing direction, as best done closer to the playback deadline.
In its simplest form,
a client separates the prefetching process into two (or more) modules that operate in parallel.
The first module is responsible for prefetching an initial base
layer (e.g., based on SVC technology) for all viewing directions
and can use a large buffer (e.g., 20-120 seconds).
The latter prefetching module(s) then prefetch additional enhancement layers
for each tile based on more up-to-date view-direction predictions.

For simplicity, let us assume that we use two modules.
For this case,
the optimization problem that must be solved by the second module,
can easily be derived using a modified
\revtwo{variation}{version}
of the optimization
problem (and solution) described and evaluated in Section 5.  In particular,
assuming that the client has tile quality $l'$ in direction $n$,
we can simply use $b_{n,l}=0$ for all $l \le l'$,
capturing that we already have tile quality $l'$ for this direction.
For the other tile qualities, the size will depend on whether some form of layered coding is implemented or not.
For example,
in the ideal case,
assuming ``perfect'' SVC without any overhead,
we would have $b_{n,l} \approx \Delta(q_{n,l}-q_{n,l'})$.
At the other extreme,
assuming that tiles would need to be completely re-downloaded at a higher quality level,
we would have $b_{n,l} \approx \Delta q_{n,l}$.  Naturally,
different implementations will fall somewhere in between these two extremes.

\begin{figure}[t]
    \centering
    \includegraphics[trim = 0mm 32mm 0mm 0mm, clip, width=0.48\textwidth]{{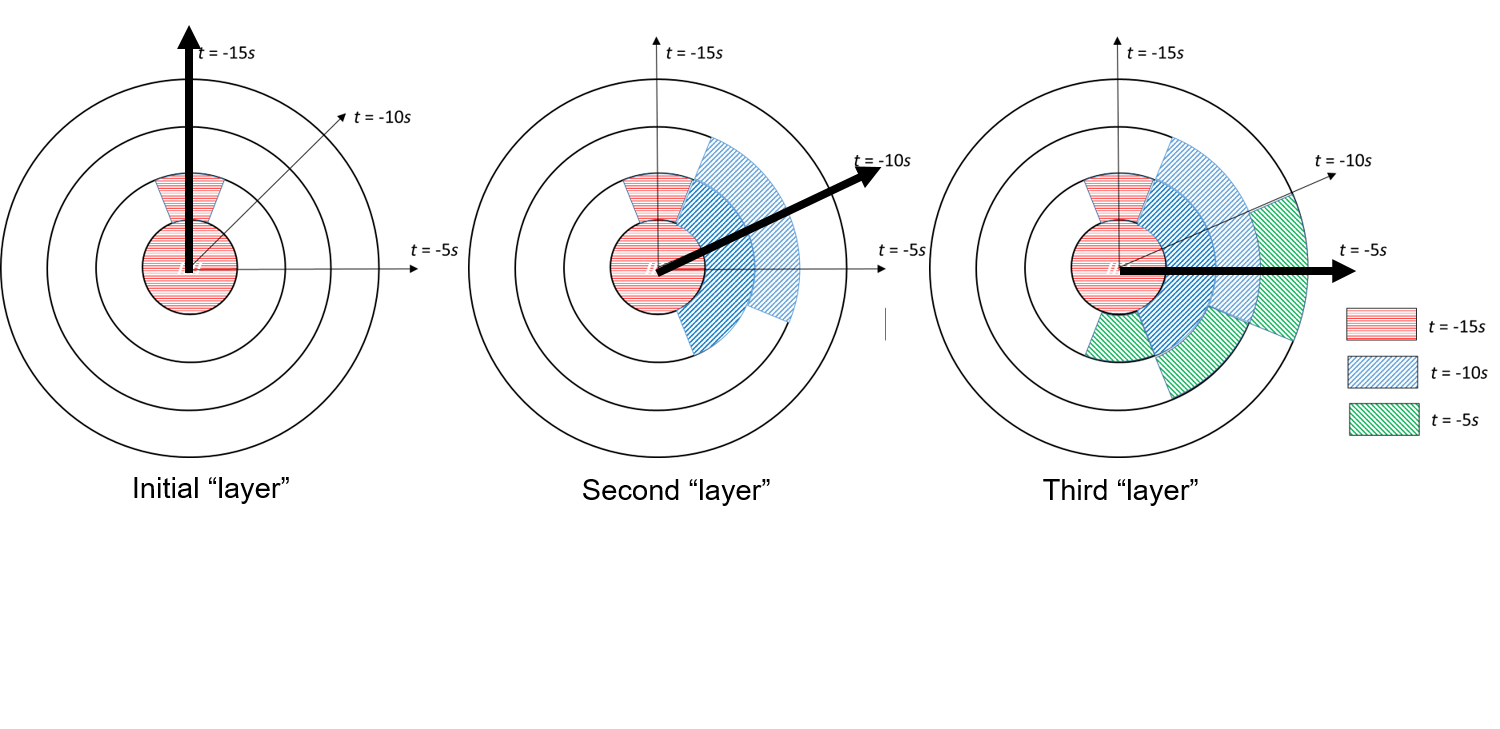}}
    \vspace{-10pt}
    \caption{Example of personalized ``layers'' based on viewing direction and downloaded tiles.}
    \label{fig:example}
    \vspace{-6pt}
\end{figure}

Figure~\ref{fig:example} shows the example layers that a client has prefetched
in each step, when using SVC together with three prefetching modules.
Here,
the first module ($A$) made its decision at time $T_{A}$ (e.g., 15 seconds before the playback deadline)
based on viewing direction $(0,1)$, the second module ($B$)
based on a direction
$(\frac{\sqrt{3}}{2},\frac{1}{2})$
and the final module ($C$)
based on viewing direction $(1,0)$.
In each step, the previously prefetched tile qualities
are leveraged so to best use the available prefetching
capacity of that module,
each module solving the above optimization problem.

We next use our measurements to
study
additional
biases that can be used to further optimize the final
quality selections.

\begin{figure}[t]
  \centering
  \includegraphics[trim = 0mm 2mm 0mm 0mm, clip, width=0.44\textwidth]{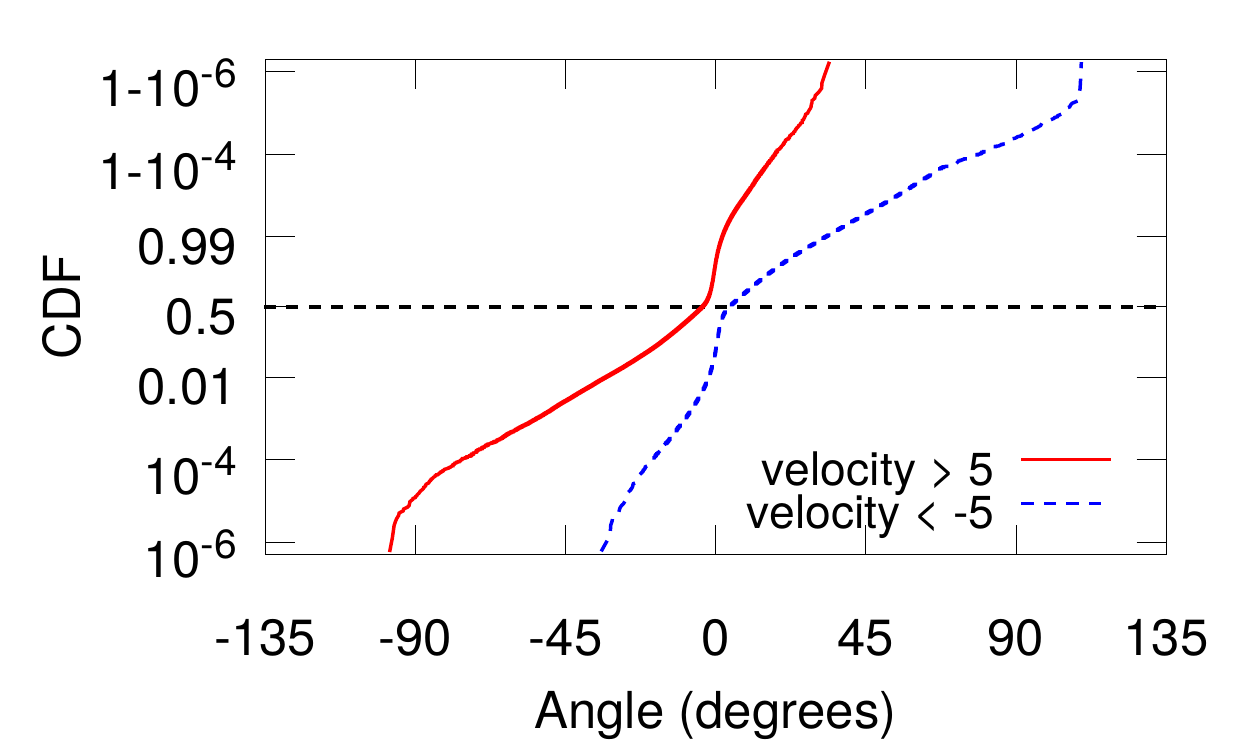}
  \vspace{-10pt}
  \caption{Change in angle over the next $T$=200ms when velocity is greater than $\pm$5$\degree$/s.}
  \label{fig:fig48}
  \vspace{-6pt}
\end{figure}

{\bf Short-term biases based on velocity:}
Not surprisingly, over short time scales,
there is strong correlation between the direction
of head movement and the
future viewing direction.
Figure~\ref{fig:fig48} shows the change in yaw-angle after 200 ms for all videos,
when the user's current velocity was higher than $\pm$ 5$\degree$ per second ($\pm$5$\degree$/s, for short).
Here, a positive (negative) velocity means that a user is turning to the left (right).

\begin{figure}[t]
  \centering
  \includegraphics[trim = 0mm 2mm 0mm 0mm, clip, width=0.44\textwidth]{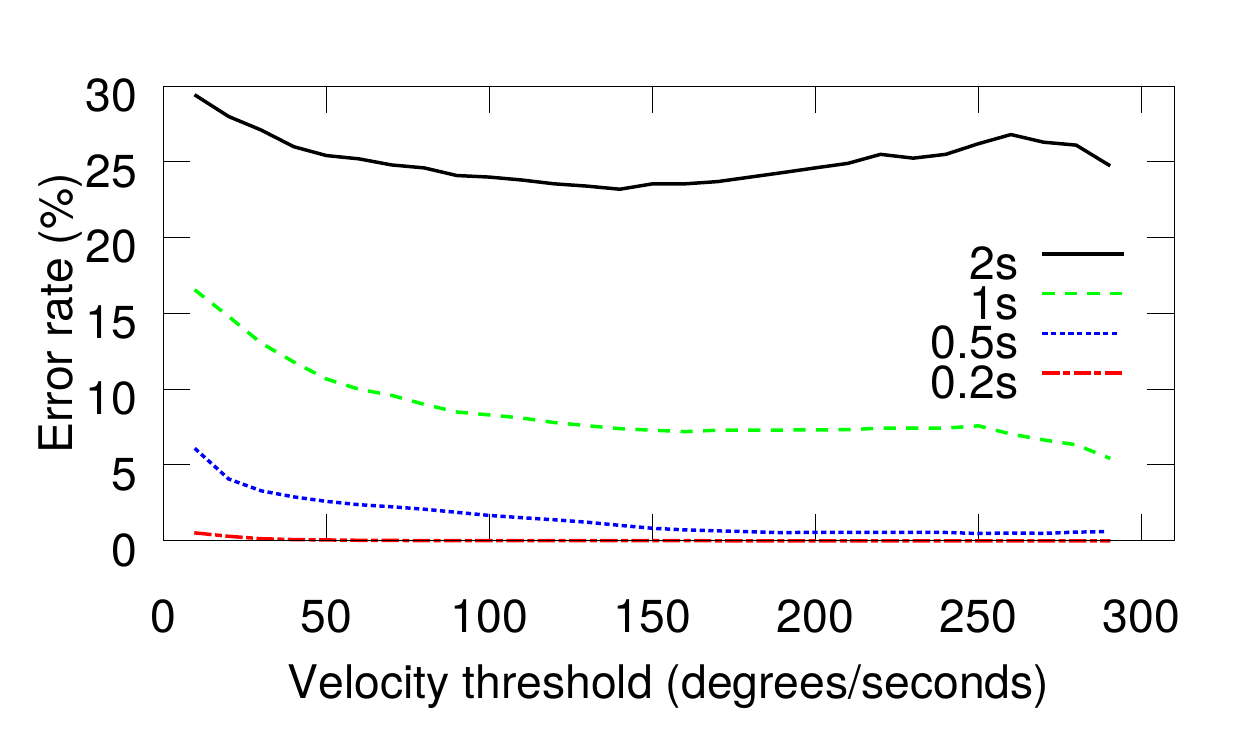}
  \vspace{-10pt}
  \caption{Prediction error rate when using current velocity to predict directional change; different prediction intervals.}
  \label{fig:fig49}
  \vspace{-6pt}
\end{figure}

Even with a small velocity threshold of $\pm$5$\degree$/s
(with 55\% of all readings having a larger directional velocity),
in 97\% of these cases
the eventual view angle is strictly in the direction suggested by the head movement direction 200 ms earlier.
The results are even stronger if using a small (additional) safety angle.
For example, in 99.9\% of the cases,
the direction is no further in the opposite direction than 9$\degree$ and 7$\degree$, respectively.
This is due to head movements being relatively smooth
and shows that there is potential for shrinking the range of angles that the point-of-view is likely to be rotated to next.
For example, referring back to Figure~\ref{fig:yaw-CDFs} we note that the user's view is expected to change no more than $\pm$46$\degree$
99\% of the time.  However, as per the above examples, in more than half of the cases
this range can be cut by 40.2\% and 42.4\%, respectively,
if also taking into account the current velocity.
This shows that for the final enhancements (prefetched over shorter time scales),
it is possible to further improve prediction accuracy.

We have observed clear biases for time scales up to 2 seconds.
Figure~\ref{fig:fig49} shows the prediction error rates when predicting left or right movement based on the
current velocity direction over different time intervals $T$: 0.2s, 0.5s, 1s and 2s.
Here, the error rate is measured as the fraction of cases that the user's viewpoint did not end up in
the same relative direction as suggested by its velocity $T$ seconds earlier.
For these results we varied the velocity threshold.  As expected, the accuracy typically improves with larger thresholds.
(The bump for the 2 second curve can partially be explained by instances where users turn more than 180 degrees.)
While
the tighter thresholds result in smaller error rates,
there are fewer instances that meet these criteria.
Given the relatively flat curves,
a relatively small threshold may therefore often be beneficial,
using a small extra safety margin (as in the example in Figure~\ref{fig:fig49}), of course,
so to protect against fast back-and-forth directional changes.

\begin{figure}[t]
  \centering
  \subfigure[$T=200$ms]{
    \includegraphics[trim = 8mm 2mm 8mm 0mm, width=0.38\textwidth]{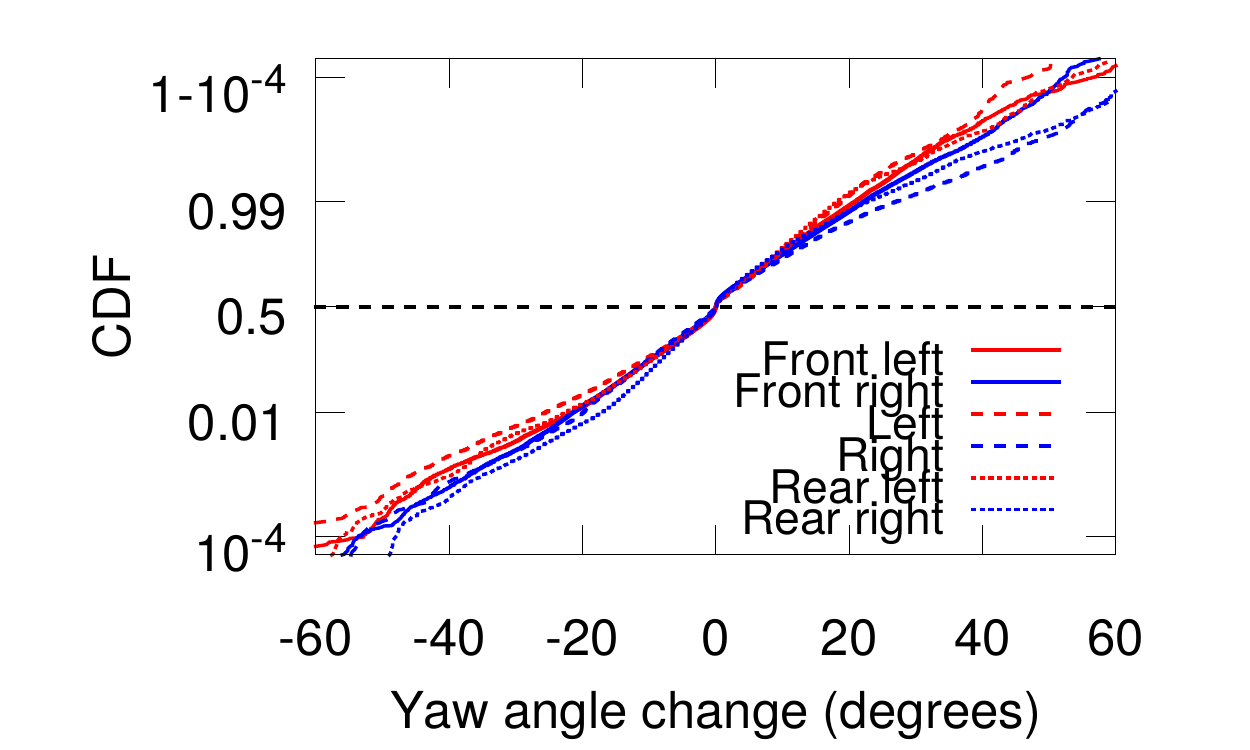}}
  \hspace{-6pt}
  \subfigure[$T=2$s]{
    \includegraphics[trim = 8mm 2mm 8mm 0mm, width=0.38\textwidth]{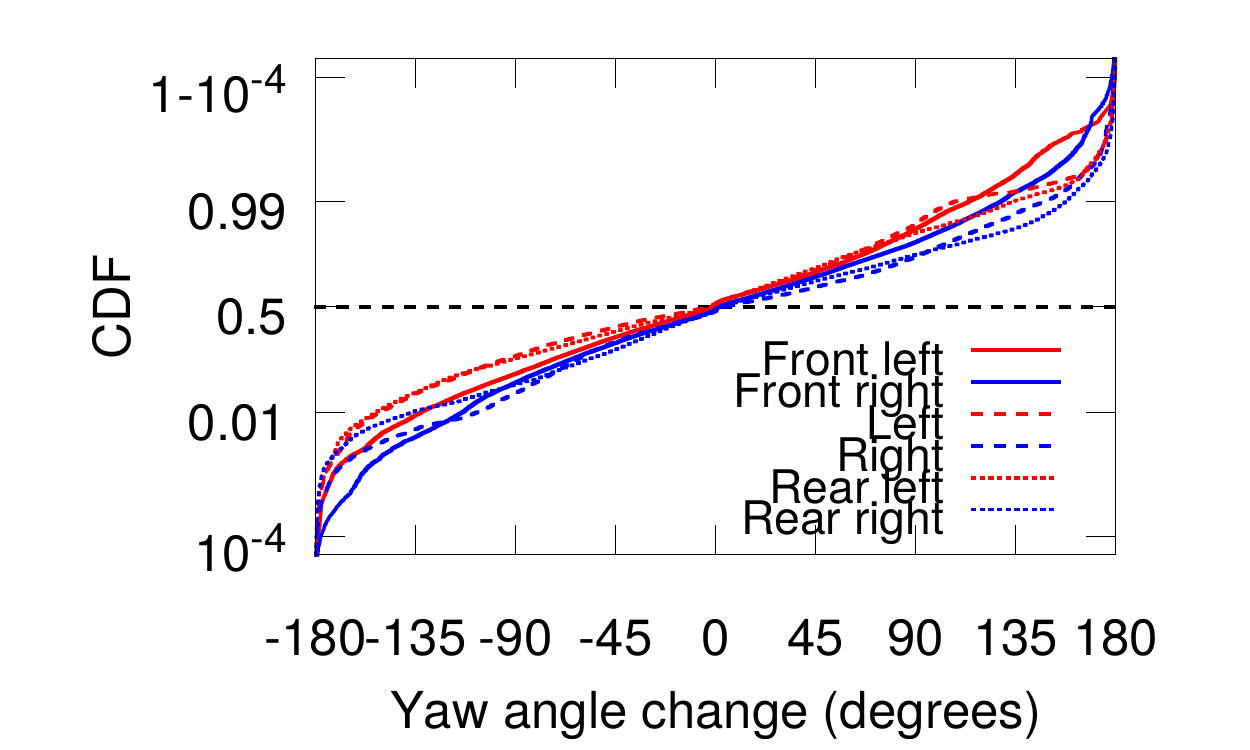}}
    \vspace{-14pt}
  \caption{Change in yaw angle, conditioned on the viewers' location in the sphere. Exploration category.}
  \label{fig:fig410}
  \vspace{-6pt}
\end{figure}

{\bf Biases towards the origin:}
Perhaps the most challenging category to predict is the exploration category.
Interestingly, even for this category,
viewers are more likely to look
in the same direction as when playback began,
also later during the playback (e.g., in Figure~\ref{fig:heat-duration}).
This may suggest that a user that currently
looking to the right of the original starting direction
may be more likely to perform large
rotations to the left, and vice versa.
To glean some insight whether this allows us to improve prediction for the exploration videos,
Figure~\ref{fig:fig410} plots the changes in yaw angle when the viewpoint was
initially located in a certain part of the sphere. Here, the sphere was
divided into 60$\degree$ parts, starting from the 0$\degree$ line.
Results are shown for the change over both 200ms and 2s.

The bias is perhaps easiest observed by comparing the relative ordering of the lines.
For example, the Left lines (currently looking to the left) tend to result larger negative (rightward) head movements
than the Right lines
and vice versa.
When discussing these results,
it should, however, be noted that for the exploration category,
left and right rotations are as likely to occur no matter where the user is looking.
For example, 50\% of all rotation are less than zero and the 50\% are larger than zero for all lines,
limiting the improvements that the above biases may provide.

\begin{figure}[t]
  \centering
  \includegraphics[trim = 0mm 2mm 0mm 0mm, clip, width=0.38\textwidth]{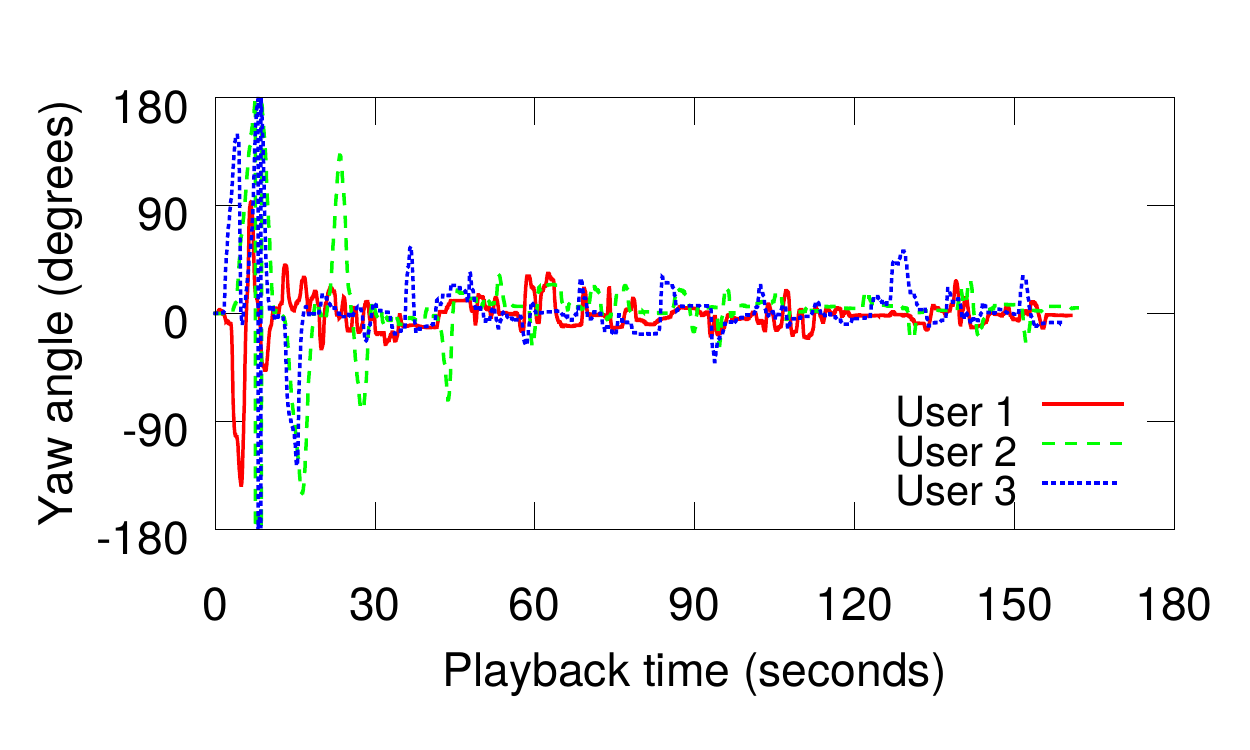}
  \vspace{-14pt}
  \caption{Viewing direction of three example users watching the Christmas scene (from the Static focus category).}
  \label{fig:fig411}
  \vspace{-6pt}
\end{figure}

{\bf Reduced exploration over time:}
Referring back to our discussion of Figure~\ref{fig:angular-diffs},
we have found that users tend to explore
more at
the beginning of a video
and that the head movements typically reduce over time.
This is also illustrated in Figure~\ref{fig:fig411},
where we show the viewing angle of three example viewers watching the Christmas scene video (from the static focus category).
During the first 20 seconds,
there is a lot of head movements
as each user explores almost every yaw angle.
Once the users have learned where the focus should be,
the viewing angle stays relatively stable, centered roughly between $\pm$30$\degree$.
This behavior has
\revtwo{also been}{been}
observed for many users for most of the static focus and rides videos.

\begin{figure}[t]
  \centering
    \subfigure[$T=200$ms]{
      \includegraphics[trim = 2mm 2mm 2mm 0mm, width=0.34\textwidth]{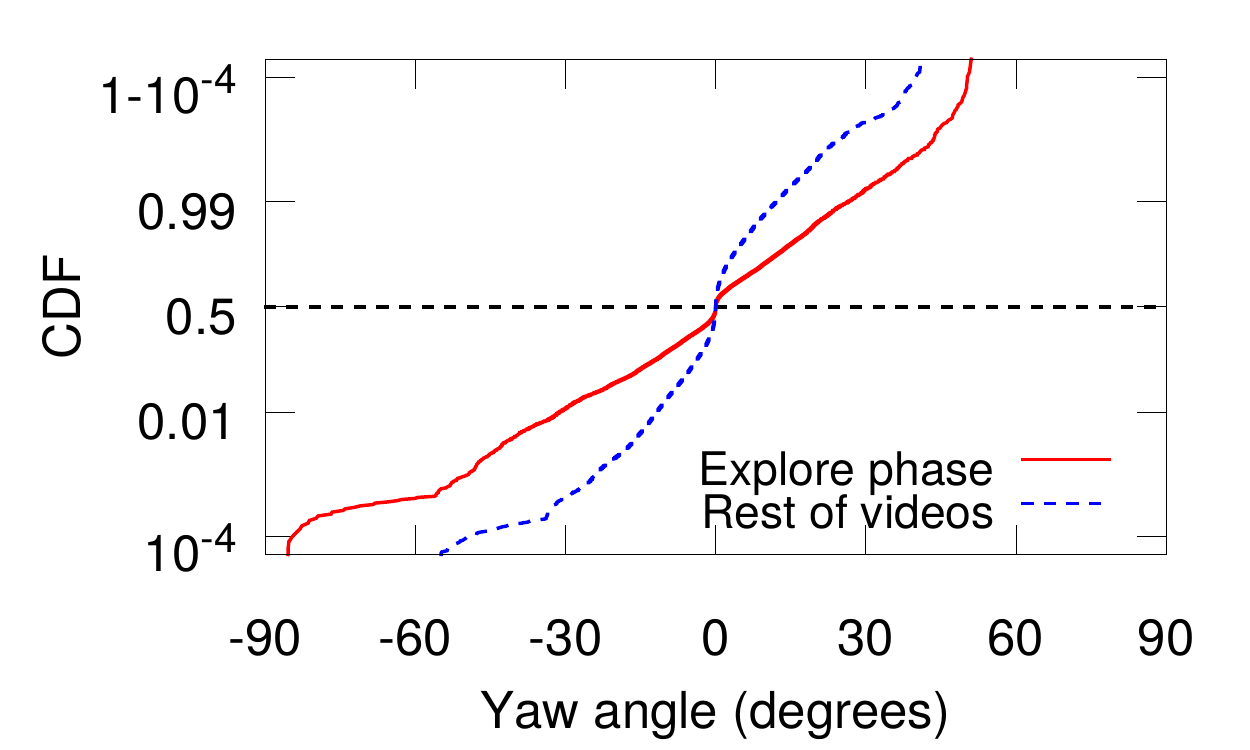}}
    \hspace{-2pt}
    \subfigure[$T=2$s]{
      \includegraphics[trim = 2mm 2mm 2mm 0mm, width=0.34\textwidth]{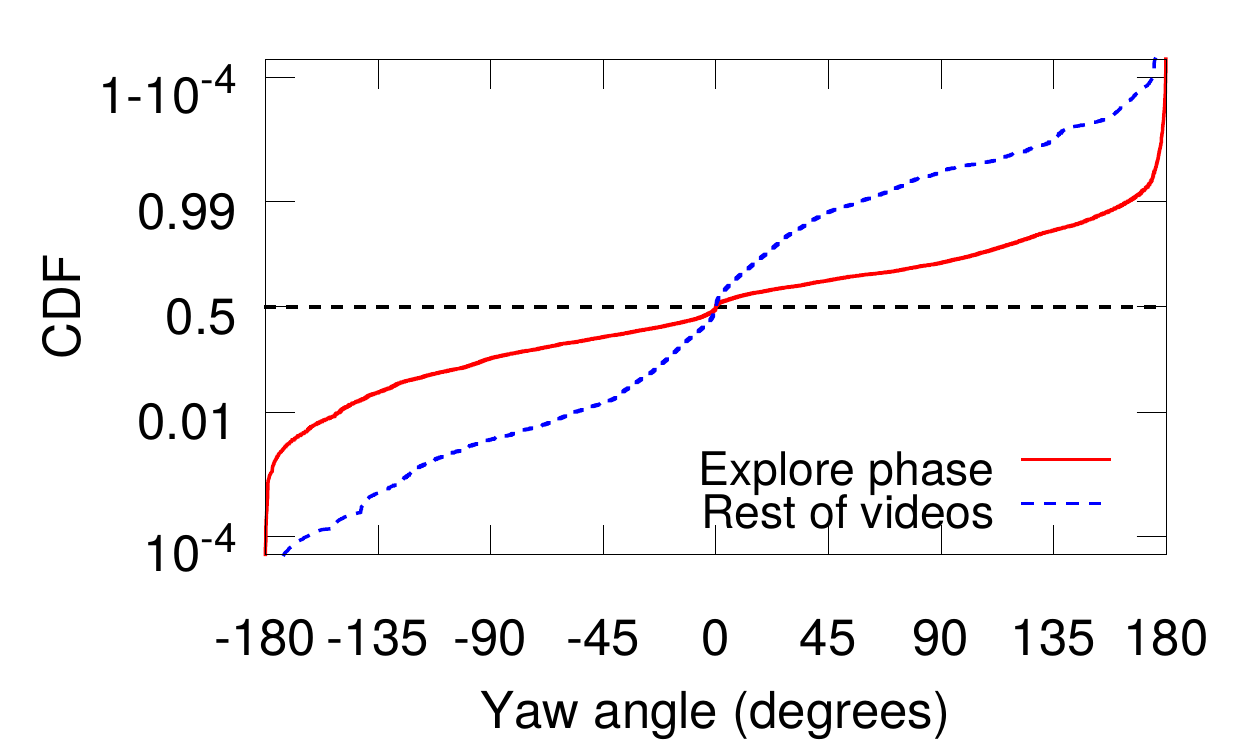}}
        \vspace{-14pt}
    \caption{Change in view angle conditioned on users being in initial 20 second exploration phase or later in the videos.}
    \label{fig:fig412}
    \vspace{-6pt}
\end{figure}

This shows that it may be possible to be more restrictive in the
\revtwo{viewing angles prefetched for later in the video.}{prefetched viewing angles.}
To quantify these effects, consider Figure~\ref{fig:fig412}.
Here, we show the CDFs (on dual-log scales) for the change in yaw angle
during the exploration phase (first 20 seconds) versus the rest of the video.
Results are shown across all users,
when the view change is measured over both 200 ms and 2s.
For the case when measuring changes over 200ms,
99\% of the changes after the exploration phase are smaller than $\pm$15$\degree$ compared to $\pm$39$\degree$ for the exploration phase.
The corresponding 99.9\% ranges are  $\pm$24$\degree$ and $\pm$49$\degree$.
\revtwo{For the case when}{When}
measuring changes over 2s,
the differences are smaller but still noticeable.
For example, the 99\% ranges reduce from $\pm$164$\degree$ to $\pm$77$\degree$
and the 99.9\% ranges from $\pm$178$\degree$ to $\pm$122$\degree$.

Although prediction schemes could
\revtwo{therefore benefit}{benefit}
from adjusting the parameters
used for predictions over the duration of a video, the video content
\revtwo{itself is an important factor here.}{is an important factor.}
Analysis
of individual videos may be required to learn how each video is best divided
so to
\revtwo{best optimize}{optimize}
the prefetching.
In this paper, we primarily compare categories of videos (rather than individual videos)
across different time scales.  Fine grained optimizations such as the ones discussed
in this paragraph (or when discussing moving focus optimizations
that take into account prior viewers of that video)
remain future work.

\section{Related Work}\label{sec:related}

While it is common that the whole 360$\degree$ view is streamed to a client
in a single chunk with a consistent encoding rate across the frame,
different projection and quality adaptive download techniques are
being used by some of the most popular 360$\degree$ HMDs~\cite{ZHLL17}.
We are also not the first to consider the possibility of adaptively downloading different qualities for
different viewing directions~\cite{vengat181,vengat179} or to characterize 360$\degree$ viewer behavior~\cite{CoSS17,LFL+17}.

Some of the most important challenges in this area include user head movement tracking/prediction~\cite{vengat178,vengat179}
and bandwidth management~\cite{vengat181,vengat182}.
Bao et al.~\cite{vengat179} propose a motion-prediction-based transmission scheme that reduces bandwidth consumption
during streaming of 360$\degree$ videos. The prediction scheme is based on viewing behavior data collected similarly
to this paper, but with substantially shorter user sessions and
without categorizing videos.
Yet, their findings suggest that view-dependent 360$\degree$ transmission schemes with motion prediction
can reduce bandwidth consumption by 45\% at the cost of only very small performance degradation.

Others have shown significant storage and bandwidth savings by converting the
equirectangular representation typically used to store 360$\degree$ videos
to a cube map layout~\cite{ref17},
performed QoE-based measurement studies~\cite{SST+17}, and
studied the impact that projection techniques, quantization parameters, and tile patterns
may have on the playback experience and resource requirements~\cite{GrTM17}.
The usage of tiles allows for independent encoding of different regions of a frame.
This helps the decoder, both in terms of potential decoding parallelism and in
selective reconstruction of parts of the frame.

Hosseini and Swaminathan~\cite{vengat181} propose an adaptive tile-based streaming approach for
bandwidth-efficient streaming of 360$\degree$ videos.
The system spatially divides a 360$\degree$ video's equirectangular representation into
several tiles, utilizes MPEG-DASH Spatial Representation Description (SRD)~\cite{ref15}
to describe the spatial relationship between the tiles,
and then prioritizes the tiles based on the user's field of view.
Again, large bandwidth savings (72\%) are demonstrated with only small quality degradation. 
Hamza and Hefeeda~\cite{ref16} illustrate how a client can be implemented to make use of the SRD
to find the available resolution layers, select the most appropriate ones and enabling a
seamless switch when panning between spatial regions of a video.

Rather than using tiles,
Kuzyakov and Pio~\cite{ref18} present a view-dependent streaming technique for 360$\degree$ video that
efficiently utilizes bandwidth
by transforming the original video
into 30 smaller sized versions,
where each version has a specific area
in high quality,
gradually decreasing the quality away from this area.

Several prior works have demonstrated interactive tiled streaming of high-resolution
videos~\cite{ref12,ref13,ref14}.
This includes
delivery of
ultra-high resolution videos based on a
user's region-of-interest~\cite{ref9,ref10}.
Others have used tile-based spatial segmentation to support pan/tilt/zoom interactions during live streaming~\cite{ref12},
for interactive 4k video delivery during the 2014 Commonwealth games~\cite{ref13},
and for a coaching/training application~\cite{ref14}.

Within the context of HAS,
the tradeoff between accurate prefetching based on expected user behavior and eventual
quality of experience
\revtwo{have}{has}
also been analyzed for other forms of interactive media,
including multi-view
video~\cite{SSY+15,CEKP17},
branched video~\cite{KCE+14},
and free-viewpoint video~\cite{XiCL12,HaHe16}.
For all these types of video the simplest method is to download
the entire video (or set of views) but significant download savings are possible by carefully
and adaptively prefetching different areas of the video at different encoding rates.

In contrast to the above 360$\degree$ works,
we consider
longer sessions in which the users watch different categories of videos,
we characterize the viewing behavior within and across these video categories,
and we use an optimization framework to both qualitatively and quantitatively characterize the
general prefetch aggressiveness tradeoff described and analyzed in this paper.

\section{Conclusions}\label{sec:conclusions}

This paper presents a data-driven characterization
of the prefetching aggressiveness tradeoff associated with
how far ahead in time from the current play point prefetching should be done.
In particular, we collect head movement data for 32 users
as they watch a set of 30 videos a total of 439 times.
Using this data, we then characterize
user behavior for 
four different categories of 360$\degree$ videos
(i.e., static focus, moving focus, rides, and exploration)
and provide both qualitative and quantitative insights
regarding how best to address
the prefetching aggressiveness tradeoff.

In general, we have observed significant differences among the video categories with
respect to the predictability of the viewpoint at different time scales,
and hence on the granularity with which view-based prefetching can be done.
As expected, the highest predictability is achieved over short time ranges.
To explore the prefetching aggressiveness tradeoff
more precisely,
we presented an optimization problem that we solved using dynamic programming,
allowing us to study the optimized tradeoff curves.
Based on the insights provided by the optimization model,
we then discuss other system optimizations
and measurement-based biases that can be used to further improve the user's quality of experience.

{\bf Acknowledgements:}
The authors are thankful
to the participants of the study and
to our shepherd Simon Gunkel
and the anonymous reviewers for their feedback. This work was funded in part
by the Swedish Research Council (VR)
and the Natural Sciences and Engineering Research
Council (NSERC) of Canada.

{
  \bibliographystyle{ACM-Reference-Format-Emir}
  \bibliography{references}
}

\commentout{
\appendix

\section{Experiments and Dataset}\label{sec:dataset}

This appendix describes our measurement methodology
and the collected dataset.

\begin{table}
  \centering
  \caption{Summary of videos.  (To watch video, use URL of the form: \url{https://www.youtube.com/watch?v=VideoID}, where VideoID is replaced based on table.}
          {\small
            \begin{tabular}{|l|p{6.7cm}|}
              \hline
              Category & Video Name (Duration, VideoID) \\\hline
              Exploration &
              Zayed Road
              (3:00, \url{uZGrikvGen4}),
              Burj Khalifa
              (2:30, \url{bdq4H1CIehI}),
              Hadrain's Wall
              (3:36, \url{2zeKpeRZ8uA}),
              New York
              (1:59, \url{T3e-GqZ37uc}),
              White House
              (5:16, \url{98U2jdk8OGI}),
              Waldo
              (1:00, \url{hM9Tg_dQkxY}),
              Skyhub
              (4:00, \url{D9-i_F3xYhI})\\
              \hline
              Static &
              Christmas Scene
              (2:49, \url{4qLi-MnkxBY}),
              Boxing 
              (3:29, \url{raKh0OIERew}),
              Elephants
              (2:49, \url{2bpICIClAIg}),
              Mongolia
              (1:52, \url{VuOfQzt2rI0}),
              Orange
              (2:43, \url{i29ITMfLVU0})\\
              \hline
              Moving &
              Christmas Story
              (4:14, \url{XiDRZfeL_hc}),
              Assassin's Creed
              (2:31, \url{a69EoIiYqoE}),
              Clash of Clans
              (1:23, \url{wczdECcwRw0}),
              Frog
              (3:13, \url{sk8hm7DXD5w}),
              Solar System
              (4:32, \url{ZnOTprOTHc8}),
              Invasion
              (4:04, \url{gPUDZPWhiiE})\\
              \hline
              Rides &
              F1
              (1:54, \url{2M0inetghnk}),
              Le Mans
              (3:00, \url{LD4XfM2TZ2k}),
              Roller Coaster
              (2:11, \url{LhfkK6nQSow}),
              Total War
              (1:49, \url{YSBWwnOHvM8}),
              Blue Angels
              (2:30, \url{H6SsB3JYqQg}),
              Ski
              (2:48, \url{kMCYo5rO6RY})\\
              \hline
              Misc. &
              Hockey
              (2:25, \url{8DKVvb17xsM}),
              Tennis
              (4:05, \url{U-_yX4e4Z_w}),
              Avenger
              (2:58, \url{3LSf6_ROCdY}),
              Trike Bike
              (3:14, \url{jU-pZSsYhDk}),
              Temple
              (4:36, \url{Lx14NDttRWo}),
              Cats
              (1:59, \url{0RtmVnD8_XM})\\
              \hline 
          \end{tabular}}
          \label{tab:videos}
          \vspace{-0pt}
\end{table}

\begin{table}
  \centering
  \caption{Summary of videos.  (To watch video, use URL of the form: \url{https://www.youtube.com/watch?v=VideoID}, where VideoID is replaced based on table.}
          {\small
            \begin{tabular}{|l|l|l|l|}
\hline
& Video Name & Duration & VideoID \\\hline
\multirow{7}{*}{\begin{sideways}{Exploration}\end{sideways}}
& Zayed Road
& 3:00& \url{uZGrikvGen4}\\
& Burj Khalifa
& 2:30& \url{bdq4H1CIehI}\\
& Hadrain's Wall
& 3:36& \url{2zeKpeRZ8uA}\\
& New York
& 1:59& \url{T3e-GqZ37uc}\\
& White House
& 5:16 & \url{98U2jdk8OGI}\\
& Waldo
& 1:00& \url{hM9Tg_dQkxY}\\
& Skyhub
& 4:00& \url{D9-i_F3xYhI}\\
\hline
\multirow{5}{*}{\begin{sideways}{Static}\end{sideways}}
& Christmas Scene
& 2:49& \url{4qLi-MnkxBY}\\
& Boxing
& 3:29& \url{raKh0OIERew}\\
& Elephants
& 2:49& \url{2bpICIClAIg}\\
& Mongolia
& 1:52 &\url{VuOfQzt2rI0}\\
& Orange
& 2:43& \url{i29ITMfLVU0}\\
\hline
\multirow{6}{*}{\begin{sideways}{Moving}\end{sideways}}
& Christmas Story
& 4:14& \url{XiDRZfeL_hc}\\
& Assassin's Creed
& 2:31& \url{a69EoIiYqoE}\\
& Clash of Clans
& 1:23& \url{wczdECcwRw0}\\
& Frog
& 3:13& \url{sk8hm7DXD5w}\\
& Solar System
& 4:32& \url{ZnOTprOTHc8}\\
& Invasion
& 4:04& \url{gPUDZPWhiiE}\\
\hline
\multirow{6}{*}{\begin{sideways}{Rides}\end{sideways}}
& F1
& 1:54& \url{2M0inetghnk}\\
& Le Mans
& 3:00& \url{LD4XfM2TZ2k}\\
& Roller Coaster
& 2:11& \url{LhfkK6nQSow}\\
& Total War
& 1:49& \url{YSBWwnOHvM8}\\
& Blue Angels
& 2:30& \url{H6SsB3JYqQg}\\
& Ski
& 2:48& \url{kMCYo5rO6RY}\\
\hline
\multirow{6}{*}{\begin{sideways}{Misc.}\end{sideways}}
& Hockey
& 2:25& \url{8DKVvb17xsM}\\
& Tennis
& 4:05& \url{U-_yX4e4Z_w}\\
& Avenger
& 2:58& \url{3LSf6_ROCdY}\\
& Trike Bike
& 3:14& \url{jU-pZSsYhDk}\\
& Temple
& 4:36& \url{Lx14NDttRWo}\\
& Cats
& 1:59& \url{0RtmVnD8_XM}\\
\hline 
\end{tabular}}
\label{tab:videos}
\vspace{-0pt}
\end{table}
}

\end{document}